\documentclass[preprint,
superscriptaddress,
 amsmath,amssymb,
 prl,
]{revtex4-1}
\usepackage{subcaption}
\usepackage{xcolor}
\usepackage{graphicx}
\usepackage{dcolumn}
\usepackage{bm}

\begin{document}

\title{Correlated Ion Transport and the Gel Phase in Room Temperature Ionic Liquids}

\author{Michael McEldrew}
\affiliation{Department of Chemical Engineering, Massachusetts Institute of Technology, Cambridge, MA, USA}

\author{Zachary A. H. Goodwin}
\affiliation{Department of Chemistry, Imperial College of London, Molecular Sciences Research Hub, White City Campus, Wood Lane, London W12 0BZ, UK}
\affiliation{Thomas Young Centre for Theory and Simulation of Materials, Imperial College of London, South Kensington Campus, London SW7 2AZ, UK}

\author{Hongbo  Zhao}
\affiliation{Department of Chemical Engineering, Massachusetts Institute of Technology, Cambridge, MA, USA}

\author{Martin Z. Bazant}
\affiliation{Department of Chemical Engineering, Massachusetts Institute of Technology, Cambridge, MA, USA}
\affiliation{Department of Mathematics, Massachusetts Institute of Technology, Cambridge, MA, USA}

\author{Alexei A. Kornyshev}
\email{a.kornyshev@imperial.ac.uk}
\affiliation{Thomas Young Centre for Theory and Simulation of Materials, Imperial College of London, South Kensington Campus, London SW7 2AZ, UK}
\affiliation{Department of Chemistry, Imperial College of London, Molecular Sciences Research Hub, White City Campus, Wood Lane, London W12 0BZ, UK}
\affiliation{Institute of Molecular Science and Engineering, Imperial College of London, South Kensington Campus, London SW7 2AZ, UK}

\date{\today}
\begin{abstract}
Here we present a theory of ion aggregation and gelation of room temperature ionic liquids (RTILs). Based on it, we investigate the effect of ion aggregation on correlated ion transport - ionic conductivity and transference numbers - obtaining closed-form expressions for these quantities. The theory depends on the maximum number of associations a cation and anion can form, and the strength of their association. To validate the presented theory, we perform molecular dynamics simulations on several RTILs, and a range of temperatures for one RTIL. The simulations indicate the formation of large clusters, even percolating through the system under certain circumstances, thus forming a gel, with the theory accurately describing the obtained cluster distributions in all cases. We discuss the possibility of observing a gel phase in neat RTILs, which has hitherto not been discussed in any detail.

\begin{center}
    \includegraphics[clip, width=0.6\textwidth]{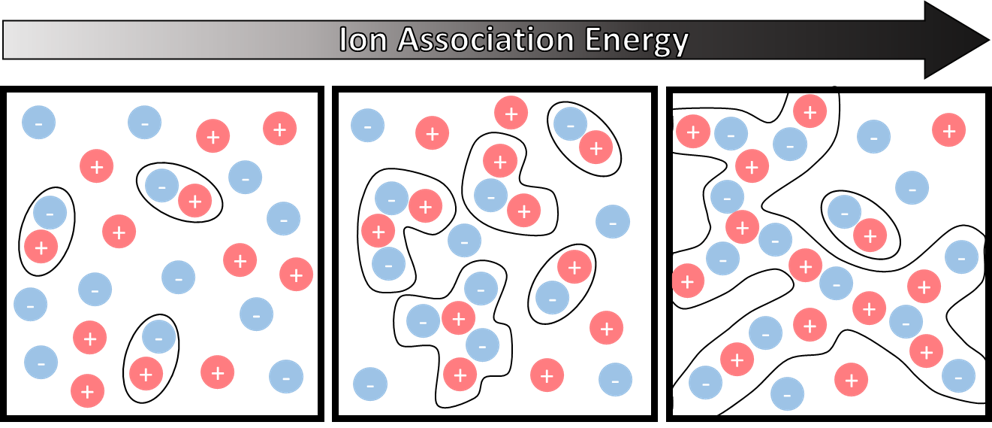}
\end{center}

\end{abstract}

\maketitle

\textit{Introduction}-Neat room temperature ionic liquids (RTILs) are electrolytes without any solvent: they are composed of molecular cations and anions that are sufficiently asymmetric and bulky to prevent solidification at room temperature~\cite{Hallett2011,Welton1999,wasserscheid2008,Fedorov2014,Hermann2008}. RTILs are excellent solvents, have extremely low vapour pressures~\cite{Hallett2011,Welton1999,wasserscheid2008}, and are of great interest in energy storage devices because of their ability to withstand larger voltages without electrochemical decomposition~\cite{Fedorov2014,Kondrat2016}. However, owing to their high ionic concentrations and complex molecular nature, \textit{simple}, chemically-specific theoretical descriptions of RTILs have remained somewhat elusive~\cite{son2020ion}. 

The conductivity of RTILs approximately follows that predicted by the Nernst-Einstein relation with some suppression factor. This has been interpreted as the ionicity of RTILs~\cite{MacFarlane2009,Kirchner2014}, with the reduction factor being attributed to the presence of ion pairs in the system which do not contribute to conductivity~\cite{Zhang2015}. In fact, pioneering molecular dynamics simulations of Feng \textit{et al.}~\cite{feng2019free} investigated the role of ion clustering in the conductivity, where it was found that ``free" ions contribute the most.

Although the picture of RTILs as a mixture of ion pairs and free ions might be useful conceptually~\cite{lee2014room,goodwin2017underscreening,goodwin2017mean,Chen2017,adar2017bjerrum}, it is natural to question the validity of such a picture in such a highly concentrated system~\cite{Hayes2015}. Surface force apparatus measurements performed in RTILs, when interpreted in terms of the DLVO theory, has reported extraordinary long screening lengths~\cite{Smith2016,Gebbie2013,Gebbie2015} consistent with $\ll$ 1\% of free ions~\cite{Gebbie2013,Gebbie2015}. While other methods of probing the number of free ions suggest significantly more free ions (15-25\%) for common RTILs~\cite{feng2019free,Chen2017}. With such significant degree of ion association, we would undoubtedly expect high order ionic clusters to be present~\cite{avni2020charge,feng2019free,france2019}. 

Thus, simple theoretical descriptions of RTILs beyond ion pairs are desired, but scarce. Recently Ref.~\citenum{mceldrew2020theory} developed a thermodynamic theory of ion aggregation with arbitrary sized ionic clusters which predicted the emergence of a percolating ionic gel in super-concentrated electrolytes. It built off the theory of aggregation and gelation in polymer physics~\cite{flory1942thermodynamics,tanaka1989,tanaka1990thermodynamic,tanaka1994,tanaka1995,ishida1997,tanaka1998,tanaka1999,tanaka2002}, where clustering and percolation of aggregates is known to occur. Simulations of super-concentrated electrolytes have shown the presence of percolating ion networks~\cite{borodin2017liquid,choi2018graph,jeon2020modeling}. Moreover, an elastic response has been measured in certain RTILs~\cite{makino2008viscoelastic,tao2015rheology,Elhamarnah2019,tao2015rheology,Shakeel2019,Shamim2010,Choi2014}, which can be indicative of the formation of a gel. Thus, it seems possible, if not probable, that extensive ion aggregation, and even gelation, is present in RTILs. 

Here we study the limiting case of that~\cite{mceldrew2020theory} theory: solvent-free RTILs modelled as an incompressible solution composed of solely ions. This naturally leads to a physically transparent framework of correlated ion-transport in RTILs, yielding coupled flux constitutive relations for diffusion and a modified Nernst-Eistein (NE) equation for ionic conductivity (also seen in Ref.~\citenum{leonard2018}). Our analysis allows us to determine the importance of clusters' contribution to ionic conductivity. We perform molecular dynamics (MD) simulations of 6 RTILs, and develop a general association criteria that is used to determine the cluster distribution of the simulated RTILs. From MD simulations, we determine the handful of parameters needed to compute the cluster distribution from our theory. We find the independently computed MD cluster distribution matches the theoretically computed distribution,with parameters derived from MD, extremely well (note we only compare quantities based off the cluster distribution between MD and theory - we do not investigate transport using MD here). Finally, we discuss the possibility of observing the gel phase in RTILs. In the Appendix, we have a table of symbols, derivations of all equations and further details of the simulations.

\begin{figure}[hbt!]
\centering
\includegraphics[width=0.49\textwidth]{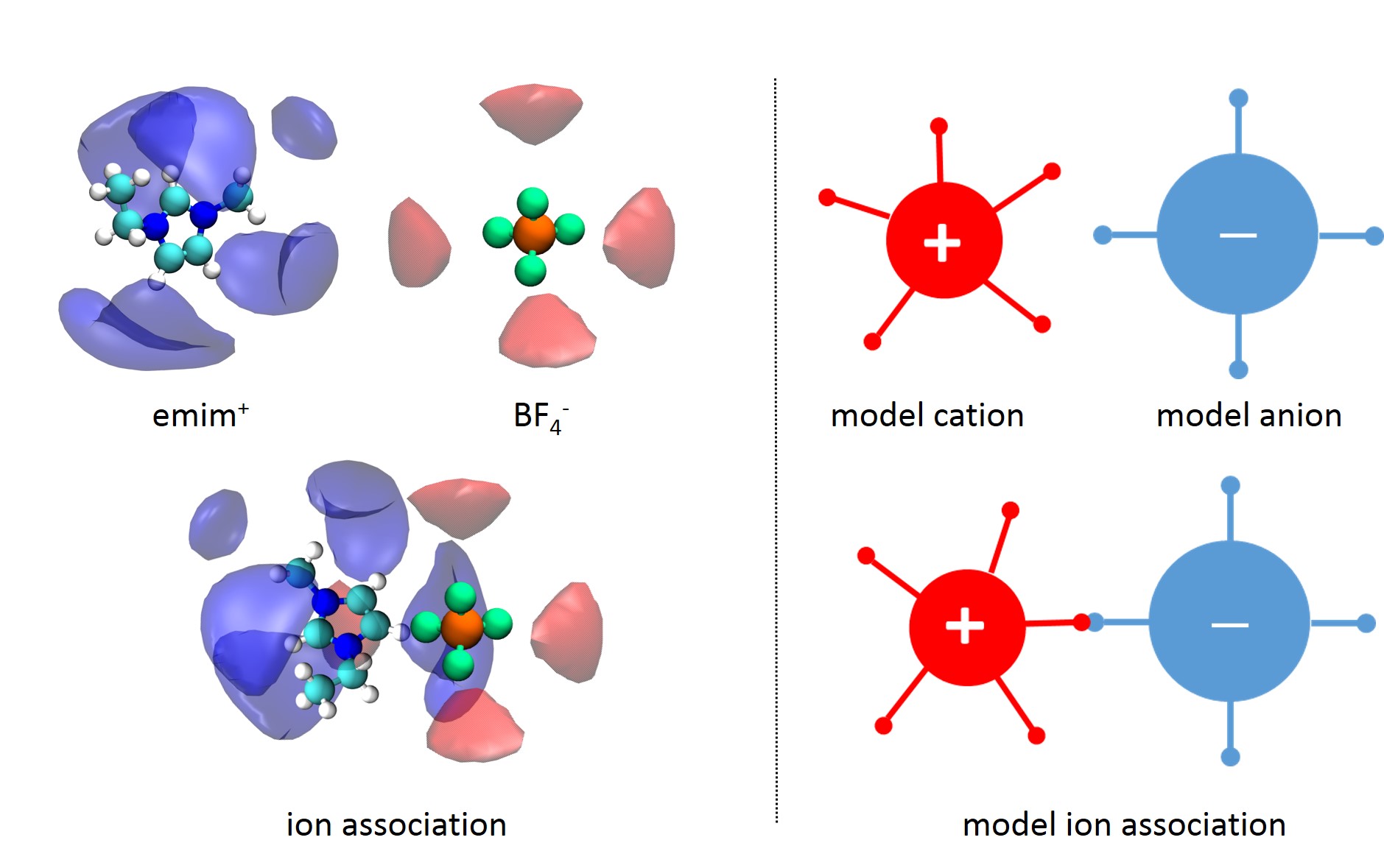}
\caption{(Left) Schematic of example RTIL ions, emim$^+$ and BF$_4^-$ with iso-density surface (corresponding to 2$\times$ the bulk density) of BF$_4^-$ around emim$^+$, and emim$^+$ around BF$_4^-$. The former shows five distinct regions of preferred location for BF$_4^-$ around emim$^+$ ($f_+=5$), and the latter four distinct regions of preferred location for emim$^+$ around BF$_4^-$ ($f_-=4$). Ionic associations, like the one shown in the bottom left, are formed when BF$_4^-$ and emim$^+$ are found in each others high density regions. (Right) A cartoon depicting the `bonding' of cations and anions shown in the left panel.}
\label{fig:clust_pic}
\end{figure}

\textit{Thermodynamics of Ion Clustering}- We employ a Flory-like lattice fluid free energy of mixing~\cite{flory1942thermodynamics} used extensively for polydisperse mixtures of thermoreversibly-associating polymer mixtures~\cite{tanaka1989,tanaka1990thermodynamic,tanaka1994,tanaka1995,ishida1997,tanaka1998,tanaka1999,tanaka2002}:
\begin{align}
\beta \Delta F &= \sum_{lm} \left[N_{lm}\ln \left( \phi_{lm} \right)+N_{lm}\Delta_{lm}\right] \nonumber\\
&+ \Delta^{gel}_+ N^{gel}_+ + \Delta^{gel}_- N^{gel}_-. 
\label{eq:F}
\end{align}
Here $N_{lm}$ and $\phi_{lm}$ are the number and volume fraction of $lm$ clusters with $l$ cations and $m$ anions, respectively; $\Delta_{lm}$ is the free energy of formation of a rank $lm$ cluster, which can have contributions from the combinatorial entropy, bonding energy, and configurational entropy; $\Delta^{gel}_i$ and $N^{gel}_i$ are the free energy changes of $i$ associating to the gel and number of $i$ in the gel, respectively~\cite{flory1942thermodynamics,flory1953principles,tanaka1989}. The free energy treats the electrolyte as an ideal mixture of non-interacting ionic clusters. In this way, the correlations beyond mean-field electrostatics are being treated via the formation of clusters. Our model does not account for correlations beyond the formation of clusters. Thus, our model relies on the assumption that the majority of the excess electrostatic energy of the mixture is modelled via the formation of the ionic clusters.

The volume fraction of an ion cluster of rank $lm$ is expressed by $\phi_{lm} = (l\xi_+ + m\xi_-)N_{lm}/\Omega$, where $\xi_i$ is the volume of an ion relative to the volume of a lattice site and $\Omega$ denotes the total number of lattice sites (and is a proxy for the total volume of the mixture). The volume of a lattice site is arbitrary, and can either be taken to be the size of a cation, i.e. $\xi_+=1$ or an anion, i.e. $\xi_-=1$. The system is considered to be incompressible (this is a reasonable approximation for RTILs, as the percentage of space occupied by voids is of the order of 10 \%~\cite{Yu2012freevolume} at ambient conditions). A cation can associate to at most $f_+$ anions, and anions can associate to at most  $f_-$ cations. $f_+$ and $f_-$ are referred to as functionalities of cations and anions, respectively.

The free energy is minimized by establishing equilibria between all ion clusters and free ions, as derived in the Appendix~\cite{mceldrew2020theory}. This process yields the following relation:
\begin{align}
    \phi_{lm}=K_{lm}\phi_{10}^l\phi_{01}^m,
    \label{eq:cd1}
\end{align}
where $\phi_{10}$ and $\phi_{01}$ are the volume fractions of free cations and anions, respectively, and the equilibrium constant is $K_{lm}=\exp \left( l+m-1-\Delta_{lm} \right)$. The explicit form of $\Delta_{lm}$ was derived in Ref.~\cite{mceldrew2020theory}, and is detailed in the Appendix. It can be inserted into Eq.~\eqref{eq:cd1} yielding the thermodynamically consistent cluster distribution
\begin{align}
    \tilde{c}_{lm}=\frac{W_{lm}}{\lambda}  \left(\lambda\psi_{10}\right)^l \left(\lambda\psi_{10}\right)^m.
    \label{eq:clust}
\end{align}
where $\tilde{c}_{lm}$ is the dimensionless concentration of clusters of rank $lm$ ($\#$ per lattice site), $\lambda = \exp(-\beta \Delta F_{+-})$ is the ionic association constant (for more on this definition see the Appendix) which depends on the free energy of ion association ($\Delta F_{+-}$), $\psi_{10}=f_+ \phi_{10}/\xi_+$ and $\psi_{01}=f_- \phi_{01}/\xi_-$ are the dimensionless concentrations of available bonding sites of cations and anions, respectively, and $W_{lm}$ is the combinatorial multiplicity of $lm$ clusters (number of different ways to form a rank $lm$ cluster), given by by~\cite{stockmayer1952molecular} 
\begin{align}
    W_{lm}=\frac{(f_+l-l)!(f_-m-m)!}{l!m!(f_+l-l-m+1)!(f_-m-m-l+1)!}.
\end{align}

In  Eq.~\eqref{eq:clust}, $\tilde{c}_{lm}$ is written in terms of the volume fraction of free cations ($\phi_{10}$) and anions ($\phi_{01}$), but, in principle, $\phi_{10}$ and $\phi_{01}$ are unknown. We would like to know the distribution of clusters in terms of the overall volume fraction of species, $\phi_+$ (volume fraction of cations) \& $\phi_-$ (volume fraction of anions). We accomplish this by introducing ion association probabilities, $p_{ij}$, which is the probability that an association site of species $i$ is bound to species $j$, where $i$ and $j$ in a binary RTIL correspond to cations ($+$) and anions ($-$). In this way, the volume fraction of free cations can be written as $\phi_{10} = \phi_+(1 - p_{+-})^{f_+}$ and free anions as $\phi_{01} = \phi_-(1 - p_{-+})^{f_-}$. Furthermore, we can determine the association probabilities through $f_+ p_{+-}=f_- p_{-+}$ (conservation of associations) and $\lambda \zeta = p_{+-}p_{-+}/(1-p_{+-})(1-p_{-+})$ (mass action law), where $\zeta=f_\pm p_{\pm\mp}\tilde{c}_{salt}$ is dimensionless concentration of associations with $\tilde{c}_{salt} = 1/(\xi_+ + \xi_-)$ denoting the dimensionless concentration of salt ($\#$ per lattice site). This closes the system of equations in the pre-gel regime, and we can solve for the cluster distribution.

If we assume that the RTIL is symmetrically associating ($f_+=f_-=f$,$p_{+-}=p_{-+}=p$), then a \textit{simple} analytical form can be obtained
\begin{align}
p=\frac{1+2\tilde{c}_{salt}f\lambda-\sqrt{1+4\tilde{c}_{salt}f\lambda}}{2\tilde{c}_{salt}f\lambda}.
\label{eq:simpp}
\end{align}
Analytical solutions for the asymmetric cases are written in the Appendix, which should be more typical for RTILs. For symmetric RTILs, the fraction of free ions, $\alpha$, is simply
\begin{equation}
\alpha = \alpha_{01} + \alpha_{10} =\dfrac{\tilde{c}_{01} + \tilde{c}_{10}}{2\tilde{c}_{salt}} = \Bigg[\dfrac{\sqrt{1 + 4\tilde{c}_{salt}f\lambda} - 1}{2\tilde{c}_{salt}f\lambda} \Bigg]^f.
\end{equation}
Here $\alpha_{01/10}$ are the fractions of free anions/cations, which are related to the dimensionless concentration of anions/cations through $1/2\tilde{c}_{salt}$. For large values of $\lambda$ the fraction of free ions reduces to $\alpha = [2/\tilde{c}_{salt}f\lambda]^{f/2}$, which tends to zero. 

If the functionalities of both ions are greater than 1 and the association probabilities exceed a certain threshold, then the RTIL can form a percolating ionic gel~\cite{mceldrew2020theory}. The criterion that determines this threshold can be seen plainly by examining the weight average degree of aggregation (the expected cluster size for a given ion), $\bar{n}$, can be expressed analytically as 
\begin{equation}
    \bar{n} = \frac{\sum_{lm}(l+m)^2c_{lm}}{\sum_{lm}(l+m)c_{lm}}=\dfrac{1 + p}{1 - (f -1)p}.
    \label{eq:barn}
\end{equation}
$\bar{n}$ diverges when $p^* = 1/(f-1)$, which defines the gel point. This critical probability for gelation corresponds to a critical association strength for gelation of
\begin{equation}
    \tilde{c}_{salt}\lambda^*=\frac{(f-1)}{f(f-2)^2}.
\end{equation}
In the Appendix, we give the general critical association strength for asymmetrically associating ionic liquids. For $\lambda>\lambda^*$, the RTIL will form a gel. The critical association constant decreases as a function of $f$ (for $f > 2$). Therefore, RTILs that can form more bonds with oppositely charged ions will tend to gel more readily. 

If the RTIL does form a gel, partitioning of species into the sol ($\phi^{sol}_\pm$) and gel ($\phi^{gel}_\pm$) phases must be performed and separate association probabilities for species in the sol ($p_{\pm\mp}^{sol}$) must be defined. These probabilities are determined from Flory's criterion that the free ion volume fractions can be written equivalently in terms of overall quantities and sol quantities~\cite{flory1941molecular,flory1941molecular2}. The specific procedure associated with determining these probabilities is detailed explicitly in the Appendix. 

\begin{figure*}[hbt!]
\centering
\includegraphics[width=1\textwidth]{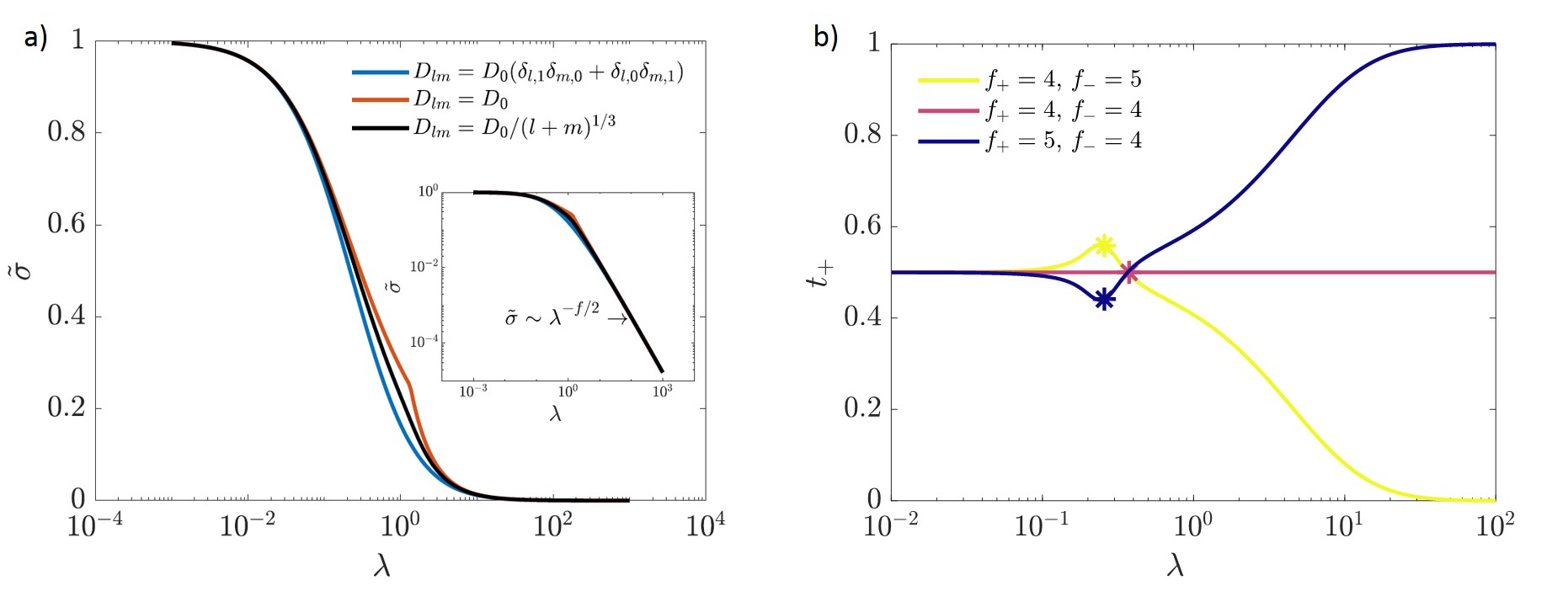}
\caption{(a) Dimensionless ionic conductivity, $\tilde{\sigma}$, as a function of the ionic association constant, $\lambda$ for $f=4$. We use three different functional forms for the cluster diffusivity, $D_{lm}$, written in the legend of the left panel. In the inset, we plot the same curves on a log-log scale. (b) Cation transference number using Eq.~\eqref{eq:tp}, with $D_{lm}=D_0/(l+m)^{1/3}$, for various ion functionalities. The sums in Eq.~\eqref{eq:tp} were cut off for clusters containing less than one hundred ions. The asterisks mark the gel point.}
\label{fig:cond}
\end{figure*}

\textit{Correlated Ion Transport}-Our thermodynamic description treats RTILs as an ideal mixture of polydisperse ionic clusters. A consistent picture of ion transport would be independent diffusion of the ionic clusters. Of course, the assumption of independent cluster diffusion neglects the electrostatic interactions between clusters, which would inevitably result in molecular ``friction" between clusters. However, as we have mentioned, the ionic associations between ions to form clusters should capture most of the electrostatic correlations beyond mean-field. Therefore, the simple presence of clusters results directly to strong correlations between ions, which is likely to dominate in strongly associated RTILs.

The molar flux of a cluster of rank $lm$, $\mathbf{j}_{lm}$, is just simply proportional to the gradient of its electrochemical potential, $\bar{\mu}_{lm}$
\begin{align}
    \mathbf{j}_{lm}=-\beta D_{lm}c_{lm}\nabla\bar{\mu}_{lm},
\end{align}
where $D_{lm}$ is the diffusivity of the rank $lm$ cluster. Note that we have excluded the ``$\sim$" on $c_{lm}$, as this is the concentration of clusters in terms of mole/volume as opposed to $\#$ per lattice site. The total flux of anions and cations can be computed by adding the contributions of all of the clusters ($\mathbf{j}_+=\sum_{lm}l\mathbf{j}_{lm}$ \& $\mathbf{j}_-=\sum_{lm}m\mathbf{j}_{lm}$).

The electrochemical potential of a rank $lm$ cluster is simply $\bar{\mu}_{lm}=\mu_{lm}+e(l-m)\Psi$, where $\Psi$ is the electrostatic potential. Moreover, assuming a local equilibrium among ion clusters, we may write the chemical potential of a rank $lm$ cluster as $\mu_{lm}=l \mu_++m\mu_-$, where $\mu_\pm$ is the chemical potential of a (free) cation or anion. Writing the flux relations in matrix vector notation, also introducing the current density [$\mathbf{i}=e(\mathbf{j}_+-\mathbf{j}_-)$], we have
\begin{widetext}
\begin{align}
    \begin{bmatrix}\mathbf{j}_+ \\ \mathbf{j}_- \\ \mathbf{i} \end{bmatrix}= -2\beta c_{salt}\begin{bmatrix}\mathcal{D}_{++} &\mathcal{D}_{+-} & e(\mathcal{D}_{++}-\mathcal{D}_{+-})\\ \mathcal{D}_{+-} &\mathcal{D}_{--} & e(\mathcal{D}_{--}-\mathcal{D}_{+-})  \\ e(\mathcal{D}_{++}-\mathcal{D}_{+-}) & e(\mathcal{D}_{+-}-\mathcal{D}_{--}) & e(\mathcal{D}_{++}-2\mathcal{D}_{+-}+\mathcal{D}_{--}) \end{bmatrix}\begin{bmatrix}\nabla\mu_+ \\ \nabla\mu_- \\ \nabla \Psi \end{bmatrix}
    \label{eq:flux}
\end{align}
\end{widetext}
where $\mathcal{D}_{++}=\sum_{lm}l^2D_{lm}\alpha_{lm}$, $\mathcal{D}_{--}=\sum_{lm}m^2D_{lm}\alpha_{lm}$, $\mathcal{D}_{+-}=\sum_{lm}lmD_{lm}\alpha_{lm}$ are binary diffusion coefficients, and $\alpha_{lm}=c_{lm}/2c_{salt}$ is the fraction clusters of rank $lm$. The binary diffusion coefficients are elements of the diffusivity tensor, $\underline{\underline{\mathcal{D}}}$, which is symmetric and positive-definite (via the Cauchy-Schwartz theorem) in accordance with the Onsager reciprocal relations~\cite{onsager1931reciprocal}.

In the absence of chemical potential gradients, the ionic current is proportional to the gradient of the electrostatic potential via the ionic conductivity
\begin{align}
    \sigma &= \frac{2e^2c_{salt}}{k_B T} \sum_{lm} (l-m)^{2} \alpha_{lm} D_{lm} \nonumber \\
    &=\frac{2e^2c_{salt}}{k_B T}(\mathcal{D}_{++}-2\mathcal{D}_{+-}+\mathcal{D}_{--})
    \label{eq:sig1}
\end{align}
In the limit of all ions being free and the diffusion coefficient of the ions being related to the unimpeded diffusion of ions (Stokes-Einstein), one would predict a much larger conductivity than is observed in simulations and experiments. Correlations between ions leads to the formation of clusters, which diminishes the concentration of free ions. Therefore, the electrical conductivity is reduced from the idealised NE relation in the presence of correlations, as has been well known for a long time in dilute electrolytes. 

Similarly, the ion transference numbers, $t_\pm$, are defined as 
\begin{align}
    t_+ = \frac{\sum_{lm} l(l-m) \alpha_{lm} D_{lm}}{\sum_{lm} (l-m)^{2} \alpha_{lm} D_{lm}}=\frac{\mathcal{D}_{++}-\mathcal{D}_{+-}}{\mathcal{D}_{++}+\mathcal{D}_{--}-2\mathcal{D}_{+-}}
    \label{eq:tp}
\end{align}
for cations, and 
\begin{align}
    t_- = \frac{\sum_{lm} m(l-m) \alpha_{lm} D_{lm}}{\sum_{lm} (l-m)^{2} \alpha_{lm} D_{lm}}=\frac{\mathcal{D}_{--}-\mathcal{D}_{+-}}{\mathcal{D}_{++}+\mathcal{D}_{--}-2\mathcal{D}_{+-}}
\end{align}
for anions. As we shall show, correlations between ions can give rise to large deviations of transference numbers between ions, provided some asymmetry is present in the system. Similar expressions (for conductivity and transference numbers) were recently stated in Refs.~\citenum{france2019} and ~\citenum{mceldrew2020theory}.

To solve for the conductivity, transference numbers and each of the elements in $\underline{\underline{\mathcal{D}}}$, we must know $D_{lm}$. This is generally not known for RTILs, but we may inspect various physical approximations of $D_{lm}$ and see how this affects the predictions for the ionic conductivity.

For example, a lower bound on the conductivity can be obtained if we assume that only free ions are mobile, i.e. the diffusion constants of all other clusters are zero: $D_{lm}=D_0(\delta_{l,1}\delta_{m,0}+\delta_{m,1}\delta_{l,0})$, where $D_0$ is the self-diffusion coefficient of a free ion. In that case, the conductivity, non-dimensionalized by the factor, $2e^2 c_{salt}D_0/k_BT$, is given by
\begin{align}
    \tilde{\sigma}_{min} = \alpha = (1 - p)^f
\end{align}
for the symmetrically associating RTIL (the asymmetric case is shown in the Appendix). On the other hand, an upper bound can be obtained if we assume the diffusion constants of all clusters are that of a free ion ($D_{lm}=D_0$), which gives
\begin{equation}
    \tilde{\sigma}_{max} = \dfrac{1 - p}{1 + (f - 1)p}.
    \label{eq:sigmax}
\end{equation}
for the symmetrically associating RTIL. Note that if gel cluster is formed, we assume it does not contribute to the conductivity, as it is finite in size. Thus, when we surpass the critical gel point, we must use the sol probability, $p^{sol}$ and multiply Eq.~\eqref{eq:sigmax} by the fraction of the ions in the sol, $w_\pm^{sol}=\phi_\pm^{sol}/\phi_\pm$. These approximations are shown in Fig.~\ref{fig:cond}, where we also include $D_{lm} = D_0/(l+m)^{1/3}$, in the spirit of Stokes-Einstein theory, which assumes that the diffusion constant is inversely proportional to the characteristic radius of the cluster (proportional to the cube root of its volume). We expect this form for $D_{lm}$ to be an optimistic approximation. Note that scaling from polymer physics would lead to $D_{lm} = D_0/\sqrt{l+m}$~\cite{tanaka2011polymer}, but we suspect such a scaling might only be valid for high order clusters in dilute solutions, where the presence of such clusters would be minuscule.

In the post-gel regime there can be very few free ions~\cite{mceldrew2020theory}, and the correlated transport model proposed here could break down. In dilute electrolytes, a Kohlrausch-like correction to mobility reflects the collective character of their motion~\cite{fouss1957}. Indeed if a small amount of free ions are moving in the sea of clusters/gel, one may have a temptation to take into account Coulomb interactions between those free ions~\cite{mceldrew2020theory}. The ions, however, are not moving unimpeded between the clusters/gel, they inter-convert between the free and clustered states. This is similar to electrons in intrinsic semiconductors that are thermally exited from valence to conduction band or of a solid electrolyte where there is a hopping mechanism for diffusion~\cite{Fedorov2014,feng2019free}. A Kohlrausch-kind correction would not be justified for RTILs then, and the simple, `single-particle' transport model of Eq.~\eqref{eq:sig1}, which accounts for collective nature of transport trough the inter-conversion of ions between free and clustered states, would be the most natural starting point. As we will see below, it is verified by molecular dynamic simulations.

In Fig.~\ref{fig:cond}a, we have plotted $\tilde{\sigma}$ as a function of $\lambda$ for symmetrically associating ions for all of the approximations outlined. We can see as the strength of the ionic association parameter is increase, it is evident that the conductivity monotonically decays for all cases. For very large association constants, we find a scaling of $\tilde{\sigma} \propto \lambda^{-f/2}_{\pm}$ for each of the models, owing to the conductivity being dominated by free ions when the RTIL has gelled significantly. Interestingly, we find that all three approximations for the cluster diffusion coefficients yield similar values of $\tilde{\sigma}$. The largest deviation between the models occurs around the gel point, as identifiable from the kink in the upper bound estimate. In Refs.~\citenum{feng2019free} and \citenum{france2019}, it was found that the free ions dominate the conductivity. We have demonstrated, through calculating the contributions from all clusters, that in fact, the free ions are always the dominant contribution~\cite{mceldrew2020theory}. The symmetrically associating RTIL approximation has a larger tendency to form neutral clusters, as opposed to charged ones. In the Appendix, we show additional examples of conductivity for various asymmetric RTILs, where charged clusters are more prevalent.

For the symmetric RTIL case, the transference numbers are trivially 1/2. Actually, many RTILs tend to have transference numbers close to 1/2 ($0.45<t_\pm<0.55$~\cite{galinski2006ionic,kowsari2008molecular}. In the asymmetric case, however, our model predicts that transference numbers deviate from 1/2 dramatically, especially when there are large correlations between ions (large $\lambda$). Sure enough, certain RTILs have been measured to have very asymmetric transference numbers ($t_\pm<0.4$ or $t_\pm>0.6$)~\cite{kowsari2008molecular}. 

In Fig.~\ref{fig:cond}b we show how the cation transference number depends on $\lambda$ for different cation functionalities [using Eq.~\eqref{eq:tp}, with $D_{lm}=D_0/(l+m)^{1/3}$]. Comments which follow apply to asymmetric functionalities, owing to the symmetric case being trivially 1/2. Prior to the gel point, the transference numbers tend to be close to 0.5, as there will be small concentrations of charged clusters and roughly equivalent dissociation of cations and anions. Close to the gel point, $t_+$ departs from 0.5 because the fraction of charged clusters drastically increases. For example, when $f_+>f_-$, stoichiometry dictates that negatively charged clusters are formed more readily than positive clusters. These negative clusters drive $t_+ < 0.5$. When $f_+<f_-$, the opposite is true. In the limit of the RTIL being significantly gelled ($\lambda \gg \lambda^*$), the transference number is governed by the fraction of free ions~\cite{mceldrew2020theory}. When $f_+>f_-$, the gel will be negatively charged, meaning there will be more free cations than anions, and $t_+$ tends to 1. 

This non-monotonicity of the transference numbers can potentially be probed experimentally, as an alternative pathway to test for gelation in RTILs. By measuring the $t_\pm$ of RTILs (via electrophoretic NMR~\cite{gouverneur2015direct}, for example) as a function of temperature (effectively varying $\lambda$), the gel temperature should coincide with the extrema in $t_\pm$. However, the precise magnitude of these extrema depends on the precise functional form of $D_{lm}$.

\textit{Comparison with Molecular Simulations}-Given that our theory is derived from polymer physics~\cite{tanaka1989,tanaka1990thermodynamic,tanaka1994,tanaka1995,ishida1997,tanaka1998,tanaka1999,tanaka2002}, it is reasonable to question if the model is representative of the microscopic behavior of RTILs, especially given that we have neglected an explicit treatment of electrostatic interactions between ions beyond ionic association~\cite{mceldrew2020theory}. To investigate this, we performed MD simulations of a number of representative imidizolium based ILs at 295~K: 1-Ethyl-3-methylimidazolium chloride (emimCl), 1-Ethyl-3-methylimidazolium bis(trifluoromethylsulfonyl)imide  (emimTFSI), 1-Ethyl-3-methylimidazolium tetrafluoroborate (emimBF$_4$), 1-Ethyl-3-methylimidazolium hexafluorophosphate (emimPF$_6$), 1-Butyl-3-methylimidazolium hexafluorophosphate (bmimPF$_6$), and 1-Hexyl-3-methylimidazolium hexafluorophosphate (hmimPF$_6$). Specific simulation details can be found in the Appendix. 

In order to compute the ion cluster distributions from  MD simulations, independent of theory, we require a criterion for ion association~\cite{feng2019free}. We require knowledge of not only free ions, but aggregates of all different ranks and sizes. In this case, a spatial criterion appears to be appropriate. The simplest spatial criterion would be a cutoff distance between the center of masses of cations and anions~\cite{feng2019free}. However, such a criterion does not account for the orientation of ions, or the specific interactions between functional groups, which mediate associations between ions. 
In the left panel of Fig.~\ref{fig:clust_pic}, we show the spatial distribution functions (SDFs) of emim$^+$ and BF$_4^-$, which were generated by the open source software, TRAVIS~\cite{brehm2011travis}. In the Appendix, we show the computed SDFs for the other simulated RTIL ions. These SDFs are visualized as iso-density surfaces corresponding to regions where the density of counter-ions (based on centers of mass) are 2$\times$ the average bulk density of counter-ions. They indicate ``hot-spot" regions around ions where counter-ions are especially stable. Therefore, for an ionic association, we require that the center of mass of two associated ions must mutually exist in each other's ``hot-spots", as depicted in Fig.~\ref{fig:clust_pic}. Here, we choose a threshold iso-density value of 2$\times$ the bulk density to define the ``hot-spot" regions. Nonetheless, in the Appendix we examine how ion association is affected by varying this threshold value. This ion-association criterion is advantageous because it is readily transferable between different RTILs, it can be used instantaneously for any MD snapshot, and it ensures that ions in close proximity have energetically favorable mutual orientations with one another. 

\begin{figure*}[hbt!]
\centering
\includegraphics[width=\textwidth]{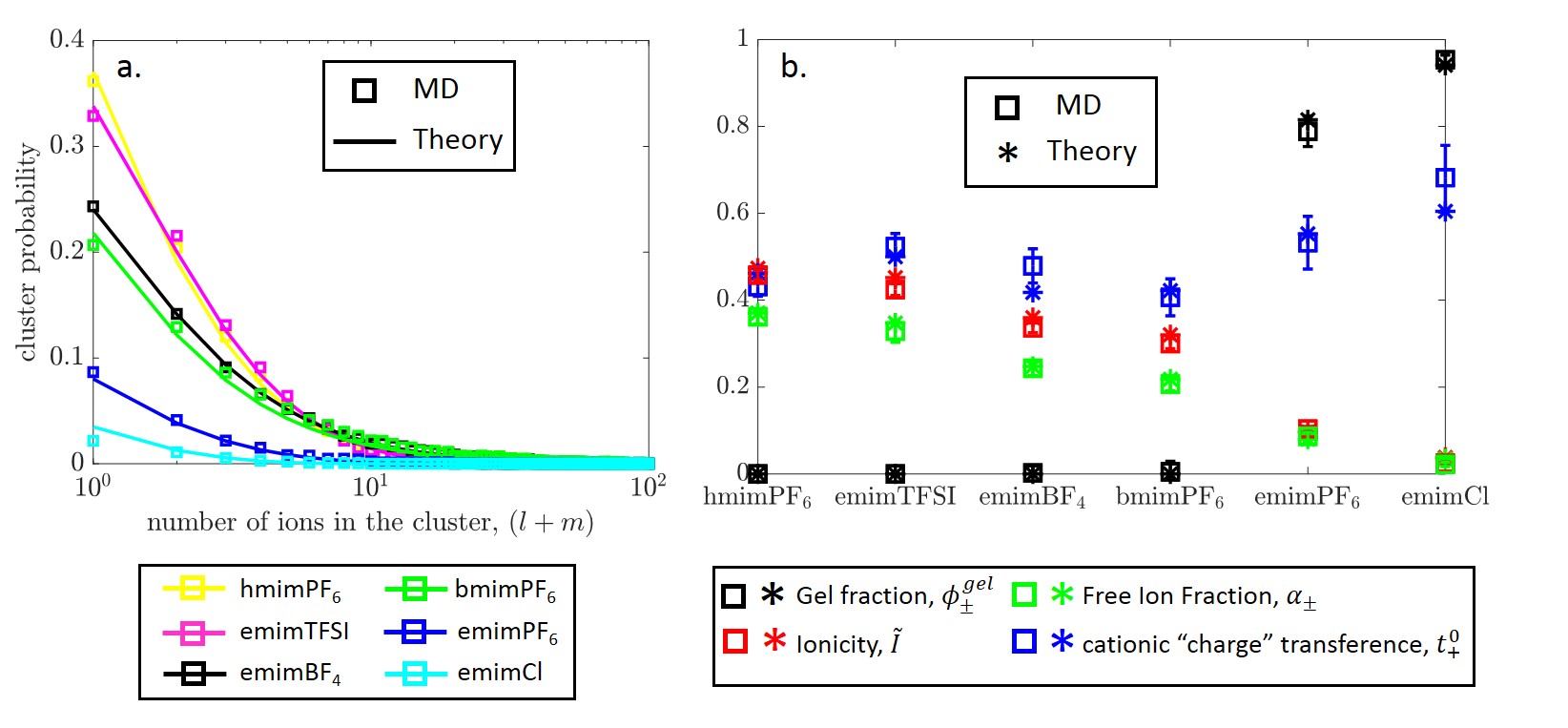}
\caption{(a) Ionic cluster distributions for six RTILs at 295~K, as computed from MD (open squares) and theory (solid lines). (b) Gel fraction, ionicity [Eq.~\eqref{eq:ionicity}], free ion fraction, and cationic charge transference [$t_+^0$, Eq.~\eqref{eq:t+0}] are plotted for all six studied RTILs, with molecular dynamics values as open squares, and theoretical values as asterisks. The sums in Eqs.~\eqref{eq:ionicity} \& \eqref{eq:t+0} were cutoff at clusters containing less than one hundred ions.}
\label{fig:mdtheory}
\end{figure*}

As we have alluded to, the SDF's for the ions displayed in Fig.~\ref{fig:clust_pic} (further examples can be found in the Appendix), can also be used to determine functionality of ions for our theory. For example, the SDF around the TFSI$^-$ and the BF$_4-$ ion shows four distinct ``hot-spot" regions, indicating that they have a functionalities of $f_-=4$. The number of ``hot-spots" itself does not always represent an ion's most apt functionality. For example, we observe eight distinct ``hot-spots" on the PF$_6^-$ ion, but we never observe more that 4 cations associated to a PF$_6^-$ ion in our simulations, leading to our choice of $f_-=4$ for PF$_6^-$. This is because imidizolium cations are bulky enough to block access to some additional ``hot-spots" when they are associated to PF$_6^-$. Thus, an ion's functionality is determined by a combination of its number of ``hot-spots", as well as the ability of the counter-ion to access those ``hot-spots".

In order to compare the MD simulation to our theory we need to determine all theoretical parameters. We explained above how ion functionality may be determined, so the only remaining parameters are $\tilde{c}_{salt}$ and $\lambda$, which may actually be contracted into a single parameter: $\Lambda = \lambda\tilde{c}_{salt}$, avoiding the additional complexity of computing $\lambda$ and $\tilde{c}_{salt}$ individually. This final parameter can be determined with knowledge of the association probabilities and ion functionalities via the aforementioned mass action law
\begin{equation}
\Lambda = \lambda \tilde{c}_{salt}=\dfrac{p_{-+}}{f_+(1-p_{+-})(1-p_{-+})}
\end{equation}
where the ion association probability is computed from MD at each time step with simple definition: $p_{\pm\mp}=(\# \text{ of associations})/((\# \text{ number of ions})f_\pm)$. The computed value of $\Lambda$ is then ensemble averaged across all time steps. The parameters computed from MD simulations for our theory are reported in Tab.~\ref{tbl:params}.

\begin{table}[!hbt]
\centering
\label{tbl:hpc_3d}
\begin{tabular}{lllllll}
\hline
 & $\quad\quad$ & $\Lambda$ & $\quad$ & $f_+$ & $\quad$ & $f_-$ \\
\hline
hmimPF$_6$ && 0.07 && 5 && 4 \\
emimTFSI   && 0.10 && 4 && 4 \\
emimBF$_4$ && 0.11 && 5 && 4 \\
bmimPF$_6$ && 0.13 && 5 && 4 \\
emimPF$_6$ && 0.30 && 5 && 4 \\
emimCl     && 0.54 && 5 && 4 \\
\hline
\end{tabular}
\caption{Summary of Model Parameters}
\label{tbl:params}
\end{table}

Overall, the cluster distribution from our theory, using parameters derived from the MD simulations, comes extremely close to the independently determined cluster distribution of the simulations, as seen in Fig.~\ref{fig:mdtheory}a. We observe the following trend in association affinity: emimCl $>$ emimPF$_6$ $>$ bmimPF$_6$ $>$ emimBF$_4$ $>$ emimTFSI $>$ hmimPF$_6$. Moreover, we also explicitly plot various quantities in Fig.~\ref{fig:mdtheory}b, including gel fraction, ionicity, free ion fraction, and the cationic charge transference. For all of these quantities, the model values nearly perfectly match those calculated from the performed MD simulations. 

It is informative to compare ionicity, $\tilde{I}$, given by
\begin{align}
    \tilde{I} = \dfrac{1}{2}\sum_{lm}(l-m)^2\alpha_{lm},
    \label{eq:ionicity}
\end{align}
to the free ion fraction, because this highlights the importance (or not) of ionic clusters in the ionic strength of the electrolyte (ionicity is essentially the dimensionless ionic strength). We see that for all RTILs, the free ion fractions are close to the ionicity, indicating that the majority of the current will be conducted by free ions~\cite{feng2019free}. In fact, for emimPF$_6$ and emimCl the ionicity and free ion fractions are essentially indistinguishable, indicating that free ions are the sole-conductors of current in these systems. For hmimPF$_6$, emimTFSI, emimBF$_4$, and bmimPF$_6$ there is a small, but noticeable difference between the ionicity and the free ion fraction. Thus, for these four RTILs we expect that ionic clusters have a non-trivial contribution to conductivity. Feng \textit{et al.}~\cite{feng2019free} determined the fraction of free ions of emimTFSI and bmimPF$_6$ to be approximately 15\% at a similar temperature, and that these free ions were the sole contributors to the conductivity. We report similar values, and to recover them exactly we would need to increase the iso-density threshold to above 2$\times$ the bulk density.

We find two of the RTILs, emimCl and emimPF$_6$, have non-zero gel fractions. It should be emphasized that our observation of percolating gel networks in emimCl and emimPF$_6$ (and lack gel in the other RTILs) depends heavily on our criterion for ionic association, specifically our iso-density threshold of 2$\times$ the bulk density. In the Appendix, we explore how varying this iso-density threshold changes the apparent association behavior of the RTIL, and we elaborate further on defining the gel in the simulations. 

It has been suggested by Gebbie \textit{et al.}~\cite{Gebbie2013,Gebbie2015} that RTILs behave as dilute electrolytes, with only 0.003\% of ions being free, as discussed in the Introduction. If such a situation were true, our model predicts that the RTIL must be gelled. In fact, with such as small proportion of free ions, the system essentially only comprises of the gel phase and free ions~\cite{mceldrew2020theory}. This raises the questions of if the gel phase, which has an elastic response, is playing a role in those surface-force measurements. 

Finally, we plot the cationic ``charge" transference number, $t_+^0$, given by the following formula
\begin{align}
t_+^0=\frac{\sum_{lm}l(l-m)\alpha_{lm}}{\sum_{lm}(l-m)^2\alpha_{lm}}.
\label{eq:t+0}
\end{align}
This is related to the cation transference number, $t_+$, in that if all clusters had equal diffusivities $t_+^0$ would correspond exactly to $t_+$. If cations and anions have equivalent functionalities the theory predicts they should form positively and negatively charged clusters with equal probability and $t_+^0=0.5$. We found (from the simulations) emimTFSI had equal functionalities and its $t_+^0$ was very close to 0.5. When the ion functionalities are asymmetric, the theory allows $t_+^0$ to deviate from 0.5. If the RTILs are in the pre-gel regime and $f_+ > f_-$, then $t_+^0$ should be smaller than 0.5, as we saw in Fig.~\ref{fig:cond}b, because there is a tendency of the RTIL to formed negatively charged clusters. The simulated hmimPF$_6$, emimBF$_4$ and bmimPF$_6$ all exhibit this behaviour. When the RTIL is substantially gelled and $f_+ > f_-$, then the theory predicts $t_+^0$ to be much larger than 0.5. In this case, there is an asymmetry in cations \& anions affinity to associate to the gel. The ion with the larger functionality tends to have a lower affinity to join the gel, leaving it significantly more free in the sol. For the most heavily gelled RTIL that we simulated, emimCl, we observe a significant deviation of $t_+^0$ from 0.5, with $t_+^0=0.65$. 

\begin{figure*}[hbt!]
\centering
\includegraphics[width=\textwidth]{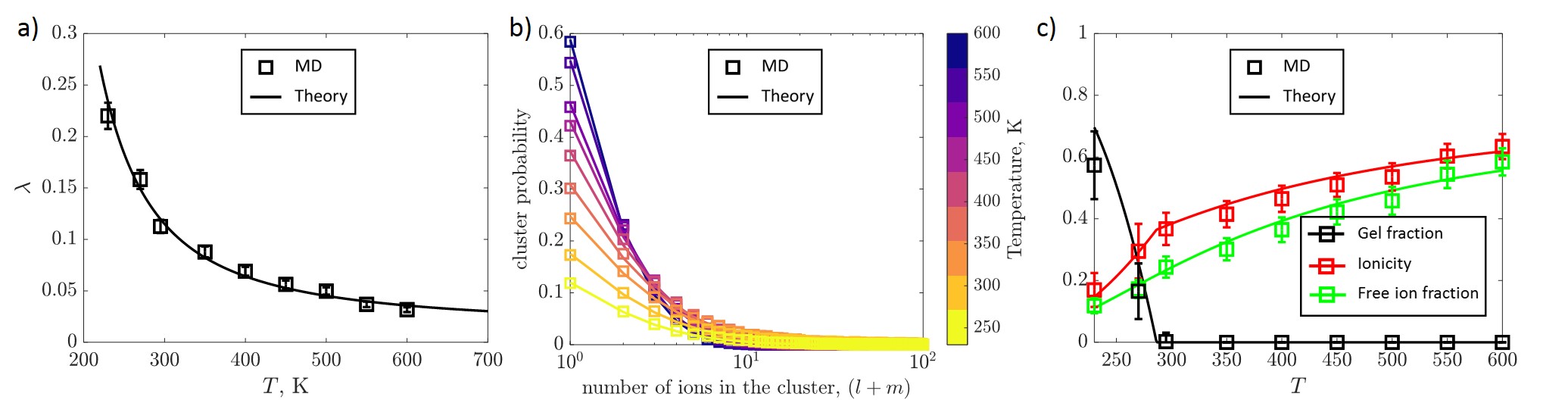}
\caption{(a) Computed (from MD simulations of emimBF$_4$) ionic association constants, $\lambda$, as a function of temperature. The theoretical curve was generated via Eq.~\eqref{eq:LTS} by fitting to the simulation data. (b) Ionic cluster distribution (cluster probability as a function of size) for various temperatures ranging from 230-600K. (c) Gel fraction, ionicity [Eq.~\eqref{eq:ionicity}], and free ion fraction are plotted as a function of temperature.}
\label{fig:temp}
\end{figure*}

\textit{Temperature dependence}--In order to further probe our model, we performed a series of MD simulations of emimBF$_4$ at various temperatures (230-600~K). In Fig.~\ref{fig:temp}a, we plot the computed $\lambda$ values as a function of temperature. As we show in the Appendix, $\lambda$ can be expressed as
\begin{align}
    \Lambda=\lambda \tilde{c}_{salt} = \frac{e^{-\beta(\Delta U_{+-}-T\Delta S_{+-})}}{\xi_++\xi_-},
    \label{eq:LTS}
\end{align}
where $\Delta U_{+-}$ and $\Delta S_{+-}$ are the energy and entropy of association, respectively. The relative molecular volumes of emim$^+$ and BF$_4^-$ give $\xi_+=2.4$ and $\xi_-=1$, respectively. The solid curve in Fig.~\ref{fig:temp}a, was generated from Eq.~\eqref{eq:LTS} with fitted values of $\Delta U_{+-}=-2.3k_BT_0$ ($T_0= $300~K), and $\Delta S_{+-}=-3.3k_B$. Thus, the ionic associations are driven by the exothermic energy of association, but at the cost of a strong decrease in entropy. The observed entropy of association is quite negative, implying that the associations are very stiff, if not completely inflexible (further explained in Appendix). Interestingly, imidizolium RTILs are known to form glasses upon cooling~\cite{fredlake2004thermophysical,leys2008temperature}. It is well-established that the stiffness or lack of flexibility in molecular chains leads to glass transitions in polymeric systems~\cite{gibbs1958nature}. Thus, the observed negative entropy of association for emimBF$_4$ in this work, implying inflexible ionic clusters, is in agreement with its ability to form a glass. In fact, the increase in ionic association upon cooling may be the principle driving force for emimBF$_4$ (and likely other RTILs) to undergo glass transitions. Future work may leverage this model to predict the glass transitions of various RTILs, or other super-concentrated electrolytes. 

In Fig.~\ref{fig:temp}b, we plot the cluster distribution of emimBF$_4$ for the studied temperatures, and once again we find quantitative agreement with the theory and simulations. Furthermore, in Fig.~\ref{fig:temp}c we plot the ionicity, free ion fraction, and gel fraction as a function of temperature. Overall the theory strongly agrees with the molecular simulations. Decreasing the temperature decreases both the ionicity and free ion fraction of emimBF$_4$, owing to the exothermic and entropy lowering nature of the ion associations. We also see that a gel network forms in the simulation upon cooling below room temperature, but remember the gel transition temperature will depend strongly on the iso-density threshold used (2$\times$ the bulk density in Fig.~\ref{fig:temp}).

It appears that, despite our neglect of explicit electrostatic interactions between ions beyond ionic associations, our theory performs adequately. We credit this to our previous statement that the majority of the electrostatic energy of the mixture is captured via the formation of the clusters. Excess contributions, such as the electrostatic interactions between ionic clusters, especially free ions, do not seem to strongly affect the cluster equilibria in RTILs~\cite{mceldrew2020theory}. Furthermore, this indicates that our model can capture significantly different RTIL chemistries via the association constant ($\lambda$) and ion functionality parameters ($f_\pm$). Recently, in Ref.~\cite{son2020ion}, the development of ``chemically specific theories" was identified as a key challenge for guiding electrolyte design in the next generation of battery electrolytes. We believe our model is a major advance in regards to this challenge. 

\textit{Discussion}-The thermodynamic theory presented here suggests that a gel could form in RTILs at temperatures around room temperature~\cite{mceldrew2020theory,Dupont2004,Hermann2008,Hayes2015}. In addition, the performed molecular dynamics simulations, and a number of others~\cite{molinari2019transport,Lopes2006,Wang2005,Bernardes2011}, indicate that a percolating cluster (a gel) may be present RTILs depending on the temperature and association criteria. This, alone, as we have mentioned, is not conclusive evidence that the gel phase is observable in RTILs. 

It is widely accepted that many RTILs are fragile glass-formers upon cooling~\cite{Hermann2008}. It could be the case that the glass transition occurs before the gel transition, which would mask the observation of the gel transition. This does not render the presented thermodynamic theory mute because it can still describe the cluster distribution at temperatures above the glass transition and, as previously mentioned, the formation of inflexible clusters could be the mechanism that drives the system toward its glass transition. 

If the gel transition occurs at a higher temperature than the glass transition, then one would expect to observe signatures of the gel phase. One form of evidence for the presence of a gel phase in RTILs is the observation of a viscoelastic rheological response. In fact, viscoelastic responses of several RTILs have been measured before the onset of the glass transition~\cite{makino2008viscoelastic,tao2015rheology,Elhamarnah2019,tao2015rheology,Shakeel2019,Shamim2010,Choi2014}. These measurements suggest that the gel phase forms before the glass phase in some RTILs. In fact, the thermodynamic onset of the gel phase could trigger the kinetic onset of the glass phase, due to the heavy entropic cost to form ionic associations.  

Furthermore, RTILs are extremely concentrated ionic systems. In model patchy particle systems~\cite{Russo2009,Smallenburg2013}, it has been shown that when the concentration of particles is sufficiently large, the system forms a glass before a gel. This suggests that, upon diluting RTILs with a solvent, the gel phase might appear before the glass phase. In fact, numerous experiments show the presence of a gel phase upon the addition of solvents (such as water)~\cite{Sturlaugson2012,Sturlaugson2013,Bernardes2011,chaban2012acetonitrile} or gelator molecules~\cite{Shakeel2019,Singh2019}. This suggests that some RTILs are in proximity to the gel transition, but further work is required to fully understand what neat RTILs (if any) are capable of forming gels.

\textit{Conclusion}-In this letter, we outlined a simple, general theory for the formation of arbitrarily large ionic clusters and the onset of a percolating infinite cluster (gel) in room temperature ionic liquids (RTILs). Such a theory is much needed for RTILs, for which the idealized picture of ion pairs is perhaps too simplified. Using the developed thermodynamics of ion clustering, we also developed a simple theory of coupled fluxes directly resulting from the presence of charged clusters in RTILs. We found that the contribution to conductivity from finite-sized charged clusters was only significant compared to that from free ions near the gel point, and that highly asymmetric transference numbers are a consequence of the asymmetry in ion functionalities. This provides a valuable connection between the molecular properties of an RTIL ion and its transport behaviour. 

We performed molecular dynamics (MD) simulations of 6 RTILs. From the spatial distribution functions of bulk simulations, we developed a general association criteria for ions based off their relative positions and orientations. This allowed the cluster distributions to be obtained for MD simulations. The parameters required for the theory are the ion functionalities and the strength of their bond. As we outlined in detail, these can be independently determined from MD and used in the theory, resulting in outstanding agreement between the theoretical and MD cluster distributions for all studied RTILs.

Some experimental signatures of clustering and gelation in RTILs have been previously reported, as we highlighted, and we believe that the presented theory should guide further experimental work on the extent of clustering and gelation in RTILs. We expect that such a theory may be extended to more complex ionic mixtures~\cite{li2020new}, such as polymer-based semi-solid electrolytes~\cite{di2011polymer} or perhaps even super-concentrated electrolyte mixtures with multiple salts ~\cite{lui2011salts,Suo2016,molinari2019transport,molinari2019general,chen202063,jiang2020high} or multiple solvents~\cite{wang2018hybrid,zhang2018aqueous,dou2018safe,Dou2019,molinari2020chelation}.


\textit{Acknowledgements}- All authors would like to acknowledge the Imperial College-MIT seed fund. MM and MZB acknowledge support from a Amar G. Bose Research Grant. ZG was supported through a studentship in the Centre for Doctoral Training on Theory and Simulation of Materials at Imperial College London funded by the EPSRC (EP/L015579/1) and from the Thomas Young Centre under grant number TYC-101. AK would like to acknowledge the research grant by the Leverhulme Trust (RPG-2016- 223). This work used the Extreme Science and Engineering Discovery Environment (XSEDE), which is supported by National Science Foundation grant number ACI-1548562.
\newpage

\section{Appendix}

\begin{table}[h]
\centering
\caption{List of Variables}
\begin{tabular}{llll}
\hline
$N_{lm}$         & Number of $lm$ clusters    & $N^{gel}_i$     & Number of species $i$ in gel \\
$f_i$              & Functionality of species $i$ & $\xi_i$    & Scaled volume of species $i$ \\
$Z$     & Coordination number of lattice & $g$ & Fraction of gauche associations \\
$\Omega$     & Total number of lattice sites  & $V$         & Total volume of mixture \\
$\tilde{c}_{lm}$       & Dimensionless concentration of  & $\tilde{c}_{i}^{gel}$    & Dimensionless concentration of \\
&cluster&&species $i$ in gel \\
$\tilde{c}_{salt}$ & Dimensionless salt concentration & $\tilde{c}_{tot}$  & Total dimensionless concentration \\
$\phi_i$         & Total volume fraction of species $i$ & $\phi_{lm}$    & Volume fraction of an $lm$ cluster\\
$\phi^{sol}_i$   & Volume fraction of species $i$ in sol & $\phi^{gel}_i$   & Volume fraction of species $i$ in gel\\
$\Delta_{lm}$  & Free energy of formation of a rank & $\psi_{i}$     & Concentration of i association sites\\
&$lm$ cluster && \\
$\Delta^{gel}_{i}$ & Free energy change of species $i$ & $\Delta_{lm}^{comb}$  & Combinatorial free energy of \\
&associating to the gel && formation of a rank $lm$ cluster\\
$\Delta_{lm}^{bind}$  & Bonding free energy of & $\Delta_{lm}^{conf}$  & Configurational free energy of \\
&  formation of an $lm$ cluster &&formation of an $lm$ cluster\\
$\mu_{lm}$    & Chemical potential of an $lm$ cluster &$\lambda$  & Ionic association strength \\
$K_{lm}$      & Equilibrium constant & $W_{lm}$      & Combinatorial enumeration \\
$\Delta U_{+-}$ & Association energy  & $\Delta S_{+-}$      & Association entropy \\
$p_{ij}$        & Association probabilities & $p^{sol}_{ij}$  & Association probabilities in the sol\\
$\bar{n}$     & Weight average degree of aggregation & $\alpha_{lm}$   & Fraction of ions in $lm$ clusters \\
$w^{sol}_i$     & Fraction of species in the sol & $w^{gel}_i$           & Fraction of species in the gel\\
$\sigma$ & Conductivity & $t_{+/-}$ & Transference numbers \\
$D_{lm}$ & Diffusivity of rank $lm$ cluster & $D_0$ & Diffusivity of free ion\\
\hline
\end{tabular}
\label{tab:my_label}
\end{table}

\newpage
\clearpage

\subsection{Derivation of Cluster Distribution and $\Delta_{lm}$}

We consider a room temperature ionic liquid (RTIL) as only comprising of cations and anions - no additional solvent shall be considered here~\cite{mceldrew2020theory}. There are $N_+$ cations and $N_-$ anions. The cations can bond to a maximum of $f_+$ anions, and the anions can bond to a maximum of $f_-$ cations (these are referred to as the functionality of the ions). These cations and anions can either exist in the sol phase or the gel phase. In the sol phase, the cations and anions can associate to form a cluster, denoted with the label $lm$ with the $l$ corresponding to the number of cations and $m$ corresponding to the number of anions. The number of each cluster in the sol phase is given by $N_{lm}$. There can then be $N_{+}^{gel}$ cations in the gel phase and $N_{-}^{gel}$ anions in the gel phase. The total number of cations and anions must be conserved, such that we have
\begin{equation}
    N_+ + N_- = \sum_{lm}(l + m)N_{lm} + N_{+}^{gel} + N_{-}^{gel}.
\end{equation}
Overall the system remains electroneutral ($N_+ = N_-$), but this does not mean that the sol and gel phase need to independently be electroneutral. In fact, very interesting transport phenomena arise when this is not the case.

The volumes of ions in RTILs are typically never the same, and the number of voids is quite small. Therefore, we neglect the presence of voids, but retain the ability for ions to be different sizes. Since we are using a lattice model, we can scale the lattice size to whatever dimension we wish. We shall keep it general here. The number of lattice sites occupied by a cation is $\xi_+$ and the number of lattice sites occupied by an anion is $\xi_-$. The total number of lattice sites, $\Omega$, is given by 
\begin{equation}
\Omega = \sum_{lm}(\xi_+l + \xi_-m)N_{lm} + \xi_+N_{+}^{gel} + \xi_-N_{-}^{gel}.
\end{equation}
Note, one can convert this expression to the total volume by multiplying by the volume of an individual lattice site. Dividing the above expression by the total number of lattice sites yields
\begin{equation}
1 = \sum_{lm}(\xi_+l + \xi_-m)\tilde{c}_{lm} + \xi_+\tilde{c}_{+}^{gel} + \xi_-\tilde{c}_{-}^{gel},
\end{equation}
where the dimensionless concentration of clusters is $\tilde{c}_{lm} = N_{lm}/\Omega$ and the concentration of cations/anions in the gel is $\tilde{c}_{+/-} = N_{+/-}/\Omega$. 

The volume fraction of each species can now be readily determined. The volume fraction of a cluster of rank $lm$ is given by
\begin{equation}
    \phi_{lm} = (\xi_+l + \xi_-m)\tilde{c}_{lm}.
\end{equation}
The volume fraction of cations in the sol phase is 
\begin{equation}
    \phi_+^{sol} = \sum_{lm}\xi_+l\tilde{c}_{lm},
\end{equation}
and for anions
\begin{equation}
    \phi_-^{sol} = \sum_{lm}\xi_+m\tilde{c}_{lm}.
\end{equation}
The volume fraction of anions in the gel is simply
\begin{equation}
    \phi_+^{gel} = \xi_+\tilde{c}^{gel}_{+},
\end{equation}
and analogously the volume fraction of anions
\begin{equation}
    \phi_-^{gel} = \xi_+\tilde{c}^{gel}_{-}.
\end{equation}
Naturally, we can express the incompressibility condition as
\begin{equation}
    1 = \phi_+^{sol} + \phi_-^{sol} + \phi_+^{gel} + \phi_-^{gel}.
\end{equation}

There are also individual relations for the conservation of cations/anions~\cite{mceldrew2020theory}. The total volume fraction of cations is
\begin{equation}
    \phi_+ = \phi_+^{sol} + \phi_+^{gel} = \dfrac{\xi_+}{\xi_+ + \xi_-}
\end{equation}
and for anions it is
\begin{equation}
    \phi_- = \phi_-^{sol} + \phi_-^{gel} = \dfrac{\xi_-}{\xi_+ + \xi_-}
\end{equation}

Recall our free energy from the main text:
\begin{equation}
\beta \Delta F = \sum_{lm} \left[N_{lm}\ln \left( \phi_{lm} \right)+N_{lm}\Delta_{lm}\right] \nonumber
+ \Delta^{gel}_+ N^{gel}_+ + \Delta^{gel}_- N^{gel}_-. 
\label{eq:F}
\end{equation}
Here $\Delta_{lm}$ and $\Delta_{+/-}^{gel}$ are the free energy of formation of a cluster of rank $lm$ from free cations and anions, and the free energy change of cations/anions associating with the gel. 

We may determine the chemical potential of a rank $lm$ cluster by differentiating the free energy with respect to $N_{lm}$:
\begin{equation}
\beta \mu_{lm}=\ln \phi_{lm} + 1 - (\xi_+ l + \xi_-m) \tilde{c}_{tot}+\Delta_{lm}  
\label{eq:muclust1}
\end{equation}
where $\tilde{c}_{tot}=\sum_{lm}\tilde{c}_{lm}$ is the total dimensionless concentration for species ($\#$ per lattice site). Note, the explicit form of $\Omega$ has to be used when differentiating. Establishing an equilibrium between all cluster requires the following condition to be satisfied:
\begin{align}
    \mu_{lm}=l\mu_{10}+m\mu_{01}.
    \label{eq:clusteq}
\end{align}
Note the indices 01 and 10 correspond to free cations and anions, respectively. 

Plugging in Eq.~\eqref{eq:muclust1} into Eq.~\eqref{eq:clusteq}, we obtain Eq. (2) from the main text:
\begin{align}
    \phi_{lm}=K_{lm}\phi_{10}^{l}\phi_{01}^{m}
    \label{eq:clustp}
\end{align}
where $K_{lm}=\exp\left(l+m-1-\Delta_{lm}\right)$ and $\Delta_{lm}$ is the free energy of formation for rank $lm$ clusters from free cations and free anions. As in Ref.~\citenum{mceldrew2020theory}, $\Delta_{lm}$ contains three major contributions: 1) combinatorial, 2) binding and 3) configurational. 

The combinatorial contribution, $\Delta_{lm}^{comb}$ is purely entropic, and given by the expression
\begin{align}
    \Delta_{lm}^{comb}=- \log f_+^lf_-^mW_{lm}
\end{align}
where $W_{lm}$, is  the enumeration of the ways that a rank $lm$ cluster can be formed. For Cayley tree type clusters (no intracluster loops), Stockmayer\cite{stockmayer1952molecular} determined the exact expression to be 
\begin{align}
    W_{lm}=\frac{(f_+l-l)!(f_-m-m)!}{l!m!(f_+l-l-m+1)!(f_-m-m-l+1)!}.
\end{align}
The binding contribution may be determined by counting the number of associations in a cluster. For a general rank $lm$ cluster with no intracluster loops the number of associations is $l+m-1$. Thus, if we assign a constant binding energy, $\Delta u_{\pm}$ for each association, then the binding contribution, $\Delta_{lm}^{bind}$ is written explicitly as
\begin{align}
    \Delta_{lm}^{bind}=(l+m-1)\Delta u^{bind}_{\pm}.
    \label{eq:bind}
\end{align}
We note that the binding energy $\Delta u^{bind}_{\pm}$ is largely electrostatic in nature and assumed to be a constant, despite if there are other species associated to the same ion (this independent bonding assumption also comes in through the bonding probabilities later, and future work might wish to relax this assumption). Indeed, it is likely that in a real RTIL system the binding energy of an ion to a counter-ion will be dependent on the local chemical environment of the associating species. Actually, there have been some work  incorporating the local chemical environment, electrostatics, and fluid structure into the association energy for ions forming ions pairs in RTILs~\cite{lee2014room} and concentrated electrolytes~\cite{ebeling1980analytical,levin1996criticality} to improve upon the Bjerrum's theory of ion pairing~\cite{bjerrum1926k}. However, here our focus is on the formation of higher order clusters involving potentially many ions, and the assumption of constant binding energy allows us to undertake this additional complexity with some analytical tractability. 

The final contribution to $\Delta_{lm}$ is the configurational part, $\Delta_{lm}^{conf}$ associated with the entropy of placing a rank $lm$ cluster on a lattice with coordination number $Z$. As RTILs are known glass-formers, it stands to reason that they form semi-flexible to inflexible associations. In this case, we allow the associations to partition between trans and gauche orientations, which  differ by an energy (units $k_BT$) of $\epsilon$ (trans conformation has the reference energy of 0, and gauche $\epsilon$). For this, we use augment Flory's expression for the so-called entropy of disorientation  used in lattice fluid  theory\cite{flory1942thermodynamics,flory1953principles,tanaka1989,tanaka1999} to account for the partitioning of associations into trans and gauche orientations \cite{flory1956statistical,tanaka2011polymer}:
\begin{align}
    \Delta_{lm}^{conf} =-\ln\left(\frac{(\xi_+l+\xi_-m)\left[\frac{(Z-2)^{2g}}{Z2eg^{2g}(1-g)^{2(1-g)}}\right]^{l+m-1}}{\xi_+^l\xi_-^m}\right)+g\epsilon
\end{align}
where $g$ is the fraction of associations in the gauche conformation given by\cite{flory1956statistical}
\begin{align}
    g=\frac{(Z-2)e^{-\epsilon}}{1+(Z-2)e^{-\epsilon}}
\end{align}
In the high temperature limit ($\epsilon \rightarrow 0$), we obtain $g\rightarrow (Z-2)/(Z-1)$ for completely flexible chains. In the low temperature limit ($\epsilon\rightarrow\infty$), we obtain $g \rightarrow 0$.

Thus, in total, $\Delta_{lm}$ is given simply by
\begin{align}
    \Delta_{lm}=\Delta^{comb}_{lm}+\Delta^{bind}_{lm}+\Delta^{conf}_{lm}
\end{align}
If we plug this into Eq.~\eqref{eq:clustp}, also recalling that $(\xi_+l+\xi_-m)\tilde{c}_{lm}=\phi_{lm}$, then we obtain the thermodynamically consistent cluster distribution
\begin{align}
    \tilde{c}_{lm}=\frac{W_{lm}}{\lambda}  \left(\lambda\psi_{10}\right)^l \left(\lambda\psi_{10}\right)^m.
\end{align}
where $\tilde{c}_{lm}$ is the dimensionless concentration of clusters of rank $lm$ ($\#$ per lattice site), $\tilde{c}_{salt}$ is the dimensionless concentration of salt molecules, $\psi_{10}=f_+ \phi_{10}/\xi_+$ and $\psi_{01}=f_- \phi_{01}/\xi_-$ are the dimensionless concentrations of available bonding sites of cations and anions, respectively, and $\lambda$ is the ionic association.

In the theory which partitions associations into trans and gauche orientations, the constant given by  
\begin{align}
    \lambda=\frac{(Z-2)^{2g}}{Z g^{2g} (1-g)^{2(1-g)}}\exp \left( -\Delta u^{bind}_{\pm} -g \epsilon \right).
\end{align}
For completely flexible chains, we obtain
\begin{align}
    \lambda^{flex}=\frac{(Z-1)^{2}}{Z}\exp \left( -\Delta u^{bind}_{\pm}\right).
\end{align}
Conversely, for completely inflexible chains, we obtain
\begin{align}
    \lambda^{inflex}=\frac{1}{Z}\exp \left( -\Delta u^{bind}_{\pm}\right).
\end{align}
The interpretation of $\lambda$ as an association constant also lends the following definition:
\begin{align}
    \lambda=\exp(-\Delta f_{+-})=\exp(-\Delta u_{+-}+\Delta s_{+-})
\end{align}
where $\Delta u_{+-}$  is the total dimensionless energy (non-dimensionalized by $k_BT$) involved in ion associations (binding energy + gauche/trans conformation) and $\Delta s_{+-}$ is the dimensionless entropy (non-dimensionalized by $k_B$) of association associated with the conformational entropy of the clusters. It is clear upon inspection that generally $\Delta u_{+-}=\Delta u^{bind}_{+-}+g\epsilon$, with $\Delta u^{flex}_{+-}=\Delta u^{inflex}_{+-}=\Delta u^{bind}_{+-}$ high (flexible) and low (inflexible) temperature limits.  We also can evaluate $\Delta s_{+-}$ in the various temperature/flexibility limits
\begin{align}
    \Delta s_{+-}=\ln \left( \frac{(Z-2)^{2g}}{Z g^{2g} (1-g)^{2(1-g)}}\right),
\end{align}
\begin{align}
    \Delta s^{flex}_{+-}=\ln \left( \frac{(Z-1)^{2}}{Z}\right),
\end{align}
and 
\begin{align}
    \Delta s^{inflex}_{+-}=-\ln \left(Z\right)
\end{align}
The flexibility of the ionic clusters must be evaluated via the temperature dependence of association, as was studies for the emimBF$_4$ system in Fig. 6 of the main paper. There, it was found that the temperature dependence of $\lambda$ was found to agree with completely inflexible clusters, with $\Delta s^{inflex}_{+-}=-3.3$. The inflexibility of the ion associations in emimBF$_4$ seems to be an agreement with its ability to form a glass. Chain flexibility is thought to be a principle characteristic of glass-forming polymeric analogues.

\subsection{General Association Probabilities for Asymmetric Room Temperature RTILs}

In the main text we explicitly wrote the solutions to the ionic association probability when the ions are symmetric (equal functionalities and sizes). In general, RTIL ions will be potentially highly asymmetric. In this case the association probabilities ($p_{+-}$ \& $p_{-+}$) satisfy the following equations. First we have
\begin{align}
    f_+p_{+-}=f_-p_{-+}
    \label{eq:p1}
\end{align}
which is conservation of ionic associations (there an equal number of cation-anion associations as anion-cation associations). Second, we have
\begin{align}
    \lambda\zeta_{+-}=\frac{p_{+-}p_{-+}}{(1-p_{+-})(1-p_{-+})}
    \label{eq:p2}
\end{align}
where $\zeta_{+-}=f_+p_{+-}\tilde{c}_{salt}=f_+p_{-+}\tilde{c}_{salt}$ is the number of ion associations per lattice site ($\tilde{c}_{salt}$ is the number of salt molecules per lattice site), and $\lambda$ is the ionic association constant. These two equations may be solved explicitly 
\begin{align}
    f_+p_{+-}=f_-p_{-+}=\frac{1 + \tilde{c}_{salt} (f_- + f_+) \lambda - \sqrt{
  1 + \tilde{c}_{salt} \lambda \left[ 2(f_- + f_+) + \tilde{c}_{salt} (f_- - f_+)^2 \lambda\right]}}{2 \tilde{c}_{salt}\lambda}
\end{align}
We plot the an example case of asymmetric ionic association probabilities in the left panel of Fig.~\ref{fig:sol_gel_as}. Up to the gel point, the association probabilities increase as the strength of the ionic association is increased. The probability of an anion binding to a cation is larger than that of a cation binding to an anion, owing to the larger functionality of cations in Fig.~\ref{fig:sol_gel_as}. In the right panel of Fig.~\ref{fig:sol_gel_as}, we plot the fraction of free cations and anions, respectively. It can be seen that the fraction of free ions decrease monotonically towards the gel point with increasing ionic association. 

Furthermore, we may solve for the critical gel point using the same criterion as in the main text $\left[p^*_{+-}p^*_{-+}=1/(f_+-1)(f_--1)\right]$ for the general asymmetrically associating RTIL, obtaining
\begin{align}
    \lambda^*=\frac{\tilde{c}_{salt}^{-1}\sqrt{f_+f_-(f_+-1)(f_--1)}}{f_+f_-(1+(f_+-1)(f_--1))-(f_++f_-)\sqrt{f_+f_-(f_+-1)(f_--1)}}.
\end{align}

\subsection{Post Gel Association Probabilities}

Once the critical gel threshold has been surpassed, we must determine the volume fractions of cations and anions in the gel, $\phi_+^{gel}$ and $\phi_-^{sol}$, respectively. In order to do this, we employ Flory's treatment of the post-gel regime in which the volume fraction of free ions can be written equivalently in terms of overall association probabilities, $p_{ij}$, and association probabilities taking into account only the species residing in the sol, $p^{sol}_{ij}$

\begin{align}
    \phi_+(1-p_{+-})^{f_+}=\phi_+^{sol}(1-p^{sol}_{+-})^{f_+}
    \label{eq:gel1}
\end{align}
\begin{align}
    \phi_-(1-p_{-+})^{f_-}=\phi_-^{sol}(1-p^{sol}_{-+})^{f_-}
    \label{eq:gel2}
\end{align}
Where $\phi_\pm^{sol}=1-\phi_\pm^{gel}$ is the volume fraction of cations or anions in the sol. We may determine each of the two unknown $\phi_i^{sol}$ variables, as well as the two unknown sol association probabilities, $p_{ij}^{sol}$, using Eqs.~\eqref{eq:gel1}-\eqref{eq:gel2} in addition to Eqs.~\eqref{eq:p1}-\eqref{eq:p2}, however in this case we use sol-specific quantities.

Thus, we have four equations and four unknowns (two sol association probabilities and two sol species volume fractions). The fraction of species, $i$, in the gel is simply given by 
\begin{align}
    w_i^{gel}=1-\phi^{sol}_i/\phi_i
\end{align}
Note that prior to the critical gel concentration, we have the trivial solution that $p_{\pm\mp}=p^{sol}_{\pm\mp}$ and $\phi_i=\phi^{sol}_i$, yielding a gel fraction of $w_i^{gel}=0$. However, beyond the gel point, there is a non-trivial solution yielding $w_i^{gel}>0$.

\begin{figure*}[hbt!]
\centering
\includegraphics[width=1\textwidth]{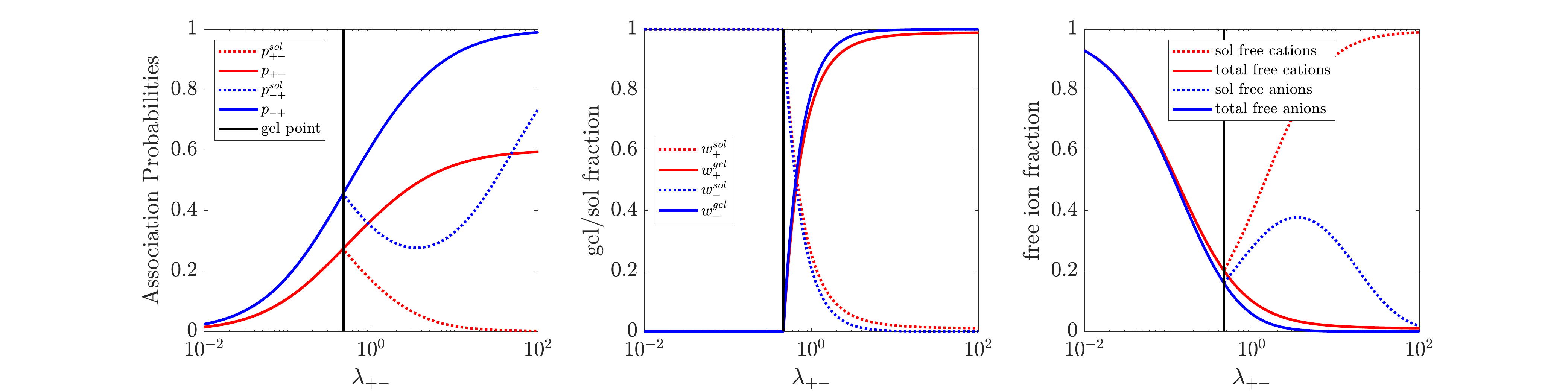}
\caption{Association probabilities, $p_{ij}$ and $p_{ij}^{sol}$, (left), the gel and sol fractions ($w_\pm^{sol}=\phi_\pm^{sol}/\phi_\pm$  \& $w_\pm^{gel}=1-w_\pm^{sol}$, respectively) (middle), and the overall fraction of free ions, $\alpha_\pm$, and fraction of the free ions in the sol, $\alpha_\pm^{sol}$ (right) as functions of the ionic association constant, $\lambda$. The curves are generated for an asymmetric RTIL, with $f_+=5$ \& $f_-=3$  ($\xi_+=\xi_-=1$).}
\label{fig:sol_gel_as}
\end{figure*}

In the left panel of Fig.~\ref{fig:sol_gel_as}, we display how the association probabilities change as a function of $\lambda$ past the gel point. The overall association probabilities continue to increase with the ionic association strength, as one might expect. The sol association probabilities initially start to decrease as one increases the value of $\lambda$ past the gel point. This is a reflection of the larger clusters binding to increase the fraction of the gel phase. The sol association probability of cations binding to anions decreases towards zero in the limit of very large ionic association strengths. In such a limit, there are free cations surrounded by a negatively charged gel. This is important implications in the transference numbers, as outlined in the main text.

In the middle panel of Fig.~\ref{fig:sol_gel_as}, the fraction of gel and sol is shown as a function of ionic association strength. As expected, for values of $\lambda > \lambda^*$, the fraction of the ions bound into the gel monotonically increases, and the fraction of ions in the sol monotonically decreases.

Finally, in the right panel of Fig.~\ref{fig:sol_gel_as}, the free ion fraction is displayed. Past the gel point, the total free ion fraction continues to decrease with increasing ionic association. However, the fraction of free cations and anions in the sol can be qualitatively different depending on the functionalities. As eluded to previously, the cations association probability decreases to zero in the limit of large $\lambda$, which causes all of the free ions to be cations. In contrast, the fraction of free anions decreases to zero as the anions are in excess in the gel when their functionality is lower than the cations.

\subsection{Cluster Distributions and Weight-Average Degree of aggregation}

\begin{figure*}[hbt!]
\centering
\begin{subfigure}{0.5\textwidth}
\includegraphics[width=\textwidth]{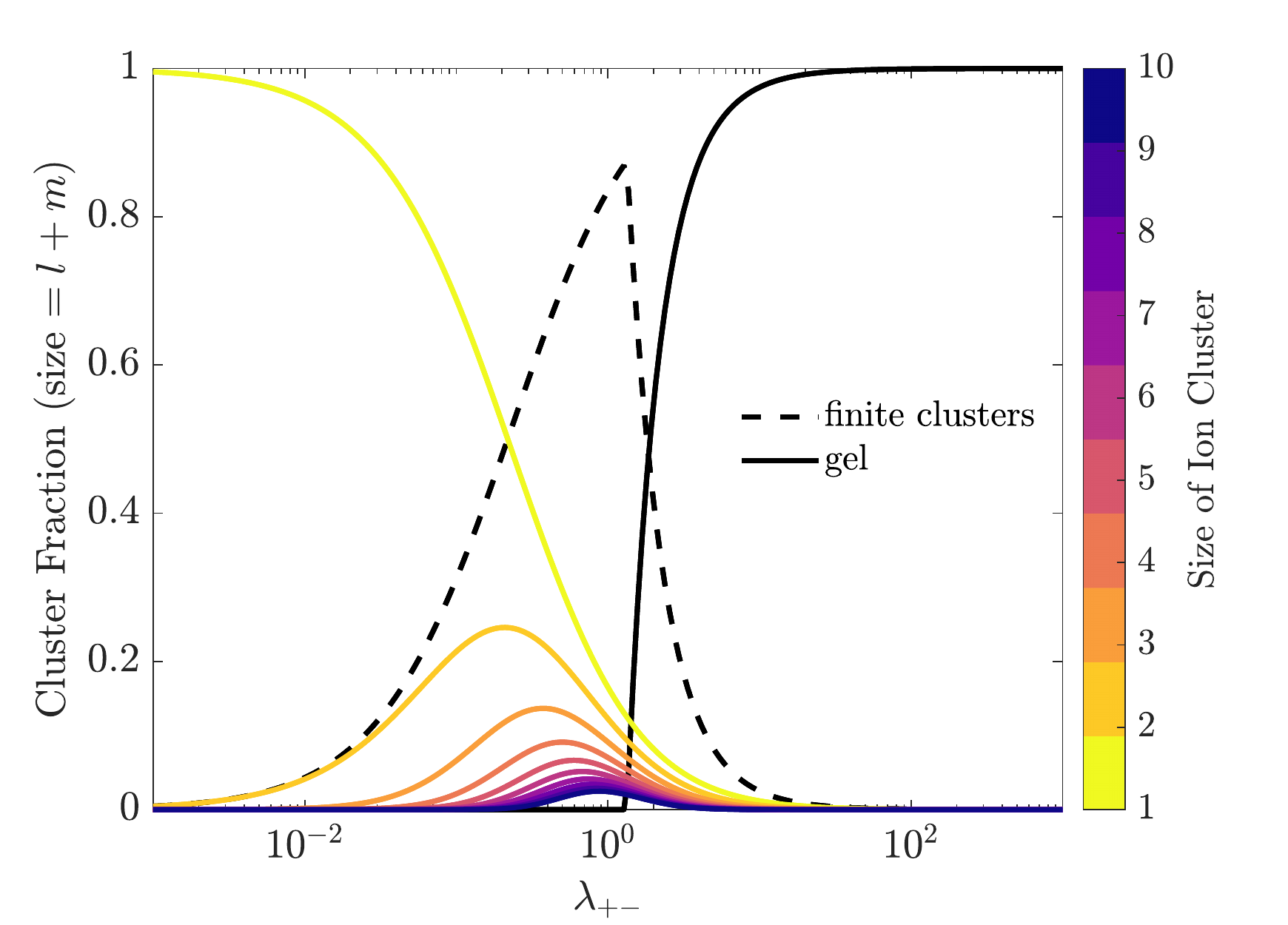}
\end{subfigure}
\begin{subfigure}{0.49\textwidth}
\includegraphics[width=\textwidth]{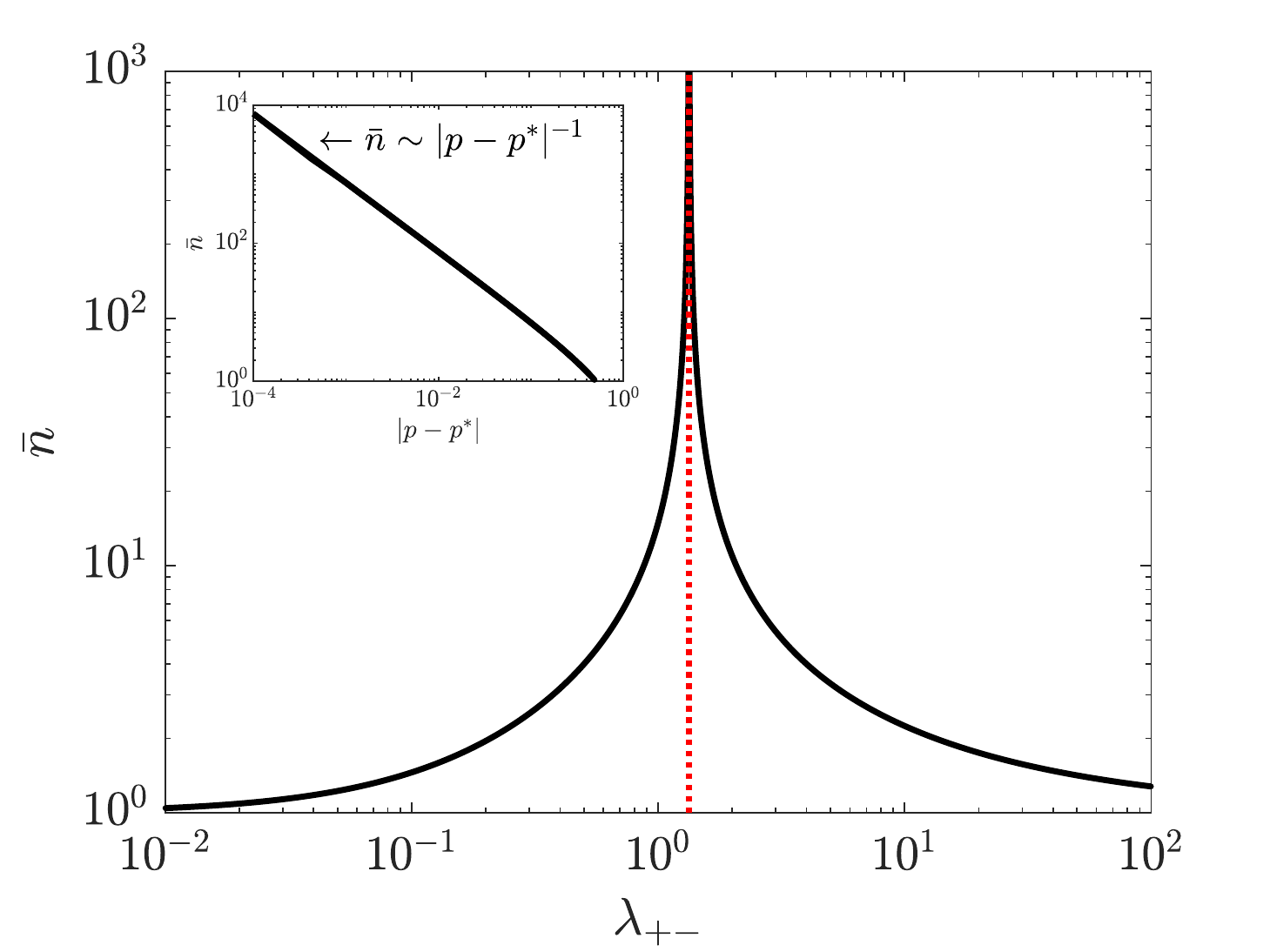}
\end{subfigure}
\caption{(Left)- Cluster fractions corresponding to clusters containing one to ten ions as a function of the ionic association constant (left), $\lambda$. (Right)- Weight average degree of aggregation, $\bar{n}$, as a function of $\lambda$. The weight average degree of aggregation is a measure of the average size of clusters. Beyond the gel point, indicated by the red dotted line, we plot the weight averaged degree of clusters in the sol. The inset in the right panel plots weight average degree of aggregation as a function of absolute difference between the ionic association probability, $p$, and the critical association probability, $p^*$ on a log-log scale, which shows that $\bar{n}$ diverges at the gel point with a critical exponent of -1.  All curves are generated for the symmetric RTIL, with $f=3$.}
\label{fig:cluster_frac}
\end{figure*}

In the left panel of Fig.~\ref{fig:cluster_frac}, the fractions of each size of cluster is shown as a function of $\lambda$. The dependence of the free ion fraction and the fraction of the gel as a function of the ionic association were described in detail in the previous section, and can also be seen in Fig.~\ref{fig:sol_gel_as}. As the ionic association strength increases from small values, the fraction of finite clusters initially increases. This is most pronounced for ion pairs, which can be seen as the most prevalent species after free ions. Clusters of larger sizes start to make up more significant fractions of the RTIL until  $\lambda^*$ is reached. At which point, the finite clusters decay quite rapidly. This can be seen in the dashed line of the left panel of Fig.~\ref{fig:cluster_frac}.

In the main text we reported the weight average degree of aggregation, $\bar{n}$, for RTILs with symmetric ions. In 1952, Stockmayer derived a closed form for the weight average degree of aggregation - expected size of a cluster for any given ion - of a mixture of asymmetrically associating units:
\begin{align}
    \bar{n}&=\frac{\sum_l\sum_m(l+m)^2\tilde{c}_{lm}}{\sum_l\sum_m(l+m)\tilde{c}_{lm}} \nonumber \\
    &=
    1+\frac{f_\mp}{2 p^{sol}_{\pm\mp}}\left( \frac{(f_+-1)(p^{sol}_{+-})^2p^{sol}_{-+}+(f_--1)(p^{sol}_{-+})^2p^{sol}_{+-}+2p_{+-}^{sol}p_{-+}^{sol}}{1-(f_+-1)(f_--1)p^{sol}_{+-}p^{sol}_{-+}} \right)
    \label{eq:n_as}
\end{align}
Stockmayer's summation is also extremely useful in deriving a closed-form maximum for the ionic conductivity. 

In the right panel of Fig.~\ref{fig:cluster_frac}, $\bar{n}$ is displayed as a function of the ionic association strength. It is clear that there is a divergence in $\bar{n}$ at the gel point, which is denoted by the vertical line in Fig.~\ref{fig:cluster_frac}. More details can be found in Ref.~\citenum{mceldrew2020theory}, where this was studied in more detail.

\subsection{Conductivity Bounds of Asymmetric Ionic Liquids}

Recall, our definition of conductivity from the main text:
\begin{align}
    \sigma = \frac{e^2c_{salt}}{k_B T} \sum_{lm} (l-m)^{2} \tilde{c}_{lm} D_{lm}=\frac{e^2c_{salt}}{k_B T}(\mathcal{D}_{++}+\mathcal{D}_{--}-2\mathcal{D}_{+-})
    \label{eq:sig1}
\end{align}
To solve for the conductivity, one must know $D_{lm}$. However, we do not know $D_{lm}$. Therefore, we wish to generate bounds on the conductivity based on the cluster distribution for a given $\lambda$. A minimum conductivity can be obtained from just counting the free ions. While a maximum bound for the conductivity can be achieved through taking $D_{lm}$ as a constant for all $lm$. It is \textit{stressed that the maximum bound of the conductivity is not physical}, but it can provide information about how clusters contribute to conductivity. 

\begin{figure*}[hbt!]
\centering
\includegraphics[width=0.5\textwidth]{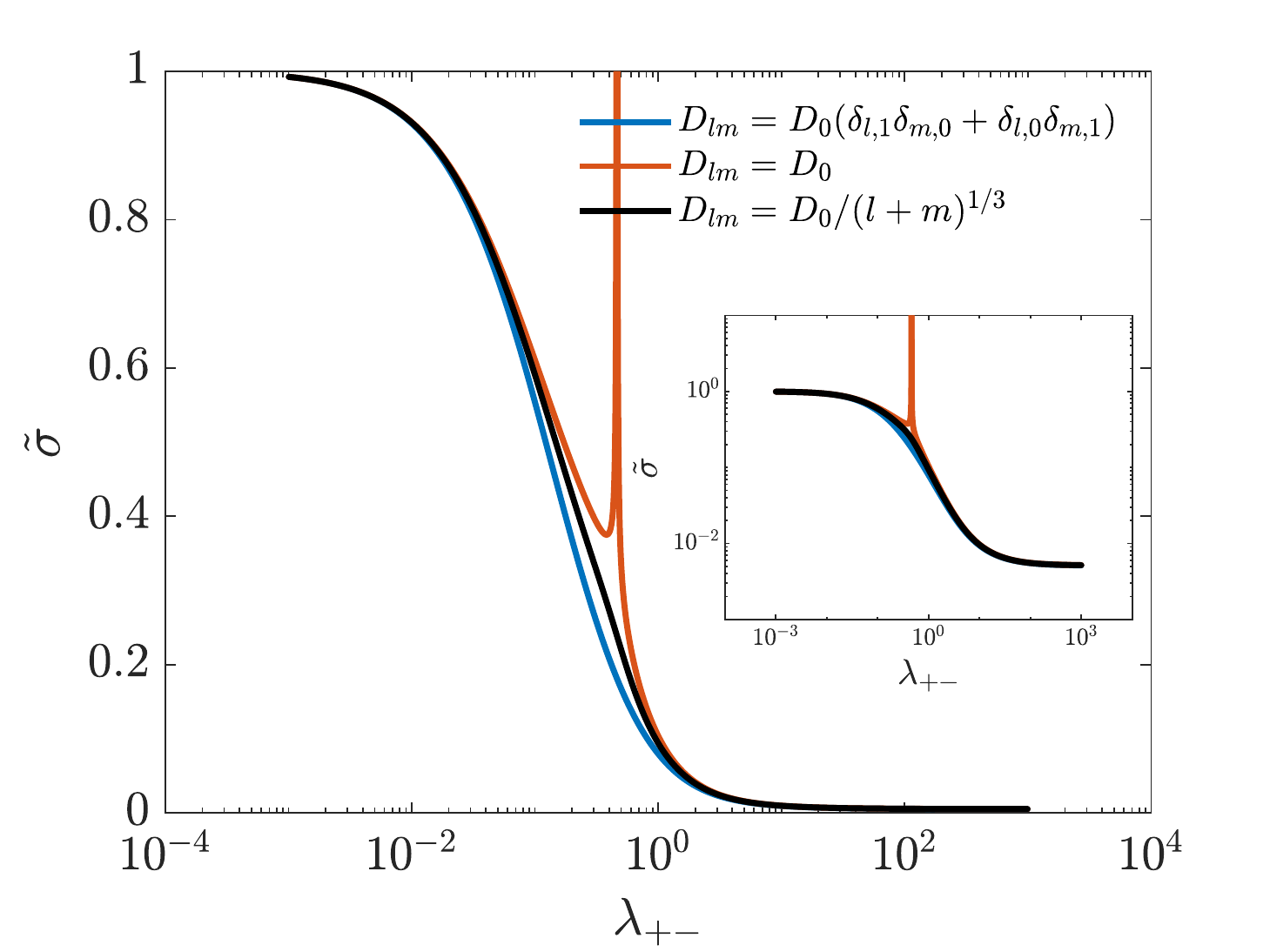}
\caption{The dimensionless conductivity (made dimensionless with NE conductivity) is plotted as a function of $\lambda$ for $f_-=4$, $f_+=5$, using three different approximations for the cluster diffusivity:$D_{lm}=D_0(\delta_{l,1}\delta_{m,0}+\delta_{l,0}\delta_{m,1})$, $D_{lm}=D_0$, and $D_{lm}=D_0/(l+m)^{1/3}$. Note, the $D_{lm}=D_0$ approximation is an upper bound for the conductivity. In the asymmetric case there is an unphysical divergence in the conductivity near the gel point because of very large, highly charged clusters also have large diffusivities in this approximation. Further explanation can be found in the text.}
\label{fig:condas}
\end{figure*}

If we assume the all clusters have the same conductivity (that of a free ion) and non-dimensionalize the conductivity ($\tilde{\sigma}$) with the factor, $2 e^2c_{salt}D_0/k_BT$, then our summation simply becomes 
\begin{align}
    \tilde{\sigma}_{max} &= \sum_{lm} (l-m)^{2} \tilde{c}_{lm} \nonumber \\
\end{align}
This summation is quite similar to the numerator in Eq.~\eqref{eq:n_as}. In fact, we need only change one sign of the $2p^{sol}_{+-}p^{sol}_{-+}$ term in the numerator of Eq.~\eqref{eq:n_as}, as well as multiply the whole equation by $\sum_l\sum_m(l+m)\tilde{c}_{lm}=(w^{sol}_++w^{sol}_-)/2$:
\begin{align}
    \tilde{\sigma}_{max}=\frac{w^{sol}_++w^{sol}_-}{2}\left[1+\frac{f_\mp}{2 p^{sol}_{\pm\mp}}\left( \frac{(f_+-1)(p^{sol}_{+-})^2p^{sol}_{-+}+(f_--1)(p^{sol}_{-+})^2p^{sol}_{+-}-2p_{+-}^{sol}p_{-+}^{sol}}{1-(f_+-1)(f_--1)p^{sol}_{+-}p^{sol}_{-+}} \right) \right]
\end{align}
In the asymmetric case, $\tilde{\sigma}_{max}$ diverges at the gel point, as shown in Fig.~\ref{fig:condas}. This is not physical, and we do not claim it to be so, as this approximation is meant as the upper most bound of the conductivity. This divergence is a result of the diverging weight average degree of aggregation (average size of the clusters). In the asymmetric case, as the clusters become larger, there is an increased tendency for the cluster to be highly charged. It should be noted that these highly charged clusters are causing this divergence will be tending towards macroscopic length scales involving thousands of ions, and therefore, should actually have a \textit{much smaller} diffusion coefficient than $D_0$. Note that for a symmetric RTIL we arrive at the much simpler formula that was written in the main text:
\begin{align}
    \tilde{\sigma}_{max} =  w^{sol}_\pm\left(\frac{1-p^{sol}}{1+(f-1)p^{sol}} \right)
\end{align}

The assumption that clusters all have equivalent diffusivities is obviously not true, but it does provide us with a closed form solution for a maximum conductivity. We may also obtain a very simple closed form for the minimum. A reasonable approximation is that free ions solely contribute to the ionic current~\cite{feng2019free}. This is effectively proposing that $D_{lm}=D_0\delta_{l+m,1}$, i.e., the diffusion coefficient goes to zero for any charged clusters containing more than one ion. Thus, we obtain
\begin{align}
    \sigma_{min} = \frac{e^2c_{salt}D_0}{k_BT}(\alpha_++\alpha_-)    
\end{align}
where $\alpha_\pm$ is the fraction of free cations or anions, given by
\begin{align}
    \alpha_\pm=(1-p_{\pm\mp})^{f_\pm}
\end{align}
For the symmetric RTIL we obtain
\begin{align}
    \tilde{\sigma}_{min} = (1-p)^f.
\end{align}

In Fig.~\ref{fig:condas} we display $\tilde{\sigma}$ as a function of $\lambda$ for the outlined approximations of $D_{lm}$ in the asymmetric case ($f_+ \neq f_-$). Specifically, we show the upper bound, the lower bound and also when $D_{lm} = D_0/(l + m)^{1/3}$. The latter case is simply the Stokes-Einstein relation for clusters. We note that the Stokes-Einstein form is likely to be an overestimate for the diffusitvities, as RTILs are a crowded environment. Interestingly, away from the gel point, despite the drastic differences in $D_{lm}$, all approximations are quite similar. At very small $\lambda$, practically all ions are free, and only a small number of ion pairs and clusters exist. For very large $\lambda$, again there is essentially only free ions present, with the rest of the ions being associated to the gel phase. It is at the intermediate region, close to the gel point, where the approximations become quite different. Actually, it is only the upper bound in the asymmetric case which becomes drastically different from the other two approximations. Remember, this upper bound for the conductivity is not physical.

There is another approach that can be taken to obtain an analytical expression for the conductivity. If we assume that the diffusion coefficient can be approximated as $D_{lm} = D_{\pm}/f_{\pm}W_{lm}$, then closed-form analytical expressions can be obtained (note we shall choose to work in the pre-gel regime for notational convenience). This choice of $D_{lm}$ is mostly chosen for mathematical convenience, but it holds certain limiting behaviour, which means that it might turn out to work well. The chosen form of $D_{lm}$ states that there is a strong suppression in the diffusion coefficients of large clusters. 

Using this, we can separate the sums out over all cluster sizes of one charge ($\pm z$), and perform the sum over all possible clusters. We shall only investigate the case of singly charged clusters here for $f > 2$ (smaller values of $f$ are trivial cases). 
In the sum for the conductivity, $l$ and $m$ are constrained, such that we only need to solve
\begin{multline}
\sigma = \frac{e^2c_{salt}}{k_BT}(1-p_{+-})(1-p_{-+})\Bigg\{\sum_{n}\dfrac{D_+}{p_{-+}}\left[\frac{p_{-+}}{1-p_{-+}}(1-p_{+-})^{f_+-1}\right]^{n+1}\left[\frac{p_{+-}}{1-p_{+-}}(1-p_{-+})^{f_--1}\right]^{n} + \\ \sum_{n}\dfrac{D_-}{p_{+-}}\left[\frac{p_{-+}}{1-p_{-+}}(1-p_{+-})^{f_+-1}\right]^{n}\left[\frac{p_{+-}}{1-p_{+-}}(1-p_{-+})^{f_--1}\right]^{n+1}\Bigg\}.
\end{multline}
This sum starts at $n = 0$ and ends at $n = N$ (note that in practice $N$ is so large that it can effectively be taken as $\infty$ without any loss of accuracy, but the $\infty$ is technically a contribution from the gel phase), and it can be readily solved to give
\begin{align}
\sigma = \frac{e^2c_{salt}}{k_BT}\dfrac{D_{+}(1 - p_{+-})^{f_+} + D_{-}(1 - p_{-+})^{f_-}}{1 - [p_{-+}(1 - p_{+-})^{f_+ - 2}][p_{+-}(1 - p_{-+})^{f_- - 2}]}.
\label{eq:sig1}
\end{align}
It is interesting to note that the presence of all singly charged species tends to simply rescale the diffusion coefficient
\begin{align}
\sigma = \frac{e^2c_{salt}}{k_BT}(D^{'}_{+}\alpha_+ + D^{'}_{-}\alpha_-).
\end{align}
In the limit of the symmetric RTIL case, we obtain the following
\begin{align}
\tilde{\sigma} = \dfrac{\alpha}{1 - p^{2}(1-p)^{2(f - 2)}}.
\label{eq:sig1}
\end{align}
The expression states that there is an increase over the free ion case, as charged clusters contribute to the conductivity. It is also clear that the increase predominantly occurs when $p$ is at intermediate values (not near 0 or 1). This approximation can be systematically improved upon, by adding clusters or higher charges, but the results become cumbersome quite quickly and depend on the functionalities more strongly.

\subsection{Asymmetric transference numbers}

Recall from the main text, the expression for the cationic transference number
\begin{equation}
    t_+ = \frac{\sum_{lm} l(l-m) \alpha_{lm} D_{lm}}{\sum_{lm} (l-m)^{2} \alpha_{lm} D_{lm}}.
    \label{eq:tp}
\end{equation}
We shall only show the cationic transference numbers, as obtaining the anionic ones are essentially identical. If we wish to obtain expressions for the transference numbers, we must know $D_{lm}$. We can again, use different approximations of $D_{lm}$ and study how this affects the transference numbers, as we do not know the exact $D_{lm}$.

\begin{figure*}[hbt!]
\centering
\includegraphics[width=\textwidth]{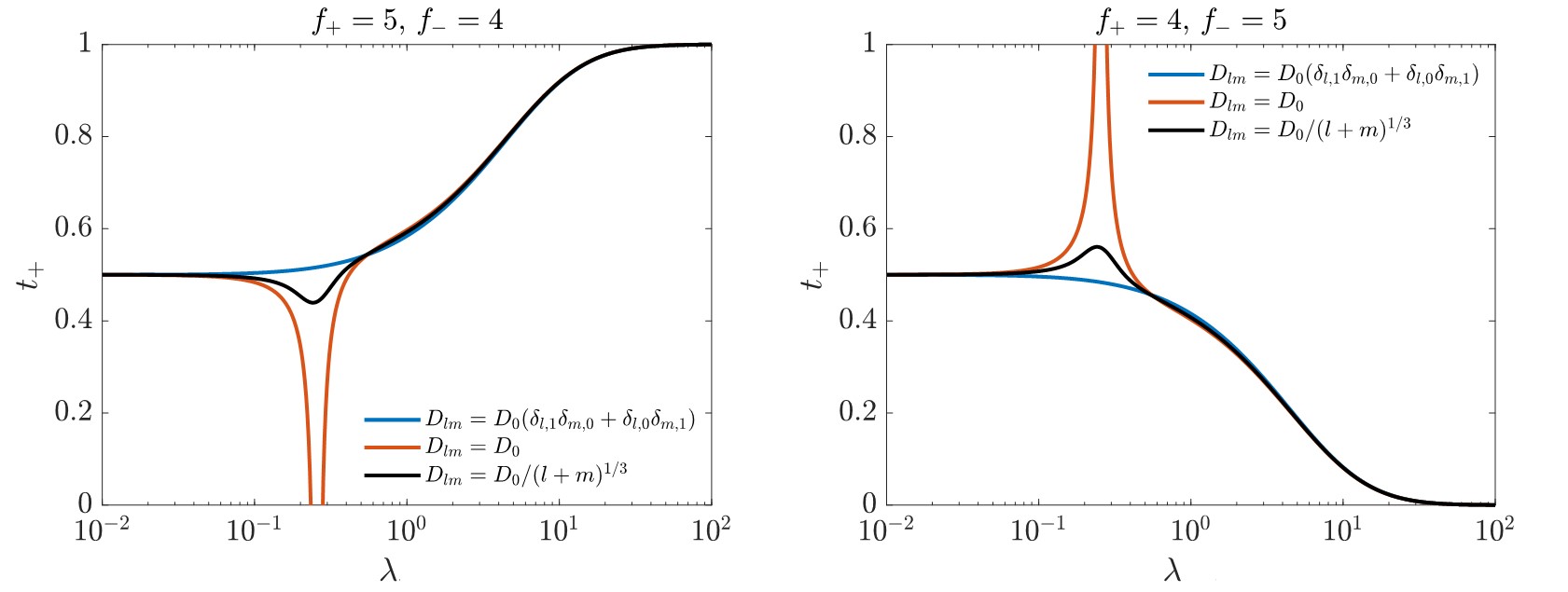}
\caption{The cation transference number is plotted as a function of $\lambda$ for a) $f_-=4$, $f_+=5$ and b)  $f_-=5$, $f_+=4$, using three different approximations for the cluster diffusivity: $D_{lm}=D_0(\delta_{l,1}\delta_{m,0}+\delta_{l,0}\delta_{m,1})$, $D_{lm}=D_0$, and $D_{lm}=D_0/(l+m)^{1/3}$. Note that for the case of $D_{lm}=D_0$ there is unphysical behaviour near the gel point. This is not expected to happen, as the divergence originates from extremely large clusters with high charges which, in the approximate, diffuse like free ions.}
\label{fig:trans}
\end{figure*}

Assuming $D_{lm}=D_0$ again, we can obtain closed-form expressions again
\begin{align}
    t_+=\frac{p_{-+}^{sol}/f_+ +\left[(f_+-1)(p^{sol}_{+-})^2p_{-+}^{sol}-p^{sol}_{+-}p^{sol}_{-+}\right]/\mathcal{B}}{p_{-+}^{sol}/f_+ + p_{+-}^{sol}/f_-+\left[(f_+-1)(p^{sol}_{+-})^2p^{sol}_{-+}+(f_--1)(p^{sol}_{-+})^2p^{sol}_{+-}-2p_{+-}^{sol}p_{-+}^{sol}\right]/\mathcal{B}}
\end{align}
Where $\mathcal{B}=1-p^{sol}_{+-}p^{sol}_{+-}(f_+-1)(f_--1)$. Note that the transference numbers also have unphysical behaviour near the gel point in the asymmetric case, as shown in Fig.~\ref{fig:trans}. The cation transference number tends to very negative values when the functionality of cations is larger than anions. This is because very large negative clusters are being formed, and with $D_{lm} = D_0$ these dominate in this approximation. Remember, this is a upper/lower bound (depending on the functionality) for $t_+$. We do not expect such behaviour to actually occur. 

Trivially, if we only count the free ions and they have the same diffusion coefficient
\begin{equation}
t_+ = \dfrac{\alpha_+}{\alpha_+ + \alpha_-}.
\end{equation}

In Fig.~\ref{fig:trans} we display the approximations for the cation transference numbers outlined before for two sets of functionalities. We also display a more reasonable physical case is for when $D_{lm} = D_0/(l + m)^{1/3}$, which was motivated by Stokes-Einstein. This is still likely to be an overestimate for the diffusivities of large clusters. Similar to conductivity, we find that at very small and large $\lambda$, all curves remain very close. This can be attributed to the transference numbers being dominated by free ions in these limits. We observe the largest deviations near the gel point. The $D_{lm} = D_0/(l + m)^{1/3}$ case is explained in detail in the main text, so we refrain from repeating that explanation here. The approximation of only free ions contributing does not exhibit any interesting features near the gel point, owing to the fraction of free ions changing continuously through the gel point. In the case of $D_{lm} = D_0$, there is an unphysical strong divergence in the transference number. We do not expect the transference numbers to diverge in this way, as the approximation of $D_{lm} = D_0$ does not physically make sense.

It does, however, raise an interesting question about transference numbers being negative and larger than 1. Within the approximations outlined here, it appears as though negative transference numbers appear near the gel point only if $D_{lm} = D_0$. In all approximations, however, the transference number approaches to 1 or 0 in the limit of very large $\lambda$. This corresponds to essentially only one type of ion being free, with all of the other ions being bound up in the gel. If the transference numbers are close to 0, then only small perturbations are required to get negative transference numbers.

Again, a perturbative approach can also be used to investigate the transference numbers. If we once again only assume singly charged clusters contribute and take the sharply decaying diffusivity. Naturally, we must investigate the asymmetric case, otherwise we trivially obtain symmetric transference numbers. The denominator for the transference number is given by the conductivity expression, so we shall only focus on the numerator. Using the outlined approximations, it becomes

\begin{multline}
t_+\tilde{\sigma} = \dfrac{(1-p_{+-})(1-p_{-+})}{2}\Bigg\{\sum_{n}\dfrac{nD_+}{p_{-+}}\left[\frac{p_{-+}}{1-p_{-+}}(1-p_{+-})^{f_+-1}\right]^{n}\left[\frac{p_{+-}}{1-p_{+-}}(1-p_{-+})^{f_--1}\right]^{n-1} - \\ \sum_{n}\dfrac{nD_-}{p_{+-}}\left[\frac{p_{-+}}{1-p_{-+}}(1-p_{+-})^{f_+-1}\right]^{n}\left[\frac{p_{+-}}{1-p_{+-}}(1-p_{-+})^{f_--1}\right]^{n+1}\Bigg\}.
\end{multline}
Again, this sum is easy to solve, and it yields
\begin{multline}
t_+ = \dfrac{D_+(1-p_{+-})^{f_+} - D_-(1-p_{-+})^{f_-}[p_{-+}(1 - p_{+-})^{f_+ - 2}][p_{+-}(1 - p_{-+})^{f_- - 2}]}{\Big\{1 - [p_{-+}(1 - p_{+-})^{f_+ - 2}][p_{+-}(1 - p_{-+})^{f_- - 2}]\Big\}\Big\{D_{+}(1 - p_{+-})^{f_+} + D_{-}(1 - p_{-+})^{f_-}\Big\}}
\end{multline}
In the symmetric RTIL limit, these expression reduce to 1/2. In the limit of only free ions, we have 

\begin{equation}
t_+ = \dfrac{D_+\alpha_+}{D_{+}\alpha_+ + D_{-}\alpha_-}
\end{equation}








\subsection{Gel Definition in MD Simulations}

\begin{figure*}[hbt!]
\centering
\includegraphics[width=0.5\textwidth]{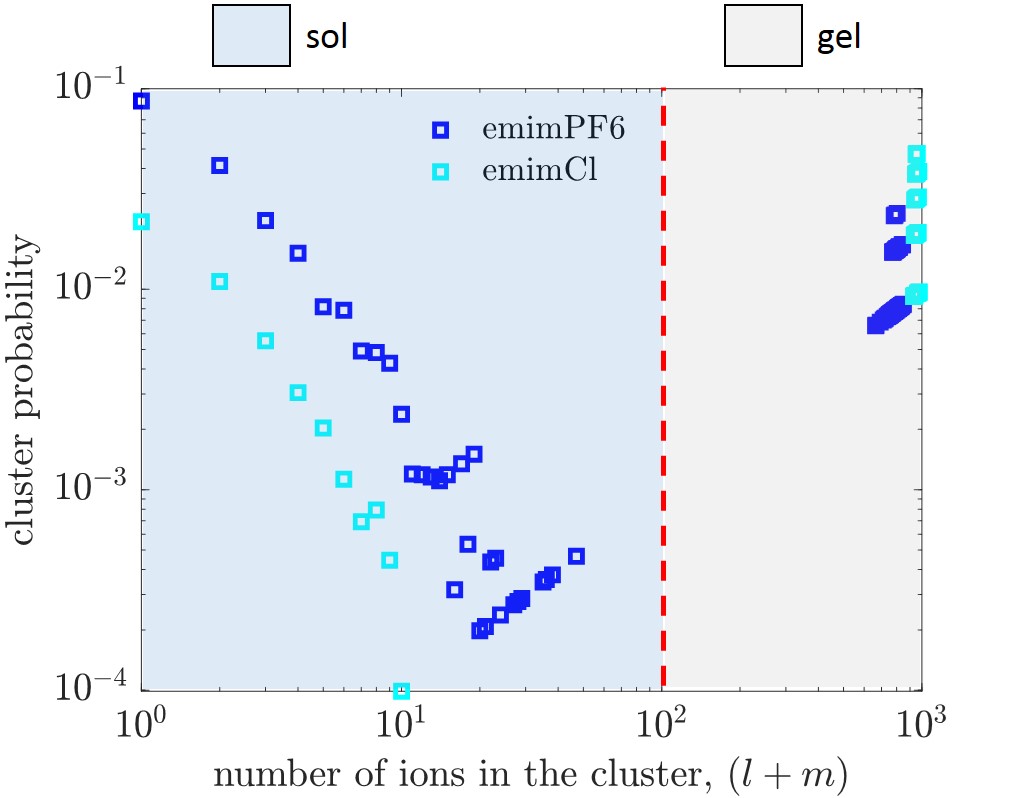}
\caption{The probability distribution of clusters observed from the molecular dynamics simulations is plotted for both of the gel-forming RTILs: emimCl and emimPF$_6$. The discontinuity in the probability distribution provides a natural cutoff for defining the gel network and finite sol clusters in the molecular dynamics  simulations. In practice we define a cluster size threshold of one hundred ions (red dashed line) to  define the gel/sol cutoff.}
\label{fig:gel}
\end{figure*}

We define ions to be a part of the gel if it was incorporated in a cluster that is larger than one hundred ions. Ostensibly this seems like an arbitrary cutoff. The precise distinction between gel and sol can be seen explicitly in Fig.~\ref{fig:gel}, where we plot (on a log-log scale) the probability distribution of ion clusters in emimCl and emimPF$_6$. We clearly see that there is a distinct discontinuity in the probability distribution of clusters. The probability distribution of clusters decays monotonically with cluster size up to a cluster size of about 100 ions. We then observe a cluster probability peak for both RTILs at cluster sizes of around 500 ions. This peak corresponds to the gel. In principle, We could define the sol to be any cluster outside of this peak. In practice, we simply define clusters containing less than one hundred ions, to be in the sol, and everything else to be in the gel.

\subsection{Varying Iso-Density Threshold for Ion association}

\begin{figure*}[hbt!]
\centering
\includegraphics[width=\textwidth]{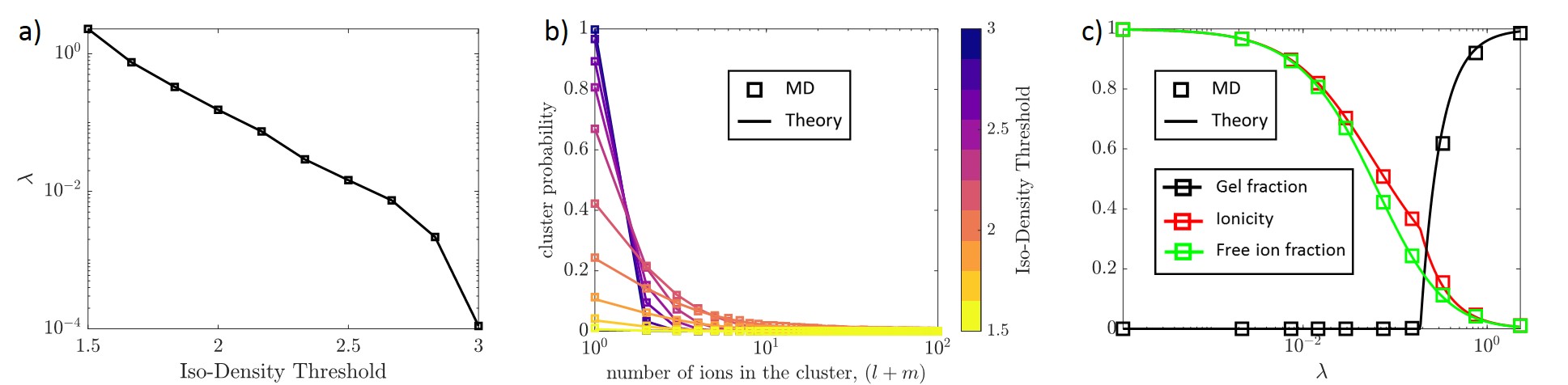}
\caption{(a) Computed (from MD simulations) ionic association constants, $\lambda$, as a function of the chosen iso-density threshold value that defines the ion association. Solid line drawn for guide of the eye. (b) The ionic cluster distribution (cluster probability as a function of size) for different iso-density threshold values varying from 1.5 to 3 times the bulk density of cations/anions. (c) The gel fraction, ionicity [Eq. (16) of the main text], and free ion fraction are plotted as a function of the ionic association constant ($\lambda$), with molecular dynamics values as open squares, and theoretical values solid curves.}
\label{fig:thresh}
\end{figure*}

Varying the iso-density threshold that defines ionic association changes the apparent ionic association constant, $\lambda$, of the RTIL. Thus, by changing the threshold value of ion association for a specific RTIL, we can explore how its ion association is affected by $\lambda$ (remember, this is calculated using the ion functionalities, association probabilities and mass action law, as outlined in the main text) for a more direct theoretical comparison of MD and theory. As an example, we have analyzed the ion association of emimBF$_4$ by varying the iso-density threshold for ion association from 1.5 to 3 $\times$ the bulk density at 295~. In Fig.~\ref{fig:thresh}b \&  c, we plot the cluster distribution, gel fraction, ionicity, and free ion fraction as functions of the iso-density threshold and the corresponding ion association constants (Fig. \ref{fig:thresh}a). We see that the computed association constants decrease monotonically with increasing iso-density threshold. This is primarily because decreasing the threshold increases volume of the ionic ``hot-spots" increasing the apparent ionic association constant. The computed $\lambda$ values produce theoretical cluster distributions that nearly perfectly match the simulated distributions as seen in in Fig.~\ref{fig:thresh}b. In addition to the cluster distribution, we see in Fig.~\ref{fig:thresh}c that the apparent ionicity and free ion fraction is vastly affected by our choice of threshold, albeit in theoretically  predictable manner. Specifically, the fraction of the ionicity that is accounted for by free ions changes as a function of $\lambda$ in a manner that is exactly expected from our theory. Finally, we observe that the presence and extent of gel is strongly dependent on the choice of iso-density threshold. Thus, because the precise microscopic definition of ionic association is somewhat nebulous, we cannot strongly assert strongly that RTILs for ionic gels without further experiments or simulations. Nonetheless, our conclusion remains that even for substantially different iso-density threshold criteria, our theory is able to model the resulting ion association and gelation behavior.

\subsection{Simulation Details}

In this study, we performed all-atom classical MD simulations using LAMMPS \cite{plimpton1995} in a fully periodic geometry. For model emimTFSI, emimBF$_4$, emimCl, emimPF$_6$, bmimPF$_6$, and hmimPF$_6$ RTILs, we performed simulations containing 500 ion pairs.  The simulations were performed at fixed temperature (295 K) and pressure (1 bar), with Nose-Hoover thermostat and barostat until the density of the fluid relaxed to a constant, which required 10 ns, with 1 fs time steps. Next, we switched to constant volume simulation box still with a fixed temperature of 295 K, again using the Nose-Hoover thermostat, and equilibrate for an additional 4 ns. The production runs were performed for an additional 6 ns to obtain the average probabilities for the clusters. The initial configurations for all simulations were generated using the open-source software, PACKMOL \cite{martinez2009packmol}. We used the open source software TRAVIS to measure the spatial distribution functions of the center of mass of cations around molecular anions, and vice versa. The spatial distribution functions were then visualized using the open-source software, VMD \cite{HUMP96}.

We employed the CL$\&$P force field to model all RTILs, which was developed for RTILs, with same functional form as the OPLSAA force field\cite{CanongiaLopes2012}. Interatomic interactions are determined using Lorentz-Berthelot mixing rules. Long range electrostatic interactions were computed using the Particle-Particle Particle-Mesh (PPPM) solver (with a cut-off length of 12 ${\text \AA}$), which maps particle charge to a 3D mesh for the periodic simulations.

\subsection{Spatial Distribution Functions: Visualization}

\begin{figure*}[hbt!]
\centering
\includegraphics[width=0.75\textwidth]{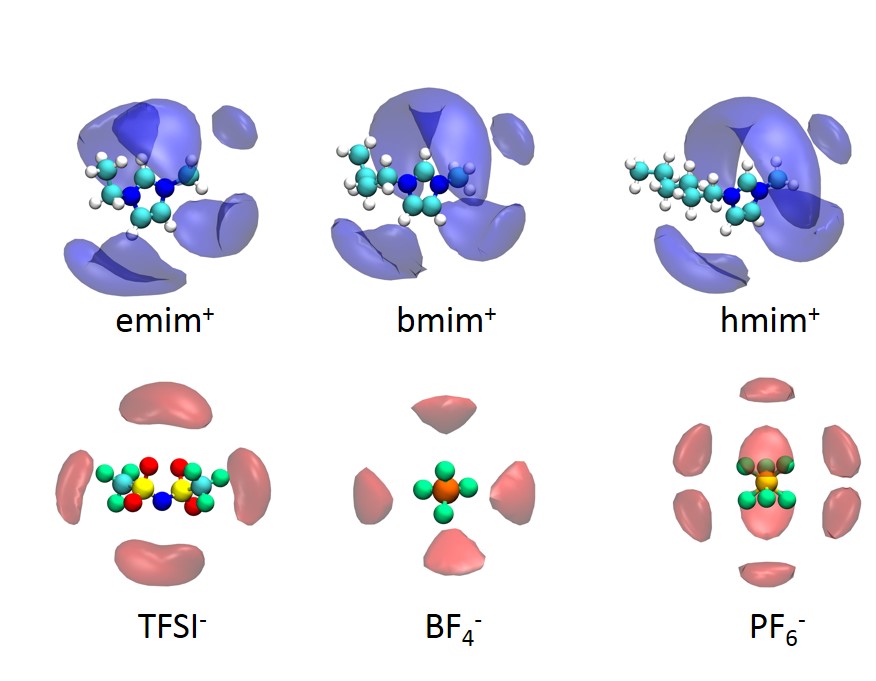}
\caption{The spatial distribution functions of 6 molecular ions visualized as iso-density surfaces corresponding to twice the bulk density. The iso-density surfaces correspond to regions of especially high density of counter-ion, indicating that there are preferred orientations between neighboring ions.}
\label{fig:sdf}
\end{figure*}

It is physically instructive to visualize the SDFs of the RTILs from each of our simulations. In Fig.~\ref{fig:sdf}, we show iso-density surfaces of counter-ions (center-of-mass) distributed around each of the simulated ions. These iso-density surfaces correspond to twice the bulk density of the counter-ions. This visualized iso-density corresponds to the threshold used to define ionic associations. For each of the cations (emim$^+$, bmim$^+$, and hmim$^+$), the visualized SDF's in Fig.~\ref{fig:sdf}, are from simulations with PF$_6^-$ as the counter-ion. Similarly, for each of the anions (TFSI$^-$, BF$_4^-$, and PF$_6^-$), the visualized SDFs are from simulations with emim$^+$ as the counter-ion.

\bibliography{main_SI.bib}

\begin{thebibliography}{88}%
\makeatletter
\providecommand \@ifxundefined [1]{%
 \@ifx{#1\undefined}
}%
\providecommand \@ifnum [1]{%
 \ifnum #1\expandafter \@firstoftwo
 \else \expandafter \@secondoftwo
 \fi
}%
\providecommand \@ifx [1]{%
 \ifx #1\expandafter \@firstoftwo
 \else \expandafter \@secondoftwo
 \fi
}%
\providecommand \natexlab [1]{#1}%
\providecommand \enquote  [1]{``#1''}%
\providecommand \bibnamefont  [1]{#1}%
\providecommand \bibfnamefont [1]{#1}%
\providecommand \citenamefont [1]{#1}%
\providecommand \href@noop [0]{\@secondoftwo}%
\providecommand \href [0]{\begingroup \@sanitize@url \@href}%
\providecommand \@href[1]{\@@startlink{#1}\@@href}%
\providecommand \@@href[1]{\endgroup#1\@@endlink}%
\providecommand \@sanitize@url [0]{\catcode `\\12\catcode `\$12\catcode
  `\&12\catcode `\#12\catcode `\^12\catcode `\_12\catcode `\%12\relax}%
\providecommand \@@startlink[1]{}%
\providecommand \@@endlink[0]{}%
\providecommand \url  [0]{\begingroup\@sanitize@url \@url }%
\providecommand \@url [1]{\endgroup\@href {#1}{\urlprefix }}%
\providecommand \urlprefix  [0]{URL }%
\providecommand \Eprint [0]{\href }%
\providecommand \doibase [0]{http://dx.doi.org/}%
\providecommand \selectlanguage [0]{\@gobble}%
\providecommand \bibinfo  [0]{\@secondoftwo}%
\providecommand \bibfield  [0]{\@secondoftwo}%
\providecommand \translation [1]{[#1]}%
\providecommand \BibitemOpen [0]{}%
\providecommand \bibitemStop [0]{}%
\providecommand \bibitemNoStop [0]{.\EOS\space}%
\providecommand \EOS [0]{\spacefactor3000\relax}%
\providecommand \BibitemShut  [1]{\csname bibitem#1\endcsname}%
\let\auto@bib@innerbib\@empty
\bibitem [{\citenamefont {Hallett}\ and\ \citenamefont
  {Welton}(2011)}]{Hallett2011}%
  \BibitemOpen
  \bibfield  {author} {\bibinfo {author} {\bibfnamefont {J.~P.}\ \bibnamefont
  {Hallett}}\ and\ \bibinfo {author} {\bibfnamefont {T.}~\bibnamefont
  {Welton}},\ }\href@noop {} {\bibfield  {journal} {\bibinfo  {journal} {Chem.
  Rev.}\ }\textbf {\bibinfo {volume} {111}},\ \bibinfo {pages} {3508} (\bibinfo
  {year} {2011})}\BibitemShut {NoStop}%
\bibitem [{\citenamefont {Welton}(1999)}]{Welton1999}%
  \BibitemOpen
  \bibfield  {author} {\bibinfo {author} {\bibfnamefont {T.}~\bibnamefont
  {Welton}},\ }\href@noop {} {\bibfield  {journal} {\bibinfo  {journal} {Chem.
  Rev.}\ }\textbf {\bibinfo {volume} {99}},\ \bibinfo {pages} {2071} (\bibinfo
  {year} {1999})}\BibitemShut {NoStop}%
\bibitem [{\citenamefont {Wasserscheid}\ and\ \citenamefont
  {Welton}(2008)}]{wasserscheid2008}%
  \BibitemOpen
  \bibfield  {author} {\bibinfo {author} {\bibfnamefont {P.}~\bibnamefont
  {Wasserscheid}}\ and\ \bibinfo {author} {\bibfnamefont {T.}~\bibnamefont
  {Welton}},\ }\href@noop {} {\emph {\bibinfo {title} {Ionic liquids in
  synthesis}}}\ (\bibinfo  {publisher} {John Wiley \& Sons},\ \bibinfo {year}
  {2008})\BibitemShut {NoStop}%
\bibitem [{\citenamefont {Fedorov}\ and\ \citenamefont
  {Kornyshev}(2014)}]{Fedorov2014}%
  \BibitemOpen
  \bibfield  {author} {\bibinfo {author} {\bibfnamefont {M.~V.}\ \bibnamefont
  {Fedorov}}\ and\ \bibinfo {author} {\bibfnamefont {A.~A.}\ \bibnamefont
  {Kornyshev}},\ }\href {\doibase 10.1021/cr400374x} {\bibfield  {journal}
  {\bibinfo  {journal} {Chem. Rev.}\ }\textbf {\bibinfo {volume} {114}},\
  \bibinfo {pages} {2978} (\bibinfo {year} {2014})}\BibitemShut {NoStop}%
\bibitem [{\citenamefont {Weing\"{a}rtner}(2008)}]{Hermann2008}%
  \BibitemOpen
  \bibfield  {author} {\bibinfo {author} {\bibfnamefont {H.}~\bibnamefont
  {Weing\"{a}rtner}},\ }\href@noop {} {\bibfield  {journal} {\bibinfo
  {journal} {Angewandte Chemie International Edition}\ }\textbf {\bibinfo
  {volume} {47}},\ \bibinfo {pages} {654} (\bibinfo {year} {2008})}\BibitemShut
  {NoStop}%
\bibitem [{\citenamefont {Kondrat}\ and\ \citenamefont
  {Kornyshev}(2016)}]{Kondrat2016}%
  \BibitemOpen
  \bibfield  {author} {\bibinfo {author} {\bibfnamefont {S.}~\bibnamefont
  {Kondrat}}\ and\ \bibinfo {author} {\bibfnamefont {A.~A.}\ \bibnamefont
  {Kornyshev}},\ }\href@noop {} {\bibfield  {journal} {\bibinfo  {journal}
  {Nanoscale Horiz.}\ }\textbf {\bibinfo {volume} {1}},\ \bibinfo {pages} {45}
  (\bibinfo {year} {2016})}\BibitemShut {NoStop}%
\bibitem [{\citenamefont {Son}\ and\ \citenamefont {Wang}(2020)}]{son2020ion}%
  \BibitemOpen
  \bibfield  {author} {\bibinfo {author} {\bibfnamefont {C.~Y.}\ \bibnamefont
  {Son}}\ and\ \bibinfo {author} {\bibfnamefont {Z.-G.}\ \bibnamefont {Wang}},\
  }\href@noop {} {\bibfield  {journal} {\bibinfo  {journal} {The Journal of
  Chemical Physics}\ }\textbf {\bibinfo {volume} {153}},\ \bibinfo {pages}
  {100903} (\bibinfo {year} {2020})}\BibitemShut {NoStop}%
\bibitem [{\citenamefont {MacFarlane}\ \emph {et~al.}(2009)\citenamefont
  {MacFarlane}, \citenamefont {Forsyth}, \citenamefont {Izgorodina},
  \citenamefont {Abbott}, \citenamefont {Annata},\ and\ \citenamefont
  {Fraser}}]{MacFarlane2009}%
  \BibitemOpen
  \bibfield  {author} {\bibinfo {author} {\bibfnamefont {D.~R.}\ \bibnamefont
  {MacFarlane}}, \bibinfo {author} {\bibfnamefont {M.}~\bibnamefont {Forsyth}},
  \bibinfo {author} {\bibfnamefont {E.~I.}\ \bibnamefont {Izgorodina}},
  \bibinfo {author} {\bibfnamefont {A.~P.}\ \bibnamefont {Abbott}}, \bibinfo
  {author} {\bibfnamefont {G.}~\bibnamefont {Annata}}, \ and\ \bibinfo {author}
  {\bibfnamefont {K.}~\bibnamefont {Fraser}},\ }\href@noop {} {\bibfield
  {journal} {\bibinfo  {journal} {Phys. Chem. Chem. Phys.}\ }\textbf {\bibinfo
  {volume} {11}},\ \bibinfo {pages} {4962} (\bibinfo {year}
  {2009})}\BibitemShut {NoStop}%
\bibitem [{\citenamefont {Holl{\'o}czki}\ \emph {et~al.}(2014)\citenamefont
  {Holl{\'o}czki}, \citenamefont {Malberg}, \citenamefont {Welton},\ and\
  \citenamefont {Kirchner}}]{Kirchner2014}%
  \BibitemOpen
  \bibfield  {author} {\bibinfo {author} {\bibfnamefont {O.}~\bibnamefont
  {Holl{\'o}czki}}, \bibinfo {author} {\bibfnamefont {F.}~\bibnamefont
  {Malberg}}, \bibinfo {author} {\bibfnamefont {T.}~\bibnamefont {Welton}}, \
  and\ \bibinfo {author} {\bibfnamefont {B.}~\bibnamefont {Kirchner}},\
  }\href@noop {} {\bibfield  {journal} {\bibinfo  {journal} {Phys. Chem. Chem.
  Phys.}\ }\textbf {\bibinfo {volume} {16}},\ \bibinfo {pages} {16880}
  (\bibinfo {year} {2014})}\BibitemShut {NoStop}%
\bibitem [{\citenamefont {Zhang}\ and\ \citenamefont
  {Maginn}(2015)}]{Zhang2015}%
  \BibitemOpen
  \bibfield  {author} {\bibinfo {author} {\bibfnamefont {Y.}~\bibnamefont
  {Zhang}}\ and\ \bibinfo {author} {\bibfnamefont {E.~J.}\ \bibnamefont
  {Maginn}},\ }\href@noop {} {\bibfield  {journal} {\bibinfo  {journal} {J.
  Phys. Chem. Lett.}\ }\textbf {\bibinfo {volume} {6}},\ \bibinfo {pages} {700}
  (\bibinfo {year} {2015})}\BibitemShut {NoStop}%
\bibitem [{\citenamefont {Feng}\ \emph {et~al.}(2019)\citenamefont {Feng},
  \citenamefont {Chen}, \citenamefont {Bi}, \citenamefont {Goodwin},
  \citenamefont {Postnikov}, \citenamefont {Brilliantov}, \citenamefont
  {Urbakh},\ and\ \citenamefont {Kornyshev}}]{feng2019free}%
  \BibitemOpen
  \bibfield  {author} {\bibinfo {author} {\bibfnamefont {G.}~\bibnamefont
  {Feng}}, \bibinfo {author} {\bibfnamefont {M.}~\bibnamefont {Chen}}, \bibinfo
  {author} {\bibfnamefont {S.}~\bibnamefont {Bi}}, \bibinfo {author}
  {\bibfnamefont {Z.~A.}\ \bibnamefont {Goodwin}}, \bibinfo {author}
  {\bibfnamefont {E.~B.}\ \bibnamefont {Postnikov}}, \bibinfo {author}
  {\bibfnamefont {N.}~\bibnamefont {Brilliantov}}, \bibinfo {author}
  {\bibfnamefont {M.}~\bibnamefont {Urbakh}}, \ and\ \bibinfo {author}
  {\bibfnamefont {A.~A.}\ \bibnamefont {Kornyshev}},\ }\href@noop {} {\bibfield
   {journal} {\bibinfo  {journal} {Phys. Rev. X}\ }\textbf {\bibinfo {volume}
  {9}},\ \bibinfo {pages} {021024} (\bibinfo {year} {2019})}\BibitemShut
  {NoStop}%
\bibitem [{\citenamefont {Lee}\ \emph {et~al.}(2014)\citenamefont {Lee},
  \citenamefont {Vella}, \citenamefont {Perkin},\ and\ \citenamefont
  {Goriely}}]{lee2014room}%
  \BibitemOpen
  \bibfield  {author} {\bibinfo {author} {\bibfnamefont {A.~A.}\ \bibnamefont
  {Lee}}, \bibinfo {author} {\bibfnamefont {D.}~\bibnamefont {Vella}}, \bibinfo
  {author} {\bibfnamefont {S.}~\bibnamefont {Perkin}}, \ and\ \bibinfo {author}
  {\bibfnamefont {A.}~\bibnamefont {Goriely}},\ }\href@noop {} {\bibfield
  {journal} {\bibinfo  {journal} {J. Phys. Chem. Lett.}\ }\textbf {\bibinfo
  {volume} {6}},\ \bibinfo {pages} {159} (\bibinfo {year} {2014})}\BibitemShut
  {NoStop}%
\bibitem [{\citenamefont {Goodwin}\ and\ \citenamefont
  {Kornyshev}(2017)}]{goodwin2017underscreening}%
  \BibitemOpen
  \bibfield  {author} {\bibinfo {author} {\bibfnamefont {Z.~A.}\ \bibnamefont
  {Goodwin}}\ and\ \bibinfo {author} {\bibfnamefont {A.~A.}\ \bibnamefont
  {Kornyshev}},\ }\href@noop {} {\bibfield  {journal} {\bibinfo  {journal}
  {Electrochem. Commun.}\ }\textbf {\bibinfo {volume} {82}},\ \bibinfo {pages}
  {129} (\bibinfo {year} {2017})}\BibitemShut {NoStop}%
\bibitem [{\citenamefont {Goodwin}\ \emph {et~al.}(2017)\citenamefont
  {Goodwin}, \citenamefont {Feng},\ and\ \citenamefont
  {Kornyshev}}]{goodwin2017mean}%
  \BibitemOpen
  \bibfield  {author} {\bibinfo {author} {\bibfnamefont {Z.~A.}\ \bibnamefont
  {Goodwin}}, \bibinfo {author} {\bibfnamefont {G.}~\bibnamefont {Feng}}, \
  and\ \bibinfo {author} {\bibfnamefont {A.~A.}\ \bibnamefont {Kornyshev}},\
  }\href@noop {} {\bibfield  {journal} {\bibinfo  {journal} {Electrochim.
  Acta}\ }\textbf {\bibinfo {volume} {225}},\ \bibinfo {pages} {190} (\bibinfo
  {year} {2017})}\BibitemShut {NoStop}%
\bibitem [{\citenamefont {Chen}\ \emph {et~al.}(2018)\citenamefont {Chen},
  \citenamefont {Goodwin}, \citenamefont {Feng},\ and\ \citenamefont
  {Kornyshev}}]{Chen2017}%
  \BibitemOpen
  \bibfield  {author} {\bibinfo {author} {\bibfnamefont {M.}~\bibnamefont
  {Chen}}, \bibinfo {author} {\bibfnamefont {Z.~A.~H.}\ \bibnamefont
  {Goodwin}}, \bibinfo {author} {\bibfnamefont {G.}~\bibnamefont {Feng}}, \
  and\ \bibinfo {author} {\bibfnamefont {A.~A.}\ \bibnamefont {Kornyshev}},\
  }\href {\doibase 10.1016/j.jelechem.2017.11.005} {\bibfield  {journal}
  {\bibinfo  {journal} {J. Electroanal. Chem.}\ }\textbf {\bibinfo {volume}
  {819}},\ \bibinfo {pages} {347} (\bibinfo {year} {2018})}\BibitemShut
  {NoStop}%
\bibitem [{\citenamefont {Adar}\ \emph {et~al.}(2017)\citenamefont {Adar},
  \citenamefont {Markovich},\ and\ \citenamefont {Andelman}}]{adar2017bjerrum}%
  \BibitemOpen
  \bibfield  {author} {\bibinfo {author} {\bibfnamefont {R.~M.}\ \bibnamefont
  {Adar}}, \bibinfo {author} {\bibfnamefont {T.}~\bibnamefont {Markovich}}, \
  and\ \bibinfo {author} {\bibfnamefont {D.}~\bibnamefont {Andelman}},\
  }\href@noop {} {\bibfield  {journal} {\bibinfo  {journal} {J. Chem. Phys.}\
  }\textbf {\bibinfo {volume} {146}},\ \bibinfo {pages} {194904} (\bibinfo
  {year} {2017})}\BibitemShut {NoStop}%
\bibitem [{\citenamefont {Hayes}\ \emph {et~al.}(2015)\citenamefont {Hayes},
  \citenamefont {Warr},\ and\ \citenamefont {Atkin}}]{Hayes2015}%
  \BibitemOpen
  \bibfield  {author} {\bibinfo {author} {\bibfnamefont {R.}~\bibnamefont
  {Hayes}}, \bibinfo {author} {\bibfnamefont {G.~G.}\ \bibnamefont {Warr}}, \
  and\ \bibinfo {author} {\bibfnamefont {R.}~\bibnamefont {Atkin}},\
  }\href@noop {} {\bibfield  {journal} {\bibinfo  {journal} {Chem. Rev.}\
  }\textbf {\bibinfo {volume} {115}},\ \bibinfo {pages} {6357} (\bibinfo {year}
  {2015})}\BibitemShut {NoStop}%
\bibitem [{\citenamefont {Smith}\ \emph {et~al.}(2016)\citenamefont {Smith},
  \citenamefont {Lee},\ and\ \citenamefont {Perkin}}]{Smith2016}%
  \BibitemOpen
  \bibfield  {author} {\bibinfo {author} {\bibfnamefont {A.~M.}\ \bibnamefont
  {Smith}}, \bibinfo {author} {\bibfnamefont {A.~A.}\ \bibnamefont {Lee}}, \
  and\ \bibinfo {author} {\bibfnamefont {S.}~\bibnamefont {Perkin}},\ }\href
  {\doibase 10.1021/acs.jpclett.6b00867} {\bibfield  {journal} {\bibinfo
  {journal} {J. Phys. Chem. Lett.}\ }\textbf {\bibinfo {volume} {7}},\ \bibinfo
  {pages} {2157} (\bibinfo {year} {2016})}\BibitemShut {NoStop}%
\bibitem [{\citenamefont {Gebbie}\ \emph {et~al.}(2013)\citenamefont {Gebbie},
  \citenamefont {Valtiner}, \citenamefont {Banquy}, \citenamefont {Fox},
  \citenamefont {Henderson},\ and\ \citenamefont {Israelachvili}}]{Gebbie2013}%
  \BibitemOpen
  \bibfield  {author} {\bibinfo {author} {\bibfnamefont {M.~A.}\ \bibnamefont
  {Gebbie}}, \bibinfo {author} {\bibfnamefont {M.}~\bibnamefont {Valtiner}},
  \bibinfo {author} {\bibfnamefont {X.}~\bibnamefont {Banquy}}, \bibinfo
  {author} {\bibfnamefont {E.~T.}\ \bibnamefont {Fox}}, \bibinfo {author}
  {\bibfnamefont {W.~A.}\ \bibnamefont {Henderson}}, \ and\ \bibinfo {author}
  {\bibfnamefont {J.~N.}\ \bibnamefont {Israelachvili}},\ }\href {\doibase
  10.1073/pnas.1307871110} {\bibfield  {journal} {\bibinfo  {journal} {Proc.
  Natl. Acad. Sci. U.S.A}\ }\textbf {\bibinfo {volume} {110}},\ \bibinfo
  {pages} {9674} (\bibinfo {year} {2013})}\BibitemShut {NoStop}%
\bibitem [{\citenamefont {Gebbie}\ \emph {et~al.}(2015)\citenamefont {Gebbie},
  \citenamefont {Dobes}, \citenamefont {Valtiner},\ and\ \citenamefont
  {Israelachvili}}]{Gebbie2015}%
  \BibitemOpen
  \bibfield  {author} {\bibinfo {author} {\bibfnamefont {M.~A.}\ \bibnamefont
  {Gebbie}}, \bibinfo {author} {\bibfnamefont {H.~A.}\ \bibnamefont {Dobes}},
  \bibinfo {author} {\bibfnamefont {M.}~\bibnamefont {Valtiner}}, \ and\
  \bibinfo {author} {\bibfnamefont {J.~N.}\ \bibnamefont {Israelachvili}},\
  }\href@noop {} {\bibfield  {journal} {\bibinfo  {journal} {Proc. Natl. Acad.
  Sci. U.S.A}\ }\textbf {\bibinfo {volume} {112}},\ \bibinfo {pages} {7432}
  (\bibinfo {year} {2015})}\BibitemShut {NoStop}%
\bibitem [{\citenamefont {Avni}\ \emph {et~al.}(2020)\citenamefont {Avni},
  \citenamefont {Adar},\ and\ \citenamefont {Andelman}}]{avni2020charge}%
  \BibitemOpen
  \bibfield  {author} {\bibinfo {author} {\bibfnamefont {Y.}~\bibnamefont
  {Avni}}, \bibinfo {author} {\bibfnamefont {R.~M.}\ \bibnamefont {Adar}}, \
  and\ \bibinfo {author} {\bibfnamefont {D.}~\bibnamefont {Andelman}},\
  }\href@noop {} {\bibfield  {journal} {\bibinfo  {journal} {Phys. Rev. E}\
  }\textbf {\bibinfo {volume} {101}},\ \bibinfo {pages} {010601} (\bibinfo
  {year} {2020})}\BibitemShut {NoStop}%
\bibitem [{\citenamefont {France-Lanord}\ and\ \citenamefont
  {Grossman}(2019)}]{france2019}%
  \BibitemOpen
  \bibfield  {author} {\bibinfo {author} {\bibfnamefont {A.}~\bibnamefont
  {France-Lanord}}\ and\ \bibinfo {author} {\bibfnamefont {J.~C.}\ \bibnamefont
  {Grossman}},\ }\href@noop {} {\bibfield  {journal} {\bibinfo  {journal}
  {Physical review letters}\ }\textbf {\bibinfo {volume} {122}},\ \bibinfo
  {pages} {136001} (\bibinfo {year} {2019})}\BibitemShut {NoStop}%
\bibitem [{\citenamefont {McEldrew}\ \emph {et~al.}(2020)\citenamefont
  {McEldrew}, \citenamefont {Goodwin}, \citenamefont {Bi}, \citenamefont
  {Bazant},\ and\ \citenamefont {Kornyshev}}]{mceldrew2020theory}%
  \BibitemOpen
  \bibfield  {author} {\bibinfo {author} {\bibfnamefont {M.}~\bibnamefont
  {McEldrew}}, \bibinfo {author} {\bibfnamefont {Z.~A.}\ \bibnamefont
  {Goodwin}}, \bibinfo {author} {\bibfnamefont {S.}~\bibnamefont {Bi}},
  \bibinfo {author} {\bibfnamefont {M.~Z.}\ \bibnamefont {Bazant}}, \ and\
  \bibinfo {author} {\bibfnamefont {A.~A.}\ \bibnamefont {Kornyshev}},\
  }\href@noop {} {\bibfield  {journal} {\bibinfo  {journal} {J. Chem. Phys.}\
  }\textbf {\bibinfo {volume} {152}},\ \bibinfo {pages} {234506} (\bibinfo
  {year} {2020})}\BibitemShut {NoStop}%
\bibitem [{\citenamefont {Flory}(1942)}]{flory1942thermodynamics}%
  \BibitemOpen
  \bibfield  {author} {\bibinfo {author} {\bibfnamefont {P.~J.}\ \bibnamefont
  {Flory}},\ }\href@noop {} {\bibfield  {journal} {\bibinfo  {journal} {J.
  Chem. Phys.}\ }\textbf {\bibinfo {volume} {10}},\ \bibinfo {pages} {51}
  (\bibinfo {year} {1942})}\BibitemShut {NoStop}%
\bibitem [{\citenamefont {Tanaka}(1989)}]{tanaka1989}%
  \BibitemOpen
  \bibfield  {author} {\bibinfo {author} {\bibfnamefont {F.}~\bibnamefont
  {Tanaka}},\ }\href@noop {} {\bibfield  {journal} {\bibinfo  {journal}
  {Macromolecules}\ }\textbf {\bibinfo {volume} {22}},\ \bibinfo {pages} {1988}
  (\bibinfo {year} {1989})}\BibitemShut {NoStop}%
\bibitem [{\citenamefont {Tanaka}(1990)}]{tanaka1990thermodynamic}%
  \BibitemOpen
  \bibfield  {author} {\bibinfo {author} {\bibfnamefont {F.}~\bibnamefont
  {Tanaka}},\ }\href@noop {} {\bibfield  {journal} {\bibinfo  {journal}
  {Macromolecules}\ }\textbf {\bibinfo {volume} {23}},\ \bibinfo {pages} {3784}
  (\bibinfo {year} {1990})}\BibitemShut {NoStop}%
\bibitem [{\citenamefont {Tanaka}\ and\ \citenamefont
  {Stockmayer}(1994)}]{tanaka1994}%
  \BibitemOpen
  \bibfield  {author} {\bibinfo {author} {\bibfnamefont {F.}~\bibnamefont
  {Tanaka}}\ and\ \bibinfo {author} {\bibfnamefont {W.~H.}\ \bibnamefont
  {Stockmayer}},\ }\href@noop {} {\bibfield  {journal} {\bibinfo  {journal}
  {Macromolecules}\ }\textbf {\bibinfo {volume} {27}},\ \bibinfo {pages} {3943}
  (\bibinfo {year} {1994})}\BibitemShut {NoStop}%
\bibitem [{\citenamefont {Tanaka}\ and\ \citenamefont
  {Ishida}(1995)}]{tanaka1995}%
  \BibitemOpen
  \bibfield  {author} {\bibinfo {author} {\bibfnamefont {F.}~\bibnamefont
  {Tanaka}}\ and\ \bibinfo {author} {\bibfnamefont {M.}~\bibnamefont
  {Ishida}},\ }\href@noop {} {\bibfield  {journal} {\bibinfo  {journal} {J.
  Chem. Soc. Faraday Trans.}\ }\textbf {\bibinfo {volume} {91}},\ \bibinfo
  {pages} {2663} (\bibinfo {year} {1995})}\BibitemShut {NoStop}%
\bibitem [{\citenamefont {Ishida}\ and\ \citenamefont
  {Tanaka}(1997)}]{ishida1997}%
  \BibitemOpen
  \bibfield  {author} {\bibinfo {author} {\bibfnamefont {M.}~\bibnamefont
  {Ishida}}\ and\ \bibinfo {author} {\bibfnamefont {F.}~\bibnamefont
  {Tanaka}},\ }\href@noop {} {\bibfield  {journal} {\bibinfo  {journal}
  {Macromolecules}\ }\textbf {\bibinfo {volume} {30}},\ \bibinfo {pages} {3900}
  (\bibinfo {year} {1997})}\BibitemShut {NoStop}%
\bibitem [{\citenamefont {Tanaka}(1998)}]{tanaka1998}%
  \BibitemOpen
  \bibfield  {author} {\bibinfo {author} {\bibfnamefont {F.}~\bibnamefont
  {Tanaka}},\ }\href@noop {} {\bibfield  {journal} {\bibinfo  {journal}
  {Physica A: Statistical Mechanics and its Applications}\ }\textbf {\bibinfo
  {volume} {257}},\ \bibinfo {pages} {245} (\bibinfo {year}
  {1998})}\BibitemShut {NoStop}%
\bibitem [{\citenamefont {Tanaka}\ and\ \citenamefont
  {Ishida}(1999)}]{tanaka1999}%
  \BibitemOpen
  \bibfield  {author} {\bibinfo {author} {\bibfnamefont {F.}~\bibnamefont
  {Tanaka}}\ and\ \bibinfo {author} {\bibfnamefont {M.}~\bibnamefont
  {Ishida}},\ }\href@noop {} {\bibfield  {journal} {\bibinfo  {journal}
  {Macromolecules}\ }\textbf {\bibinfo {volume} {32}},\ \bibinfo {pages} {1271}
  (\bibinfo {year} {1999})}\BibitemShut {NoStop}%
\bibitem [{\citenamefont {Tanaka}(2002)}]{tanaka2002}%
  \BibitemOpen
  \bibfield  {author} {\bibinfo {author} {\bibfnamefont {F.}~\bibnamefont
  {Tanaka}},\ }\href@noop {} {\bibfield  {journal} {\bibinfo  {journal} {Polym.
  J.}\ }\textbf {\bibinfo {volume} {34}},\ \bibinfo {pages} {479} (\bibinfo
  {year} {2002})}\BibitemShut {NoStop}%
\bibitem [{\citenamefont {Borodin}\ \emph {et~al.}(2017)\citenamefont
  {Borodin}, \citenamefont {Suo}, \citenamefont {Gobet}, \citenamefont {Ren},
  \citenamefont {Wang}, \citenamefont {Faraone}, \citenamefont {Peng},
  \citenamefont {Olguin}, \citenamefont {Schroeder}, \citenamefont {Ding},
  \citenamefont {Gobroggem}, \citenamefont {Cresce}, \citenamefont {Munoz},
  \citenamefont {Dura}, \citenamefont {Greenbaum}, \citenamefont {Wang},\ and\
  \citenamefont {Xu}}]{borodin2017liquid}%
  \BibitemOpen
  \bibfield  {author} {\bibinfo {author} {\bibfnamefont {O.}~\bibnamefont
  {Borodin}}, \bibinfo {author} {\bibfnamefont {L.}~\bibnamefont {Suo}},
  \bibinfo {author} {\bibfnamefont {M.}~\bibnamefont {Gobet}}, \bibinfo
  {author} {\bibfnamefont {X.}~\bibnamefont {Ren}}, \bibinfo {author}
  {\bibfnamefont {F.}~\bibnamefont {Wang}}, \bibinfo {author} {\bibfnamefont
  {A.}~\bibnamefont {Faraone}}, \bibinfo {author} {\bibfnamefont
  {J.}~\bibnamefont {Peng}}, \bibinfo {author} {\bibfnamefont {M.}~\bibnamefont
  {Olguin}}, \bibinfo {author} {\bibfnamefont {M.}~\bibnamefont {Schroeder}},
  \bibinfo {author} {\bibfnamefont {M.~S.}\ \bibnamefont {Ding}}, \bibinfo
  {author} {\bibfnamefont {E.}~\bibnamefont {Gobroggem}}, \bibinfo {author}
  {\bibfnamefont {A.~v.~W.}\ \bibnamefont {Cresce}}, \bibinfo {author}
  {\bibfnamefont {S.}~\bibnamefont {Munoz}}, \bibinfo {author} {\bibfnamefont
  {J.~A.}\ \bibnamefont {Dura}}, \bibinfo {author} {\bibfnamefont
  {S.}~\bibnamefont {Greenbaum}}, \bibinfo {author} {\bibfnamefont
  {C.}~\bibnamefont {Wang}}, \ and\ \bibinfo {author} {\bibfnamefont
  {K.}~\bibnamefont {Xu}},\ }\href@noop {} {\bibfield  {journal} {\bibinfo
  {journal} {ACS nano}\ }\textbf {\bibinfo {volume} {11}},\ \bibinfo {pages}
  {10462} (\bibinfo {year} {2017})}\BibitemShut {NoStop}%
\bibitem [{\citenamefont {Choi}\ \emph {et~al.}(2018)\citenamefont {Choi},
  \citenamefont {Lee}, \citenamefont {Choi},\ and\ \citenamefont
  {Cho}}]{choi2018graph}%
  \BibitemOpen
  \bibfield  {author} {\bibinfo {author} {\bibfnamefont {J.-H.}\ \bibnamefont
  {Choi}}, \bibinfo {author} {\bibfnamefont {H.}~\bibnamefont {Lee}}, \bibinfo
  {author} {\bibfnamefont {H.~R.}\ \bibnamefont {Choi}}, \ and\ \bibinfo
  {author} {\bibfnamefont {M.}~\bibnamefont {Cho}},\ }\href@noop {} {\bibfield
  {journal} {\bibinfo  {journal} {Annu Rev Phys Chem}\ }\textbf {\bibinfo
  {volume} {69}},\ \bibinfo {pages} {125} (\bibinfo {year} {2018})}\BibitemShut
  {NoStop}%
\bibitem [{\citenamefont {Jeon}\ \emph {et~al.}(2020)\citenamefont {Jeon},
  \citenamefont {Lee}, \citenamefont {Choi},\ and\ \citenamefont
  {Cho}}]{jeon2020modeling}%
  \BibitemOpen
  \bibfield  {author} {\bibinfo {author} {\bibfnamefont {J.}~\bibnamefont
  {Jeon}}, \bibinfo {author} {\bibfnamefont {H.}~\bibnamefont {Lee}}, \bibinfo
  {author} {\bibfnamefont {J.-H.}\ \bibnamefont {Choi}}, \ and\ \bibinfo
  {author} {\bibfnamefont {M.}~\bibnamefont {Cho}},\ }\href@noop {} {\bibfield
  {journal} {\bibinfo  {journal} {The Journal of Physical Chemistry C}\ }
  (\bibinfo {year} {2020})}\BibitemShut {NoStop}%
\bibitem [{\citenamefont {Makino}\ \emph {et~al.}(2008)\citenamefont {Makino},
  \citenamefont {Kishikawa}, \citenamefont {Mizoshiri}, \citenamefont
  {Takeda},\ and\ \citenamefont {Yao}}]{makino2008viscoelastic}%
  \BibitemOpen
  \bibfield  {author} {\bibinfo {author} {\bibfnamefont {W.}~\bibnamefont
  {Makino}}, \bibinfo {author} {\bibfnamefont {R.}~\bibnamefont {Kishikawa}},
  \bibinfo {author} {\bibfnamefont {M.}~\bibnamefont {Mizoshiri}}, \bibinfo
  {author} {\bibfnamefont {S.}~\bibnamefont {Takeda}}, \ and\ \bibinfo {author}
  {\bibfnamefont {M.}~\bibnamefont {Yao}},\ }\href@noop {} {\bibfield
  {journal} {\bibinfo  {journal} {J. Chem. Phys.}\ }\textbf {\bibinfo {volume}
  {129}},\ \bibinfo {pages} {104510} (\bibinfo {year} {2008})}\BibitemShut
  {NoStop}%
\bibitem [{\citenamefont {Tao}\ and\ \citenamefont
  {Simon}(2015)}]{tao2015rheology}%
  \BibitemOpen
  \bibfield  {author} {\bibinfo {author} {\bibfnamefont {R.}~\bibnamefont
  {Tao}}\ and\ \bibinfo {author} {\bibfnamefont {S.~L.}\ \bibnamefont
  {Simon}},\ }\href@noop {} {\bibfield  {journal} {\bibinfo  {journal} {J.
  Phys. Chem. B}\ }\textbf {\bibinfo {volume} {119}},\ \bibinfo {pages} {11953}
  (\bibinfo {year} {2015})}\BibitemShut {NoStop}%
\bibitem [{\citenamefont {Elhamarnah}\ \emph {et~al.}(2019)\citenamefont
  {Elhamarnah}, \citenamefont {Nasser}, \citenamefont {Qiblawey}, \citenamefont
  {Benamor}, \citenamefont {Atilhan},\ and\ \citenamefont
  {Aparicio}}]{Elhamarnah2019}%
  \BibitemOpen
  \bibfield  {author} {\bibinfo {author} {\bibfnamefont {Y.~A.}\ \bibnamefont
  {Elhamarnah}}, \bibinfo {author} {\bibfnamefont {M.}~\bibnamefont {Nasser}},
  \bibinfo {author} {\bibfnamefont {H.}~\bibnamefont {Qiblawey}}, \bibinfo
  {author} {\bibfnamefont {A.}~\bibnamefont {Benamor}}, \bibinfo {author}
  {\bibfnamefont {M.}~\bibnamefont {Atilhan}}, \ and\ \bibinfo {author}
  {\bibfnamefont {S.}~\bibnamefont {Aparicio}},\ }\href@noop {} {\bibfield
  {journal} {\bibinfo  {journal} {J. Mol. Liq.}\ }\textbf {\bibinfo {volume}
  {277}},\ \bibinfo {pages} {932–} (\bibinfo {year} {2019})}\BibitemShut
  {NoStop}%
\bibitem [{\citenamefont {Shakeel}\ \emph {et~al.}(2019)\citenamefont
  {Shakeel}, \citenamefont {Mahmood}, \citenamefont {Farooq}, \citenamefont
  {Ullah}, \citenamefont {Yasin}, \citenamefont {Iqbal}, \citenamefont
  {Chassagne},\ and\ \citenamefont {Moniruzzaman}}]{Shakeel2019}%
  \BibitemOpen
  \bibfield  {author} {\bibinfo {author} {\bibfnamefont {A.}~\bibnamefont
  {Shakeel}}, \bibinfo {author} {\bibfnamefont {H.}~\bibnamefont {Mahmood}},
  \bibinfo {author} {\bibfnamefont {U.}~\bibnamefont {Farooq}}, \bibinfo
  {author} {\bibfnamefont {Z.}~\bibnamefont {Ullah}}, \bibinfo {author}
  {\bibfnamefont {S.}~\bibnamefont {Yasin}}, \bibinfo {author} {\bibfnamefont
  {T.}~\bibnamefont {Iqbal}}, \bibinfo {author} {\bibfnamefont
  {C.}~\bibnamefont {Chassagne}}, \ and\ \bibinfo {author} {\bibfnamefont
  {M.}~\bibnamefont {Moniruzzaman}},\ }\href@noop {} {\bibfield  {journal}
  {\bibinfo  {journal} {ACS Sustainable Chem. Eng.}\ }\textbf {\bibinfo
  {volume} {7}},\ \bibinfo {pages} {13586} (\bibinfo {year}
  {2019})}\BibitemShut {NoStop}%
\bibitem [{\citenamefont {Shamim}\ and\ \citenamefont
  {McKenna}(2010)}]{Shamim2010}%
  \BibitemOpen
  \bibfield  {author} {\bibinfo {author} {\bibfnamefont {N.}~\bibnamefont
  {Shamim}}\ and\ \bibinfo {author} {\bibfnamefont {G.~B.}\ \bibnamefont
  {McKenna}},\ }\href@noop {} {\bibfield  {journal} {\bibinfo  {journal} {J.
  Phys. Chem. B}\ }\textbf {\bibinfo {volume} {114}},\ \bibinfo {pages} {15742}
  (\bibinfo {year} {2010})}\BibitemShut {NoStop}%
\bibitem [{\citenamefont {Choi}\ \emph {et~al.}(2014)\citenamefont {Choi},
  \citenamefont {Ye}, \citenamefont {de~la Cruz}, \citenamefont {Liu},
  \citenamefont {Winey}, \citenamefont {Elabd}, \citenamefont {Runt},\ and\
  \citenamefont {Colby}}]{Choi2014}%
  \BibitemOpen
  \bibfield  {author} {\bibinfo {author} {\bibfnamefont {U.~H.}\ \bibnamefont
  {Choi}}, \bibinfo {author} {\bibfnamefont {Y.}~\bibnamefont {Ye}}, \bibinfo
  {author} {\bibfnamefont {D.~S.}\ \bibnamefont {de~la Cruz}}, \bibinfo
  {author} {\bibfnamefont {W.}~\bibnamefont {Liu}}, \bibinfo {author}
  {\bibfnamefont {K.~I.}\ \bibnamefont {Winey}}, \bibinfo {author}
  {\bibfnamefont {Y.~A.}\ \bibnamefont {Elabd}}, \bibinfo {author}
  {\bibfnamefont {J.}~\bibnamefont {Runt}}, \ and\ \bibinfo {author}
  {\bibfnamefont {R.~H.}\ \bibnamefont {Colby}},\ }\href@noop {} {\bibfield
  {journal} {\bibinfo  {journal} {Macromolecules}\ }\textbf {\bibinfo {volume}
  {47}},\ \bibinfo {pages} {777} (\bibinfo {year} {2014})}\BibitemShut
  {NoStop}%
\bibitem [{\citenamefont {Leonard}\ \emph {et~al.}(2018)\citenamefont
  {Leonard}, \citenamefont {Wei}, \citenamefont {Chen}, \citenamefont {Du},\
  and\ \citenamefont {Ji}}]{leonard2018}%
  \BibitemOpen
  \bibfield  {author} {\bibinfo {author} {\bibfnamefont {D.~P.}\ \bibnamefont
  {Leonard}}, \bibinfo {author} {\bibfnamefont {Z.}~\bibnamefont {Wei}},
  \bibinfo {author} {\bibfnamefont {G.}~\bibnamefont {Chen}}, \bibinfo {author}
  {\bibfnamefont {F.}~\bibnamefont {Du}}, \ and\ \bibinfo {author}
  {\bibfnamefont {X.}~\bibnamefont {Ji}},\ }\href@noop {} {\bibfield  {journal}
  {\bibinfo  {journal} {ACS Energy Lett.}\ }\textbf {\bibinfo {volume} {3}},\
  \bibinfo {pages} {373} (\bibinfo {year} {2018})}\BibitemShut {NoStop}%
\bibitem [{\citenamefont {Flory}(1953)}]{flory1953principles}%
  \BibitemOpen
  \bibfield  {author} {\bibinfo {author} {\bibfnamefont {P.~J.}\ \bibnamefont
  {Flory}},\ }\href@noop {} {\emph {\bibinfo {title} {Principles of polymer
  chemistry}}}\ (\bibinfo  {publisher} {Cornell University Press},\ \bibinfo
  {year} {1953})\BibitemShut {NoStop}%
\bibitem [{\citenamefont {Yu}\ \emph {et~al.}(2012)\citenamefont {Yu},
  \citenamefont {Beichel}, \citenamefont {Dlubek}, \citenamefont
  {Krause-Rehberg}, \citenamefont {Paluch}, \citenamefont {Pionteck},
  \citenamefont {Pfefferkorn}, \citenamefont {Bulut}, \citenamefont
  {Friedrich}, \citenamefont {Pogodina},\ and\ \citenamefont
  {Krossing}}]{Yu2012freevolume}%
  \BibitemOpen
  \bibfield  {author} {\bibinfo {author} {\bibfnamefont {Y.}~\bibnamefont
  {Yu}}, \bibinfo {author} {\bibfnamefont {W.}~\bibnamefont {Beichel}},
  \bibinfo {author} {\bibfnamefont {G.}~\bibnamefont {Dlubek}}, \bibinfo
  {author} {\bibfnamefont {R.}~\bibnamefont {Krause-Rehberg}}, \bibinfo
  {author} {\bibfnamefont {M.}~\bibnamefont {Paluch}}, \bibinfo {author}
  {\bibfnamefont {J.}~\bibnamefont {Pionteck}}, \bibinfo {author}
  {\bibfnamefont {D.}~\bibnamefont {Pfefferkorn}}, \bibinfo {author}
  {\bibfnamefont {S.}~\bibnamefont {Bulut}}, \bibinfo {author} {\bibfnamefont
  {C.}~\bibnamefont {Friedrich}}, \bibinfo {author} {\bibfnamefont
  {N.}~\bibnamefont {Pogodina}}, \ and\ \bibinfo {author} {\bibfnamefont
  {I.}~\bibnamefont {Krossing}},\ }\href@noop {} {\bibfield  {journal}
  {\bibinfo  {journal} {Phys. Chem. Chem. Phys.}\ }\textbf {\bibinfo {volume}
  {14}},\ \bibinfo {pages} {6856–6868} (\bibinfo {year} {2012})}\BibitemShut
  {NoStop}%
\bibitem [{\citenamefont {Stockmayer}(1952)}]{stockmayer1952molecular}%
  \BibitemOpen
  \bibfield  {author} {\bibinfo {author} {\bibfnamefont {W.~H.}\ \bibnamefont
  {Stockmayer}},\ }\href@noop {} {\bibfield  {journal} {\bibinfo  {journal} {J.
  Polym. Sci.}\ }\textbf {\bibinfo {volume} {9}},\ \bibinfo {pages} {69}
  (\bibinfo {year} {1952})}\BibitemShut {NoStop}%
\bibitem [{\citenamefont {Flory}(1941{\natexlab{a}})}]{flory1941molecular}%
  \BibitemOpen
  \bibfield  {author} {\bibinfo {author} {\bibfnamefont {P.~J.}\ \bibnamefont
  {Flory}},\ }\href@noop {} {\bibfield  {journal} {\bibinfo  {journal} {J. Am.
  Chem. Soc.}\ }\textbf {\bibinfo {volume} {63}},\ \bibinfo {pages} {3083}
  (\bibinfo {year} {1941}{\natexlab{a}})}\BibitemShut {NoStop}%
\bibitem [{\citenamefont {Flory}(1941{\natexlab{b}})}]{flory1941molecular2}%
  \BibitemOpen
  \bibfield  {author} {\bibinfo {author} {\bibfnamefont {P.~J.}\ \bibnamefont
  {Flory}},\ }\href@noop {} {\bibfield  {journal} {\bibinfo  {journal} {J. Am.
  Chem. Soc.}\ }\textbf {\bibinfo {volume} {63}},\ \bibinfo {pages} {3091}
  (\bibinfo {year} {1941}{\natexlab{b}})}\BibitemShut {NoStop}%
\bibitem [{\citenamefont {Onsager}(1931)}]{onsager1931reciprocal}%
  \BibitemOpen
  \bibfield  {author} {\bibinfo {author} {\bibfnamefont {L.}~\bibnamefont
  {Onsager}},\ }\href@noop {} {\bibfield  {journal} {\bibinfo  {journal} {Phys.
  Rev.}\ }\textbf {\bibinfo {volume} {37}},\ \bibinfo {pages} {405} (\bibinfo
  {year} {1931})}\BibitemShut {NoStop}%
\bibitem [{\citenamefont {Tanaka}(2011)}]{tanaka2011polymer}%
  \BibitemOpen
  \bibfield  {author} {\bibinfo {author} {\bibfnamefont {F.}~\bibnamefont
  {Tanaka}},\ }\href@noop {} {\emph {\bibinfo {title} {Polymer physics:
  applications to molecular association and thermoreversible gelation}}}\
  (\bibinfo  {publisher} {Cambridge University Press},\ \bibinfo {year}
  {2011})\BibitemShut {NoStop}%
\bibitem [{\citenamefont {Fuoss}\ and\ \citenamefont
  {Onsager}(1957)}]{fouss1957}%
  \BibitemOpen
  \bibfield  {author} {\bibinfo {author} {\bibfnamefont {R.~M.}\ \bibnamefont
  {Fuoss}}\ and\ \bibinfo {author} {\bibfnamefont {L.}~\bibnamefont
  {Onsager}},\ }\href@noop {} {\bibfield  {journal} {\bibinfo  {journal} {J.
  Phys. Chem.}\ }\textbf {\bibinfo {volume} {61}},\ \bibinfo {pages} {668}
  (\bibinfo {year} {1957})}\BibitemShut {NoStop}%
\bibitem [{\citenamefont {Gali{\'n}ski}\ \emph {et~al.}(2006)\citenamefont
  {Gali{\'n}ski}, \citenamefont {Lewandowski},\ and\ \citenamefont
  {Stepniak}}]{galinski2006ionic}%
  \BibitemOpen
  \bibfield  {author} {\bibinfo {author} {\bibfnamefont {M.}~\bibnamefont
  {Gali{\'n}ski}}, \bibinfo {author} {\bibfnamefont {A.}~\bibnamefont
  {Lewandowski}}, \ and\ \bibinfo {author} {\bibfnamefont {I.}~\bibnamefont
  {Stepniak}},\ }\href@noop {} {\bibfield  {journal} {\bibinfo  {journal}
  {Electrochim. acta}\ }\textbf {\bibinfo {volume} {51}},\ \bibinfo {pages}
  {5567} (\bibinfo {year} {2006})}\BibitemShut {NoStop}%
\bibitem [{\citenamefont {Kowsari}\ \emph {et~al.}(2008)\citenamefont
  {Kowsari}, \citenamefont {Alavi}, \citenamefont {Ashrafizaadeh},\ and\
  \citenamefont {Najafi}}]{kowsari2008molecular}%
  \BibitemOpen
  \bibfield  {author} {\bibinfo {author} {\bibfnamefont {M.}~\bibnamefont
  {Kowsari}}, \bibinfo {author} {\bibfnamefont {S.}~\bibnamefont {Alavi}},
  \bibinfo {author} {\bibfnamefont {M.}~\bibnamefont {Ashrafizaadeh}}, \ and\
  \bibinfo {author} {\bibfnamefont {B.}~\bibnamefont {Najafi}},\ }\href@noop {}
  {\bibfield  {journal} {\bibinfo  {journal} {J. Chem. Phys.}\ }\textbf
  {\bibinfo {volume} {129}},\ \bibinfo {pages} {224508} (\bibinfo {year}
  {2008})}\BibitemShut {NoStop}%
\bibitem [{\citenamefont {Gouverneur}\ \emph {et~al.}(2015)\citenamefont
  {Gouverneur}, \citenamefont {Kopp}, \citenamefont {van W{\"u}llen},\ and\
  \citenamefont {Sch{\"o}nhoff}}]{gouverneur2015direct}%
  \BibitemOpen
  \bibfield  {author} {\bibinfo {author} {\bibfnamefont {M.}~\bibnamefont
  {Gouverneur}}, \bibinfo {author} {\bibfnamefont {J.}~\bibnamefont {Kopp}},
  \bibinfo {author} {\bibfnamefont {L.}~\bibnamefont {van W{\"u}llen}}, \ and\
  \bibinfo {author} {\bibfnamefont {M.}~\bibnamefont {Sch{\"o}nhoff}},\
  }\href@noop {} {\bibfield  {journal} {\bibinfo  {journal} {Phys. Chem. Chem.
  Phys.}\ }\textbf {\bibinfo {volume} {17}},\ \bibinfo {pages} {30680}
  (\bibinfo {year} {2015})}\BibitemShut {NoStop}%
\bibitem [{\citenamefont {Brehm}\ and\ \citenamefont
  {Kirchner}(2011)}]{brehm2011travis}%
  \BibitemOpen
  \bibfield  {author} {\bibinfo {author} {\bibfnamefont {M.}~\bibnamefont
  {Brehm}}\ and\ \bibinfo {author} {\bibfnamefont {B.}~\bibnamefont
  {Kirchner}},\ }\href@noop {} {\bibfield  {journal} {\bibinfo  {journal} {J.
  Chem. Inf. Model.}\ }\textbf {\bibinfo {volume} {51}},\ \bibinfo {pages}
  {2007–2023} (\bibinfo {year} {2011})}\BibitemShut {NoStop}%
\bibitem [{\citenamefont {Fredlake}\ \emph {et~al.}(2004)\citenamefont
  {Fredlake}, \citenamefont {Crosthwaite}, \citenamefont {Hert}, \citenamefont
  {Aki},\ and\ \citenamefont {Brennecke}}]{fredlake2004thermophysical}%
  \BibitemOpen
  \bibfield  {author} {\bibinfo {author} {\bibfnamefont {C.~P.}\ \bibnamefont
  {Fredlake}}, \bibinfo {author} {\bibfnamefont {J.~M.}\ \bibnamefont
  {Crosthwaite}}, \bibinfo {author} {\bibfnamefont {D.~G.}\ \bibnamefont
  {Hert}}, \bibinfo {author} {\bibfnamefont {S.~N.}\ \bibnamefont {Aki}}, \
  and\ \bibinfo {author} {\bibfnamefont {J.~F.}\ \bibnamefont {Brennecke}},\
  }\href@noop {} {\bibfield  {journal} {\bibinfo  {journal} {J. Chem. Eng.
  Data}\ }\textbf {\bibinfo {volume} {49}},\ \bibinfo {pages} {954} (\bibinfo
  {year} {2004})}\BibitemShut {NoStop}%
\bibitem [{\citenamefont {Leys}\ \emph {et~al.}(2008)\citenamefont {Leys},
  \citenamefont {W{\"u}bbenhorst}, \citenamefont {Preethy~Menon}, \citenamefont
  {Rajesh}, \citenamefont {Thoen}, \citenamefont {Glorieux}, \citenamefont
  {Nockemann}, \citenamefont {Thijs}, \citenamefont {Binnemans},\ and\
  \citenamefont {Longuemart}}]{leys2008temperature}%
  \BibitemOpen
  \bibfield  {author} {\bibinfo {author} {\bibfnamefont {J.}~\bibnamefont
  {Leys}}, \bibinfo {author} {\bibfnamefont {M.}~\bibnamefont
  {W{\"u}bbenhorst}}, \bibinfo {author} {\bibfnamefont {C.}~\bibnamefont
  {Preethy~Menon}}, \bibinfo {author} {\bibfnamefont {R.}~\bibnamefont
  {Rajesh}}, \bibinfo {author} {\bibfnamefont {J.}~\bibnamefont {Thoen}},
  \bibinfo {author} {\bibfnamefont {C.}~\bibnamefont {Glorieux}}, \bibinfo
  {author} {\bibfnamefont {P.}~\bibnamefont {Nockemann}}, \bibinfo {author}
  {\bibfnamefont {B.}~\bibnamefont {Thijs}}, \bibinfo {author} {\bibfnamefont
  {K.}~\bibnamefont {Binnemans}}, \ and\ \bibinfo {author} {\bibfnamefont
  {S.}~\bibnamefont {Longuemart}},\ }\href@noop {} {\bibfield  {journal}
  {\bibinfo  {journal} {J. Chem. Phys.}\ }\textbf {\bibinfo {volume} {128}},\
  \bibinfo {pages} {064509} (\bibinfo {year} {2008})}\BibitemShut {NoStop}%
\bibitem [{\citenamefont {Gibbs}\ and\ \citenamefont
  {DiMarzio}(1958)}]{gibbs1958nature}%
  \BibitemOpen
  \bibfield  {author} {\bibinfo {author} {\bibfnamefont {J.~H.}\ \bibnamefont
  {Gibbs}}\ and\ \bibinfo {author} {\bibfnamefont {E.~A.}\ \bibnamefont
  {DiMarzio}},\ }\href@noop {} {\bibfield  {journal} {\bibinfo  {journal} {J.
  Chem. Phys.}\ }\textbf {\bibinfo {volume} {28}},\ \bibinfo {pages} {373}
  (\bibinfo {year} {1958})}\BibitemShut {NoStop}%
\bibitem [{\citenamefont {Dupont}(2004)}]{Dupont2004}%
  \BibitemOpen
  \bibfield  {author} {\bibinfo {author} {\bibfnamefont {J.}~\bibnamefont
  {Dupont}},\ }\href@noop {} {\bibfield  {journal} {\bibinfo  {journal} {J.
  Braz. Chem. Soc.}\ }\textbf {\bibinfo {volume} {3}},\ \bibinfo {pages} {341}
  (\bibinfo {year} {2004})}\BibitemShut {NoStop}%
\bibitem [{\citenamefont {Molinari}\ \emph
  {et~al.}(2019{\natexlab{a}})\citenamefont {Molinari}, \citenamefont {Mailoa},
  \citenamefont {Craig}, \citenamefont {Christensen},\ and\ \citenamefont
  {Kozinsky}}]{molinari2019transport}%
  \BibitemOpen
  \bibfield  {author} {\bibinfo {author} {\bibfnamefont {N.}~\bibnamefont
  {Molinari}}, \bibinfo {author} {\bibfnamefont {J.~P.}\ \bibnamefont
  {Mailoa}}, \bibinfo {author} {\bibfnamefont {N.}~\bibnamefont {Craig}},
  \bibinfo {author} {\bibfnamefont {J.}~\bibnamefont {Christensen}}, \ and\
  \bibinfo {author} {\bibfnamefont {B.}~\bibnamefont {Kozinsky}},\ }\href@noop
  {} {\bibfield  {journal} {\bibinfo  {journal} {J. Power Sources}\ }\textbf
  {\bibinfo {volume} {428}},\ \bibinfo {pages} {27} (\bibinfo {year}
  {2019}{\natexlab{a}})}\BibitemShut {NoStop}%
\bibitem [{\citenamefont {Lopes}\ and\ \citenamefont
  {P\'adua}(2006)}]{Lopes2006}%
  \BibitemOpen
  \bibfield  {author} {\bibinfo {author} {\bibfnamefont {J.~N. A.~C.}\
  \bibnamefont {Lopes}}\ and\ \bibinfo {author} {\bibfnamefont {A.~A.~H.}\
  \bibnamefont {P\'adua}},\ }\href@noop {} {\bibfield  {journal} {\bibinfo
  {journal} {J. Phys. Chem. B}\ }\textbf {\bibinfo {volume} {110}},\ \bibinfo
  {pages} {3330} (\bibinfo {year} {2006})}\BibitemShut {NoStop}%
\bibitem [{\citenamefont {Wang}\ and\ \citenamefont {Voth}(2005)}]{Wang2005}%
  \BibitemOpen
  \bibfield  {author} {\bibinfo {author} {\bibfnamefont {Y.}~\bibnamefont
  {Wang}}\ and\ \bibinfo {author} {\bibfnamefont {G.~A.}\ \bibnamefont
  {Voth}},\ }\href@noop {} {\bibfield  {journal} {\bibinfo  {journal} {J. Am.
  Chem. Soc.}\ }\textbf {\bibinfo {volume} {35}},\ \bibinfo {pages} {12192}
  (\bibinfo {year} {2005})}\BibitemShut {NoStop}%
\bibitem [{\citenamefont {Bernardes}\ \emph {et~al.}(2011)\citenamefont
  {Bernardes}, \citenamefont {da~Piedade},\ and\ \citenamefont
  {Lopes}}]{Bernardes2011}%
  \BibitemOpen
  \bibfield  {author} {\bibinfo {author} {\bibfnamefont {C.~E.~S.}\
  \bibnamefont {Bernardes}}, \bibinfo {author} {\bibfnamefont {M.~E.~M.}\
  \bibnamefont {da~Piedade}}, \ and\ \bibinfo {author} {\bibfnamefont
  {J.~N.~C.}\ \bibnamefont {Lopes}},\ }\href@noop {} {\bibfield  {journal}
  {\bibinfo  {journal} {J. Phys. Chem. B}\ }\textbf {\bibinfo {volume} {115}},\
  \bibinfo {pages} {2067} (\bibinfo {year} {2011})}\BibitemShut {NoStop}%
\bibitem [{\citenamefont {Russo}\ \emph {et~al.}(2009)\citenamefont {Russo},
  \citenamefont {Tartaglia},\ and\ \citenamefont {Sciortino}}]{Russo2009}%
  \BibitemOpen
  \bibfield  {author} {\bibinfo {author} {\bibfnamefont {J.}~\bibnamefont
  {Russo}}, \bibinfo {author} {\bibfnamefont {P.}~\bibnamefont {Tartaglia}}, \
  and\ \bibinfo {author} {\bibfnamefont {F.}~\bibnamefont {Sciortino}},\
  }\href@noop {} {\bibfield  {journal} {\bibinfo  {journal} {J. Chem. Phys.}\
  }\textbf {\bibinfo {volume} {131}},\ \bibinfo {pages} {014504} (\bibinfo
  {year} {2009})}\BibitemShut {NoStop}%
\bibitem [{\citenamefont {Smallenburg}\ and\ \citenamefont
  {Sciortino}(2013)}]{Smallenburg2013}%
  \BibitemOpen
  \bibfield  {author} {\bibinfo {author} {\bibfnamefont {F.}~\bibnamefont
  {Smallenburg}}\ and\ \bibinfo {author} {\bibfnamefont {F.}~\bibnamefont
  {Sciortino}},\ }\href@noop {} {\bibfield  {journal} {\bibinfo  {journal}
  {Nat. Phys.}\ }\textbf {\bibinfo {volume} {9}},\ \bibinfo {pages} {554}
  (\bibinfo {year} {2013})}\BibitemShut {NoStop}%
\bibitem [{\citenamefont {Sturlaugson}\ \emph {et~al.}(2012)\citenamefont
  {Sturlaugson}, \citenamefont {Fruchey},\ and\ \citenamefont
  {Fayer}}]{Sturlaugson2012}%
  \BibitemOpen
  \bibfield  {author} {\bibinfo {author} {\bibfnamefont {A.~L.}\ \bibnamefont
  {Sturlaugson}}, \bibinfo {author} {\bibfnamefont {K.~S.}\ \bibnamefont
  {Fruchey}}, \ and\ \bibinfo {author} {\bibfnamefont {M.~D.}\ \bibnamefont
  {Fayer}},\ }\href@noop {} {\bibfield  {journal} {\bibinfo  {journal} {J.
  Phys. Chem. B}\ }\textbf {\bibinfo {volume} {116}},\ \bibinfo {pages} {1777}
  (\bibinfo {year} {2012})}\BibitemShut {NoStop}%
\bibitem [{\citenamefont {Sturlaugson}\ \emph {et~al.}(2013)\citenamefont
  {Sturlaugson}, \citenamefont {Arima}, \citenamefont {Bailey},\ and\
  \citenamefont {Fayer}}]{Sturlaugson2013}%
  \BibitemOpen
  \bibfield  {author} {\bibinfo {author} {\bibfnamefont {A.~L.}\ \bibnamefont
  {Sturlaugson}}, \bibinfo {author} {\bibfnamefont {A.~Y.}\ \bibnamefont
  {Arima}}, \bibinfo {author} {\bibfnamefont {H.~E.}\ \bibnamefont {Bailey}}, \
  and\ \bibinfo {author} {\bibfnamefont {M.~D.}\ \bibnamefont {Fayer}},\
  }\href@noop {} {\bibfield  {journal} {\bibinfo  {journal} {J. Phys. Chem. B}\
  }\textbf {\bibinfo {volume} {117}},\ \bibinfo {pages} {14775} (\bibinfo
  {year} {2013})}\BibitemShut {NoStop}%
\bibitem [{\citenamefont {Chaban}\ \emph {et~al.}(2012)\citenamefont {Chaban},
  \citenamefont {Voroshylova}, \citenamefont {Kalugin},\ and\ \citenamefont
  {Prezhdo}}]{chaban2012acetonitrile}%
  \BibitemOpen
  \bibfield  {author} {\bibinfo {author} {\bibfnamefont {V.~V.}\ \bibnamefont
  {Chaban}}, \bibinfo {author} {\bibfnamefont {I.~V.}\ \bibnamefont
  {Voroshylova}}, \bibinfo {author} {\bibfnamefont {O.~N.}\ \bibnamefont
  {Kalugin}}, \ and\ \bibinfo {author} {\bibfnamefont {O.~V.}\ \bibnamefont
  {Prezhdo}},\ }\href@noop {} {\bibfield  {journal} {\bibinfo  {journal} {J.
  Phys. Chem. B}\ }\textbf {\bibinfo {volume} {116}},\ \bibinfo {pages} {7719}
  (\bibinfo {year} {2012})}\BibitemShut {NoStop}%
\bibitem [{\citenamefont {Singh}\ \emph {et~al.}(2019)\citenamefont {Singh},
  \citenamefont {Koch},\ and\ \citenamefont {Singh}}]{Singh2019}%
  \BibitemOpen
  \bibfield  {author} {\bibinfo {author} {\bibfnamefont {W.~P.}\ \bibnamefont
  {Singh}}, \bibinfo {author} {\bibfnamefont {U.}~\bibnamefont {Koch}}, \ and\
  \bibinfo {author} {\bibfnamefont {R.~S.}\ \bibnamefont {Singh}},\ }\href@noop
  {} {\bibfield  {journal} {\bibinfo  {journal} {Soft Mater.}\ ,\ \bibinfo
  {pages} {1}} (\bibinfo {year} {2019})}\BibitemShut {NoStop}%
\bibitem [{\citenamefont {Li}\ \emph {et~al.}(2020)\citenamefont {Li},
  \citenamefont {Wang}, \citenamefont {Chen}, \citenamefont {Xu},\ and\
  \citenamefont {Lu}}]{li2020new}%
  \BibitemOpen
  \bibfield  {author} {\bibinfo {author} {\bibfnamefont {M.}~\bibnamefont
  {Li}}, \bibinfo {author} {\bibfnamefont {C.}~\bibnamefont {Wang}}, \bibinfo
  {author} {\bibfnamefont {Z.}~\bibnamefont {Chen}}, \bibinfo {author}
  {\bibfnamefont {K.}~\bibnamefont {Xu}}, \ and\ \bibinfo {author}
  {\bibfnamefont {J.}~\bibnamefont {Lu}},\ }\href@noop {} {\bibfield  {journal}
  {\bibinfo  {journal} {Chem. Rev.}\ }\textbf {\bibinfo {volume} {120}},\
  \bibinfo {pages} {6783–} (\bibinfo {year} {2020})}\BibitemShut {NoStop}%
\bibitem [{\citenamefont {Di~Noto}\ \emph {et~al.}(2011)\citenamefont
  {Di~Noto}, \citenamefont {Lavina}, \citenamefont {Giffin}, \citenamefont
  {Negro},\ and\ \citenamefont {Scrosati}}]{di2011polymer}%
  \BibitemOpen
  \bibfield  {author} {\bibinfo {author} {\bibfnamefont {V.}~\bibnamefont
  {Di~Noto}}, \bibinfo {author} {\bibfnamefont {S.}~\bibnamefont {Lavina}},
  \bibinfo {author} {\bibfnamefont {G.~A.}\ \bibnamefont {Giffin}}, \bibinfo
  {author} {\bibfnamefont {E.}~\bibnamefont {Negro}}, \ and\ \bibinfo {author}
  {\bibfnamefont {B.}~\bibnamefont {Scrosati}},\ }\href@noop {} {\bibfield
  {journal} {\bibinfo  {journal} {Electrochim. Acta}\ }\textbf {\bibinfo
  {volume} {57}},\ \bibinfo {pages} {4} (\bibinfo {year} {2011})}\BibitemShut
  {NoStop}%
\bibitem [{\citenamefont {Lui}\ \emph {et~al.}(2011)\citenamefont {Lui},
  \citenamefont {Crowhurst}, \citenamefont {Hallett}, \citenamefont {Hunt},
  \citenamefont {Niedermeyer},\ and\ \citenamefont {Welton}}]{lui2011salts}%
  \BibitemOpen
  \bibfield  {author} {\bibinfo {author} {\bibfnamefont {M.~Y.}\ \bibnamefont
  {Lui}}, \bibinfo {author} {\bibfnamefont {L.}~\bibnamefont {Crowhurst}},
  \bibinfo {author} {\bibfnamefont {J.~P.}\ \bibnamefont {Hallett}}, \bibinfo
  {author} {\bibfnamefont {P.~A.}\ \bibnamefont {Hunt}}, \bibinfo {author}
  {\bibfnamefont {H.}~\bibnamefont {Niedermeyer}}, \ and\ \bibinfo {author}
  {\bibfnamefont {T.}~\bibnamefont {Welton}},\ }\href@noop {} {\bibfield
  {journal} {\bibinfo  {journal} {Chem. Sci.}\ }\textbf {\bibinfo {volume}
  {2}},\ \bibinfo {pages} {1491} (\bibinfo {year} {2011})}\BibitemShut
  {NoStop}%
\bibitem [{\citenamefont {Suo}\ \emph {et~al.}(2016)\citenamefont {Suo},
  \citenamefont {Borodin}, \citenamefont {Sun}, \citenamefont {Fan},
  \citenamefont {Yang}, \citenamefont {Wang}, \citenamefont {Gao},
  \citenamefont {Ma}, \citenamefont {Schroeder}, \citenamefont {von Cresce},
  \citenamefont {Russell}, \citenamefont {Armand}, \citenamefont {Angell},
  \citenamefont {Xu},\ and\ \citenamefont {Wang}}]{Suo2016}%
  \BibitemOpen
  \bibfield  {author} {\bibinfo {author} {\bibfnamefont {L.}~\bibnamefont
  {Suo}}, \bibinfo {author} {\bibfnamefont {O.}~\bibnamefont {Borodin}},
  \bibinfo {author} {\bibfnamefont {W.}~\bibnamefont {Sun}}, \bibinfo {author}
  {\bibfnamefont {X.}~\bibnamefont {Fan}}, \bibinfo {author} {\bibfnamefont
  {C.}~\bibnamefont {Yang}}, \bibinfo {author} {\bibfnamefont {F.}~\bibnamefont
  {Wang}}, \bibinfo {author} {\bibfnamefont {T.}~\bibnamefont {Gao}}, \bibinfo
  {author} {\bibfnamefont {Z.}~\bibnamefont {Ma}}, \bibinfo {author}
  {\bibfnamefont {M.}~\bibnamefont {Schroeder}}, \bibinfo {author}
  {\bibfnamefont {A.}~\bibnamefont {von Cresce}}, \bibinfo {author}
  {\bibfnamefont {S.~M.}\ \bibnamefont {Russell}}, \bibinfo {author}
  {\bibfnamefont {M.}~\bibnamefont {Armand}}, \bibinfo {author} {\bibfnamefont
  {A.}~\bibnamefont {Angell}}, \bibinfo {author} {\bibfnamefont
  {K.}~\bibnamefont {Xu}}, \ and\ \bibinfo {author} {\bibfnamefont
  {C.}~\bibnamefont {Wang}},\ }\href@noop {} {\bibfield  {journal} {\bibinfo
  {journal} {Angew. Chem. Int. Ed.}\ }\textbf {\bibinfo {volume} {55}},\
  \bibinfo {pages} {7136} (\bibinfo {year} {2016})}\BibitemShut {NoStop}%
\bibitem [{\citenamefont {Molinari}\ \emph
  {et~al.}(2019{\natexlab{b}})\citenamefont {Molinari}, \citenamefont
  {Mailoa},\ and\ \citenamefont {Kozinsky}}]{molinari2019general}%
  \BibitemOpen
  \bibfield  {author} {\bibinfo {author} {\bibfnamefont {N.}~\bibnamefont
  {Molinari}}, \bibinfo {author} {\bibfnamefont {J.~P.}\ \bibnamefont
  {Mailoa}}, \ and\ \bibinfo {author} {\bibfnamefont {B.}~\bibnamefont
  {Kozinsky}},\ }\href@noop {} {\bibfield  {journal} {\bibinfo  {journal} {J.
  Phys. Chem. Lett.}\ }\textbf {\bibinfo {volume} {10}},\ \bibinfo {pages}
  {2313} (\bibinfo {year} {2019}{\natexlab{b}})}\BibitemShut {NoStop}%
\bibitem [{\citenamefont {Chen}\ \emph {et~al.}(2020)\citenamefont {Chen},
  \citenamefont {Zhang}, \citenamefont {Li}, \citenamefont {Vatamanu},
  \citenamefont {Ji}, \citenamefont {Pollard}, \citenamefont {Cui},
  \citenamefont {Hou}, \citenamefont {Chen}, \citenamefont {Yang},
  \citenamefont {Ma}, \citenamefont {Ding}, \citenamefont {Garaga},
  \citenamefont {Greenbaum}, \citenamefont {Lee}, \citenamefont {Borodin},
  \citenamefont {Xu},\ and\ \citenamefont {Wang}}]{chen202063}%
  \BibitemOpen
  \bibfield  {author} {\bibinfo {author} {\bibfnamefont {L.}~\bibnamefont
  {Chen}}, \bibinfo {author} {\bibfnamefont {J.}~\bibnamefont {Zhang}},
  \bibinfo {author} {\bibfnamefont {Q.}~\bibnamefont {Li}}, \bibinfo {author}
  {\bibfnamefont {J.}~\bibnamefont {Vatamanu}}, \bibinfo {author}
  {\bibfnamefont {X.}~\bibnamefont {Ji}}, \bibinfo {author} {\bibfnamefont
  {T.~P.}\ \bibnamefont {Pollard}}, \bibinfo {author} {\bibfnamefont
  {C.}~\bibnamefont {Cui}}, \bibinfo {author} {\bibfnamefont {S.}~\bibnamefont
  {Hou}}, \bibinfo {author} {\bibfnamefont {J.}~\bibnamefont {Chen}}, \bibinfo
  {author} {\bibfnamefont {C.}~\bibnamefont {Yang}}, \bibinfo {author}
  {\bibfnamefont {L.}~\bibnamefont {Ma}}, \bibinfo {author} {\bibfnamefont
  {M.~S.}\ \bibnamefont {Ding}}, \bibinfo {author} {\bibfnamefont
  {M.}~\bibnamefont {Garaga}}, \bibinfo {author} {\bibfnamefont
  {S.}~\bibnamefont {Greenbaum}}, \bibinfo {author} {\bibfnamefont {H.-S.}\
  \bibnamefont {Lee}}, \bibinfo {author} {\bibfnamefont {O.}~\bibnamefont
  {Borodin}}, \bibinfo {author} {\bibfnamefont {K.}~\bibnamefont {Xu}}, \ and\
  \bibinfo {author} {\bibfnamefont {C.}~\bibnamefont {Wang}},\ }\href@noop {}
  {\bibfield  {journal} {\bibinfo  {journal} {ACS Energy Lett.}\ }\textbf
  {\bibinfo {volume} {5}},\ \bibinfo {pages} {968} (\bibinfo {year}
  {2020})}\BibitemShut {NoStop}%
\bibitem [{\citenamefont {Jiang}\ \emph {et~al.}(2020)\citenamefont {Jiang},
  \citenamefont {Liu}, \citenamefont {Yue}, \citenamefont {Zhang},
  \citenamefont {Zhou}, \citenamefont {Borodin}, \citenamefont {Suo},
  \citenamefont {Li}, \citenamefont {Chen}, \citenamefont {Xu},\ and\
  \citenamefont {Hu}}]{jiang2020high}%
  \BibitemOpen
  \bibfield  {author} {\bibinfo {author} {\bibfnamefont {L.}~\bibnamefont
  {Jiang}}, \bibinfo {author} {\bibfnamefont {L.}~\bibnamefont {Liu}}, \bibinfo
  {author} {\bibfnamefont {J.}~\bibnamefont {Yue}}, \bibinfo {author}
  {\bibfnamefont {Q.}~\bibnamefont {Zhang}}, \bibinfo {author} {\bibfnamefont
  {A.}~\bibnamefont {Zhou}}, \bibinfo {author} {\bibfnamefont {O.}~\bibnamefont
  {Borodin}}, \bibinfo {author} {\bibfnamefont {L.}~\bibnamefont {Suo}},
  \bibinfo {author} {\bibfnamefont {H.}~\bibnamefont {Li}}, \bibinfo {author}
  {\bibfnamefont {L.}~\bibnamefont {Chen}}, \bibinfo {author} {\bibfnamefont
  {K.}~\bibnamefont {Xu}}, \ and\ \bibinfo {author} {\bibfnamefont
  {Y.}~\bibnamefont {Hu}},\ }\href@noop {} {\bibfield  {journal} {\bibinfo
  {journal} {ADV MATER}\ }\textbf {\bibinfo {volume} {32}},\ \bibinfo {pages}
  {1904427} (\bibinfo {year} {2020})}\BibitemShut {NoStop}%
\bibitem [{\citenamefont {Wang}\ \emph {et~al.}(2018)\citenamefont {Wang},
  \citenamefont {Borodin}, \citenamefont {Ding}, \citenamefont {Gobet},
  \citenamefont {Vatamanu}, \citenamefont {Fan}, \citenamefont {Gao},
  \citenamefont {Edison}, \citenamefont {Liang}, \citenamefont {Sun},
  \citenamefont {Greenbaum}, \citenamefont {Xu},\ and\ \citenamefont
  {Wang}}]{wang2018hybrid}%
  \BibitemOpen
  \bibfield  {author} {\bibinfo {author} {\bibfnamefont {F.}~\bibnamefont
  {Wang}}, \bibinfo {author} {\bibfnamefont {O.}~\bibnamefont {Borodin}},
  \bibinfo {author} {\bibfnamefont {M.~S.}\ \bibnamefont {Ding}}, \bibinfo
  {author} {\bibfnamefont {M.}~\bibnamefont {Gobet}}, \bibinfo {author}
  {\bibfnamefont {J.}~\bibnamefont {Vatamanu}}, \bibinfo {author}
  {\bibfnamefont {X.}~\bibnamefont {Fan}}, \bibinfo {author} {\bibfnamefont
  {T.}~\bibnamefont {Gao}}, \bibinfo {author} {\bibfnamefont {N.}~\bibnamefont
  {Edison}}, \bibinfo {author} {\bibfnamefont {Y.}~\bibnamefont {Liang}},
  \bibinfo {author} {\bibfnamefont {W.}~\bibnamefont {Sun}}, \bibinfo {author}
  {\bibfnamefont {S.}~\bibnamefont {Greenbaum}}, \bibinfo {author}
  {\bibfnamefont {K.}~\bibnamefont {Xu}}, \ and\ \bibinfo {author}
  {\bibfnamefont {C.}~\bibnamefont {Wang}},\ }\href@noop {} {\bibfield
  {journal} {\bibinfo  {journal} {Joule}\ }\textbf {\bibinfo {volume} {2}},\
  \bibinfo {pages} {927} (\bibinfo {year} {2018})}\BibitemShut {NoStop}%
\bibitem [{\citenamefont {Zhang}\ \emph {et~al.}(2018)\citenamefont {Zhang},
  \citenamefont {Qin}, \citenamefont {Han},\ and\ \citenamefont
  {Passerini}}]{zhang2018aqueous}%
  \BibitemOpen
  \bibfield  {author} {\bibinfo {author} {\bibfnamefont {H.}~\bibnamefont
  {Zhang}}, \bibinfo {author} {\bibfnamefont {B.}~\bibnamefont {Qin}}, \bibinfo
  {author} {\bibfnamefont {J.}~\bibnamefont {Han}}, \ and\ \bibinfo {author}
  {\bibfnamefont {S.}~\bibnamefont {Passerini}},\ }\href@noop {} {\bibfield
  {journal} {\bibinfo  {journal} {ACS Energy Lett.}\ }\textbf {\bibinfo
  {volume} {3}},\ \bibinfo {pages} {1769} (\bibinfo {year} {2018})}\BibitemShut
  {NoStop}%
\bibitem [{\citenamefont {Dou}\ \emph {et~al.}(2018)\citenamefont {Dou},
  \citenamefont {Lei}, \citenamefont {Wang}, \citenamefont {Zhang},
  \citenamefont {Xiao}, \citenamefont {Guo}, \citenamefont {Wang},
  \citenamefont {Yang}, \citenamefont {Li}, \citenamefont {Shi}, ,\ and\
  \citenamefont {Yan}}]{dou2018safe}%
  \BibitemOpen
  \bibfield  {author} {\bibinfo {author} {\bibfnamefont {Q.}~\bibnamefont
  {Dou}}, \bibinfo {author} {\bibfnamefont {S.}~\bibnamefont {Lei}}, \bibinfo
  {author} {\bibfnamefont {D.-W.}\ \bibnamefont {Wang}}, \bibinfo {author}
  {\bibfnamefont {Q.}~\bibnamefont {Zhang}}, \bibinfo {author} {\bibfnamefont
  {D.}~\bibnamefont {Xiao}}, \bibinfo {author} {\bibfnamefont {H.}~\bibnamefont
  {Guo}}, \bibinfo {author} {\bibfnamefont {A.}~\bibnamefont {Wang}}, \bibinfo
  {author} {\bibfnamefont {H.}~\bibnamefont {Yang}}, \bibinfo {author}
  {\bibfnamefont {Y.}~\bibnamefont {Li}}, \bibinfo {author} {\bibfnamefont
  {S.}~\bibnamefont {Shi}}, , \ and\ \bibinfo {author} {\bibfnamefont
  {X.}~\bibnamefont {Yan}},\ }\href@noop {} {\bibfield  {journal} {\bibinfo
  {journal} {Energy Environ. Sci.}\ }\textbf {\bibinfo {volume} {11}},\
  \bibinfo {pages} {3212} (\bibinfo {year} {2018})}\BibitemShut {NoStop}%
\bibitem [{\citenamefont {Dou}\ \emph {et~al.}(2019)\citenamefont {Dou},
  \citenamefont {Lu}, \citenamefont {Su}, \citenamefont {Zhang}, \citenamefont
  {Lei}, \citenamefont {Bu}, \citenamefont {Liu}, \citenamefont {Xiao},
  \citenamefont {Chen}, \citenamefont {Shi},\ and\ \citenamefont
  {Yan}}]{Dou2019}%
  \BibitemOpen
  \bibfield  {author} {\bibinfo {author} {\bibfnamefont {Q.}~\bibnamefont
  {Dou}}, \bibinfo {author} {\bibfnamefont {Y.}~\bibnamefont {Lu}}, \bibinfo
  {author} {\bibfnamefont {L.}~\bibnamefont {Su}}, \bibinfo {author}
  {\bibfnamefont {X.}~\bibnamefont {Zhang}}, \bibinfo {author} {\bibfnamefont
  {S.}~\bibnamefont {Lei}}, \bibinfo {author} {\bibfnamefont {X.}~\bibnamefont
  {Bu}}, \bibinfo {author} {\bibfnamefont {L.}~\bibnamefont {Liu}}, \bibinfo
  {author} {\bibfnamefont {D.}~\bibnamefont {Xiao}}, \bibinfo {author}
  {\bibfnamefont {J.}~\bibnamefont {Chen}}, \bibinfo {author} {\bibfnamefont
  {S.}~\bibnamefont {Shi}}, \ and\ \bibinfo {author} {\bibfnamefont
  {X.}~\bibnamefont {Yan}},\ }\href@noop {} {\bibfield  {journal} {\bibinfo
  {journal} {Energy Storage Mater.}\ }\textbf {\bibinfo {volume} {23}},\
  \bibinfo {pages} {603} (\bibinfo {year} {2019})}\BibitemShut {NoStop}%
\bibitem [{\citenamefont {Molinari}\ and\ \citenamefont
  {Kozinsky}(2020)}]{molinari2020chelation}%
  \BibitemOpen
  \bibfield  {author} {\bibinfo {author} {\bibfnamefont {N.}~\bibnamefont
  {Molinari}}\ and\ \bibinfo {author} {\bibfnamefont {B.}~\bibnamefont
  {Kozinsky}},\ }\href@noop {} {\bibfield  {journal} {\bibinfo  {journal} {J.
  Phys. Chem. B}\ }\textbf {\bibinfo {volume} {124}},\ \bibinfo {pages} {2676}
  (\bibinfo {year} {2020})}\BibitemShut {NoStop}%
\bibitem [{\citenamefont {Ebeling}\ and\ \citenamefont
  {Grigo}(1980)}]{ebeling1980analytical}%
  \BibitemOpen
  \bibfield  {author} {\bibinfo {author} {\bibfnamefont {W.}~\bibnamefont
  {Ebeling}}\ and\ \bibinfo {author} {\bibfnamefont {M.}~\bibnamefont
  {Grigo}},\ }\href@noop {} {\bibfield  {journal} {\bibinfo  {journal} {Annalen
  der Physik}\ }\textbf {\bibinfo {volume} {492}},\ \bibinfo {pages} {21}
  (\bibinfo {year} {1980})}\BibitemShut {NoStop}%
\bibitem [{\citenamefont {Levin}\ and\ \citenamefont
  {Fisher}(1996)}]{levin1996criticality}%
  \BibitemOpen
  \bibfield  {author} {\bibinfo {author} {\bibfnamefont {Y.}~\bibnamefont
  {Levin}}\ and\ \bibinfo {author} {\bibfnamefont {M.~E.}\ \bibnamefont
  {Fisher}},\ }\href@noop {} {\bibfield  {journal} {\bibinfo  {journal}
  {Physica A: Statistical Mechanics and its Applications}\ }\textbf {\bibinfo
  {volume} {225}},\ \bibinfo {pages} {164} (\bibinfo {year}
  {1996})}\BibitemShut {NoStop}%
\bibitem [{\citenamefont {Bjerrum}(1926)}]{bjerrum1926k}%
  \BibitemOpen
  \bibfield  {author} {\bibinfo {author} {\bibfnamefont {N.}~\bibnamefont
  {Bjerrum}},\ }\href@noop {} {\enquote {\bibinfo {title} {K. danske vidensk.
  selsk.}}\ } (\bibinfo {year} {1926})\BibitemShut {NoStop}%
\bibitem [{\citenamefont {Flory}(1956)}]{flory1956statistical}%
  \BibitemOpen
  \bibfield  {author} {\bibinfo {author} {\bibfnamefont {P.-J.}\ \bibnamefont
  {Flory}},\ }\href@noop {} {\bibfield  {journal} {\bibinfo  {journal}
  {Proceedings of the Royal Society of London. Series A. Mathematical and
  Physical Sciences}\ }\textbf {\bibinfo {volume} {234}},\ \bibinfo {pages}
  {60} (\bibinfo {year} {1956})}\BibitemShut {NoStop}%
\bibitem [{\citenamefont {Plimpton}(1995)}]{plimpton1995}%
  \BibitemOpen
  \bibfield  {author} {\bibinfo {author} {\bibfnamefont {S.}~\bibnamefont
  {Plimpton}},\ }\href {\doibase https://doi.org/10.1006/jcph.1995.1039}
  {\bibfield  {journal} {\bibinfo  {journal} {J. Comput. Phys.}\ }\textbf
  {\bibinfo {volume} {117}},\ \bibinfo {pages} {1 } (\bibinfo {year}
  {1995})}\BibitemShut {NoStop}%
\bibitem [{\citenamefont {Mart{\'\i}nez}\ \emph {et~al.}(2009)\citenamefont
  {Mart{\'\i}nez}, \citenamefont {Andrade}, \citenamefont {Birgin},\ and\
  \citenamefont {Mart{\'\i}nez}}]{martinez2009packmol}%
  \BibitemOpen
  \bibfield  {author} {\bibinfo {author} {\bibfnamefont {L.}~\bibnamefont
  {Mart{\'\i}nez}}, \bibinfo {author} {\bibfnamefont {R.}~\bibnamefont
  {Andrade}}, \bibinfo {author} {\bibfnamefont {E.~G.}\ \bibnamefont {Birgin}},
  \ and\ \bibinfo {author} {\bibfnamefont {J.~M.}\ \bibnamefont
  {Mart{\'\i}nez}},\ }\href@noop {} {\bibfield  {journal} {\bibinfo  {journal}
  {J. Comput. Chem.}\ }\textbf {\bibinfo {volume} {30}},\ \bibinfo {pages}
  {2157} (\bibinfo {year} {2009})}\BibitemShut {NoStop}%
\bibitem [{\citenamefont {Humphrey}\ \emph {et~al.}(1996)\citenamefont
  {Humphrey}, \citenamefont {Dalke},\ and\ \citenamefont {Schulten}}]{HUMP96}%
  \BibitemOpen
  \bibfield  {author} {\bibinfo {author} {\bibfnamefont {W.}~\bibnamefont
  {Humphrey}}, \bibinfo {author} {\bibfnamefont {A.}~\bibnamefont {Dalke}}, \
  and\ \bibinfo {author} {\bibfnamefont {K.}~\bibnamefont {Schulten}},\
  }\href@noop {} {\bibfield  {journal} {\bibinfo  {journal} {J. Mol. Graphics}\
  }\textbf {\bibinfo {volume} {14}},\ \bibinfo {pages} {33} (\bibinfo {year}
  {1996})}\BibitemShut {NoStop}%
\bibitem [{\citenamefont {Canongia~Lopes}\ and\ \citenamefont
  {P{\'a}dua}(2012)}]{CanongiaLopes2012}%
  \BibitemOpen
  \bibfield  {author} {\bibinfo {author} {\bibfnamefont {J.~N.}\ \bibnamefont
  {Canongia~Lopes}}\ and\ \bibinfo {author} {\bibfnamefont {A.~A.~H.}\
  \bibnamefont {P{\'a}dua}},\ }\href {\doibase 10.1007/s00214-012-1129-7}
  {\bibfield  {journal} {\bibinfo  {journal} {Theor. Chem. Acc.}\ }\textbf
  {\bibinfo {volume} {131}},\ \bibinfo {pages} {1129} (\bibinfo {year}
  {2012})}\BibitemShut {NoStop}%
\end{thebibliography}%

\end{document}